\documentclass[prd,aps,floatfix]{revtex4}

\usepackage{graphicx}
\usepackage{amsmath}
\usepackage[psamsfonts]{amssymb}
\usepackage{longtable}
\newcommand{\be}{\begin{equation}}
\newcommand{\ee}{\end{equation}}
\newcommand{\bea}{\begin{eqnarray}}
\newcommand{\eea}{\end{eqnarray}}
\newcommand{\bi}{\begin{itemize}}
\newcommand{\ei}{\end{itemize}}

\newcommand{\bc}{\begin{center}}
\newcommand{\ec}{\end{center}}

\newcommand{\tr}{{\rm Tr}}
\newcommand{\Pslash}{p \kern -2mm /}
\newcommand{\Btob}{B\rightarrow b}
\newcommand{\btoB}{b\rightarrow B}
\newcommand{\XZtoSP}{\Xi^0\rightarrow \Sigma^+}
\newcommand{\XtoS}{\Xi\rightarrow \Sigma}
\newcommand{\ntop}{n\rightarrow p}

\begin{document}

%-- top page ---------------------------------------------------

%-- preprint number ---
%\PrePrint{YITP-08-86}
\vspace*{-10mm}
\begin{flushright}
TKYNT-08-27, YITP-08-86
\end{flushright}

%-- title ---
\title{
Lattice study of flavor SU(3) breaking in hyperon beta decay
}

%-- author list ---
\author{Shoichi Sasaki}
\email{ssasaki@phys.s.u-tokyo.ac.jp}
\affiliation{Department of Physics, The University of Tokyo, \\
Hongo 7-3-1, Tokyo 113-0033, Japan}

\author{Takeshi Yamazaki}
\email{tyamazak@yukawa.kyoto-u.ac.jp}
\affiliation{Physics Department, University of Connecticut, \\
Storrs, Connecticut 06269-3046, USA}   
\affiliation{Yukawa Institute for Theoretical Physics, Kyoto University, \\
Kitashirakawa-Oiwakecho, Sakyo, Kyoto 606-8502, Japan}

\date{\today}

%-- abstract ---
\begin{abstract}
We present a quenched lattice calculation of all six
form factors: vector [$f_1(q^2)$], weak magnetism [$f_2(q^2)$], 
induced scalar [$f_3(q^2)$], axial-vector [$g_1(q^2)$], 
weak electricity [$g_2(q^2)$], and induce pseudoscalar 
[$g_3(q^2)$] form factors, in hyperon semileptonic 
decay $\Xi^0 \rightarrow \Sigma^{+} l\overline{\nu}$ using 
domain wall fermions. The $q^2$ dependences of all form factors
in the relatively low $q^2$ region are examined in order to evaluate their 
values at zero momentum transfer. 
The $\XZtoSP$ transition is highly sensitive to flavor $SU(3)$ breaking 
since this decay corresponds to the direct analogue of neutron beta decay 
under the exchange of the down quark with the strange quark.
The pattern of flavor $SU(3)$ breaking effects in the hyperon beta decay is 
easily exposed in a comparison to results for neutron beta decay.
We measure $SU(3)$-breaking corrections to $f_1(0)$, $f_2(0)/f_1(0)$ 
and $g_1(0)/f_1(0)$.
A sign of the leading order corrections, of which the size is less than 
a few \%, on $f_1(0)$ is likely negative, while $f_2(0)/f_1(0)$ 
and $g_1(0)/f_1(0)$ receive positive corrections of order 16\% 
and 5\% respectively.
The observed pattern of the deviation from the values in the exact
$SU(3)$ limit 
does not support some of model estimates. We show that there are nonzero 
second-class form factors in the $\XZtoSP$ decay, measuring $f_3(0)/f_1(0)=0.14(9)$ and $g_2(0)/g_1(0)=0.68(18)$, which are 
comparable to the size of first-order $SU(3)$ breaking. 
It is also found that the $SU(3)$ breaking effect on $g_3(0)/g_1(0)$ 
agrees with the prediction of the generalized pion-pole dominance. 

\end{abstract}


\maketitle

%---------------------------------------------------------------------
%\tableofcontents
%\newpage
%---------------------------------------------------------------------

%--- main text -------------------------------------------------------
 
%%%%%%%%%%%%%%  SEC 1  %%%%%%%%%%%%%%%%%%%%%%%%
%\clearpage
\section{Introduction}
\label{Sec:intro}
The latest lattice calculations of $K\rightarrow \pi  l\nu$ ($K_{l3}$) semileptonic decays have been greatly developed 
with high precision~\cite{{Boyle:2007qe},{Becirevic:2004ya}, {Dawson:2006qc}}. 
The results for the vector form factor $f_{+}(0)$ of $K_{l3}$ decays, 
which deviates from unity due to flavor $SU(3)$ breaking, can be used
to provide a very precise determination of the element $V_{us}$ 
of the Cabibbo-Kobayashi-Maskawa (CKM) matrix~\cite{{Cabibbo:1963yz},{Kobayashi:1973fv}}. 
Those first-principles calculations contribute greatly to 
a stringent test of the CKM unitary through the first row relation $|V_{ud}|^2+|V_{us}|^2+|V_{ub}|^2=1$~\cite{Marciano:2007zz}.
On the other hand, $\Delta S=1$ semileptonic hyperon decays  
provide alternative determinations of $|V_{us}|$~\cite{Leutwyler:1984je}. 
The consistency of the values of $|V_{us}|$ determined
from different experiments needs to be confirmed.
As we will explain later, however, the determination of $|V_{us}|$ 
from the hyperon decays should be affected by larger theoretical 
uncertainties than those of $K_{l3}$ decays~\cite{Mateu:2005wi}. 
This is simply because the lack of a reliable theoretical calculation 
of the leading symmetry-breaking corrections in the hyperon beta decays. 
Indeed, the conventional Cabibbo model~\cite{Kobayashi:1973fv}, 
where flavor $SU(3)$ breaking effects are ignored, is commonly used for 
the analysis of the hyperon beta-decay data~\cite{Cabibbo:2003cu}. 

An essential difference from the case of $K_{l3}$ decays is
that the axial-vector current also contributes to the transition in the 
hyperon beta decays. Thus, the precise determination of $|V_{us}|$ in the hyperon beta decays requires information of the ratio of the axial-vector to vector form factors $g_1(0)/f_1(0)$ in addition to the vector form factor $f_1(0)$~\cite{Cabibbo:2003cu}.
Here we recall that the various ratios $g_1(0)/f_1(0)$ in the hyperon beta 
decays also provide vital information to analysis of strange quark spin fraction
of the proton spin, together with the polarized deep inelastic scattering data~\cite{Filippone:2001ux}.
However, such analysis heavily relies on the Cabibbo model.
The flavor $SU(3)$ breaking introduces the systematic 
uncertainty on the strange quark contribution to the proton spin.
This issue is still under debate~\cite{Ratcliffe:2004jt}. 
Recently, a new analysis, where the first-order corrections of both 
$SU(3)$ and $SU(2)$ symmetry breaking are properly taken into account
within the Cabibbo model, was proposed by Yamanishi~\cite{Yamanishi:2007zza}. The author has reported that flavor breaking effects significantly affect the 
evaluation of the amount of spin carried by the strange quarks inside the proton~\cite{Yamanishi:2007zza}.

In the hyperon decays, a model independent evaluation of
$SU(3)$-breaking corrections is highly demanded for both the 
CKM unitary test and the proton spin problem.
The weak matrix element of the hyperon beta decays can be 
calculated with high accuracy from first principles using the 
techniques of lattice QCD, similar to what is achieved in 
the case of $K_{l3}$ 
decays~\cite{{Boyle:2007qe},{Becirevic:2004ya}, {Dawson:2006qc}}. 
There is a single lattice study to be 
completed for a specific hyperon decay, $\Sigma^- \rightarrow n  l\overline{\nu}$.
The lattice simulation have been performed by Guadagnoli {\it et al.}
with ${\cal O}(a)$-improved  Wilson fermions in the quenched 
approximation~\cite{Guadagnoli:2006gj}~\footnote{
%%%%%%%%% footnote %%%%%%%
Their obtained value of $[g_1(0)/f_1(0)]_{\Sigma\rightarrow n}=-0.287(52)$ is rather large compared to the experimental value, $-0.340(17)$~\cite{Cabibbo:2003cu}. This discrepancy 
could be attributed to the serious finite volume effect on the axial-vector coupling $g_1(0)$, which was observed in the case of neutron beta decay. Note that 
they utilized about $(1.8\;{\rm fm})^3$ volume, which is not enough large
at least for neutron beta decay~\cite{{Sasaki:2003jh},{Sasaki:2007gw}}.}.
%%%%%%%%%%%%%%%%%%%%%
As we are interested primarily in flavor $SU(3)$ breaking effects {\it not} in the absolute values of each form factor, we choose a different
hyperon decay, $\Xi^0 \rightarrow \Sigma^{+} l\overline{\nu}$ in this study. 
This is simply because 
the $\XZtoSP$ transition is the direct analogue of neutron beta decay under the exchange of the down quark with the strange quark. 
The flavor $SU(3)$-breaking can be easily exposed through a comparison with
results of neutron beta decay. In Ref.~\cite{Sasaki:2003jh}, for the axial-vector
coupling of neutron beta decay, $g_1(0)=1.212(27)$ in the chiral limit is 
obtained from quenched lattice QCD calculations with domain wall fermions 
(DWFs). It underestimates the experimental value of 1.2695(29) by less than 5\%. 
Other relevant weak form factors have been subsequently 
investigated in Ref.~\cite{Sasaki:2007gw}.
We naturally extend the quenched DWF calculation for investigating 
$SU(3)$-breaking corrections to the hyperon semileptonic decay form factors.

Our paper is organized as follows. 
In Sec.~\ref{Sec:hbd}, we first present a brief introduction of the hyperon
beta decays. In Sec.~\ref{Sec:numeric},
details of our Monte Carlo simulations and some basic results
are given. We also describe the lattice method for calculating
the baryon beta-decay form factors. 
Section~\ref{Sec:f_1} is devoted to our determination of
the scalar form factor $f_S(q^2)$, which 
will be defined in the next section, at finite momentum transfer.
We describe the interpolation of the form factor to 
zero momentum transfer and the chiral extrapolation 
in order to evaluate the value of $f_1(0)$ at the physical point.
We also present our estimate of $|V_{us}|$. 
Section~\ref{Sec:ratio_g_1_ov_f_1} gives the result of the ratio 
of $g_1(0)/f_1(0)$ at the physical point. We then discuss flavor 
$SU(3)$-breaking effects appeared in $g_1(0)/f_1(0)$
for the $\XZtoSP$ transition in comparison to neutron beta decay. 
The results for the other form factors
including the second-class form factors $f_3$ and $g_2$ 
are presented in Sec.~\ref{Sec:other_FF}.
Finally, in Sec.~\ref{Sec:summary}, we summarize our results
and conclusions.

%%%%%%%%%%%% Section II %%%%%%%%%%%%%%%%%%%
%\clearpage
\section{Hyperon beta decays}
\label{Sec:hbd}

The general form of the baryon matrix element for semileptonic
decays $B \rightarrow bl\overline{\nu}$ is given by 
both the vector and axial-vector transitions:
%
% eq.
%
\be
%\langle b(p') | J_{\alpha}^{\rm wk}(x)| B(p) \rangle 
\langle b(p') | V_{\alpha}(x)+A_{\alpha}(x) | B(p) \rangle 
={\overline u}_{b}(p') 
\left(
{\cal O}^{V}_{\alpha}(q)+{\cal O}^{A}_{\alpha}(q) 
\right)
u_{B}(p) e^{iq\cdot x},
\ee
where $q\equiv p-p'$ is the momentum transfer between 
the initial state ($B$) and the final state ($b$) which
belong to the lightest $J^P=1/2^{+}$ $SU(3)$ octet of 
baryons ($p, n, \Lambda,\Sigma, \Xi$). The vector and axial-vector
currents are defined as $V_{\alpha}(x)={\bar u}(x)\gamma_{\alpha}d(x)$
and $A_{\alpha}(x)={\bar u}(x)\gamma_{\alpha}\gamma_5d(x)$
for $\Delta S=0$ decays, and $V_{\alpha}(x)={\bar u}(x)\gamma_{\alpha}s(x)$
and $A_{\alpha}(x)={\bar u}(x)\gamma_{\alpha}\gamma_5s(x)$
for $\Delta S=1$ decays.
Six form factors are needed to describe the hyperon beta decays: the vector ($f_1$), weak magnetism ($f_2$) and induced scalar ($f_3$) form factors 
for the vector current,
%
% eq.
%
\be
{\cal O}^{V}_{\alpha}(q) 
= \gamma_{\alpha} f_1^{\Btob}(q^2) + \sigma_{\alpha \beta}q_{\beta} \frac{f_2^{\Btob}(q^2)}{M_{B}+M_{b}}+iq_{\alpha}\frac{f_3^{\Btob}(q^2)}{M_{B}+M_{b}}
\label{Eq:VcMat}
\ee
and the axial-vector ($g_1$), weak electricity ($g_2$) and induced pseudo-scalar ($g_3$) from factors for the axial current,
%
% eq
%
\be
{\cal O}^{A}_{\alpha}(q)
= \gamma_{\alpha}\gamma_5 g_1^{\Btob}(q^2) 
+ \sigma_{\alpha \beta}q_{\beta}\gamma_5\frac{g_2^{\Btob}(q^2)}{M_{B}+M_{b}} 
+iq_{\alpha} \gamma_5 \frac{g_3^{\Btob}(q^2)}{M_{B}+M_{b}},
\label{Eq:AxMat}
\ee
which are here given in the Euclidean metric convention
(we have defined 
$\sigma_{\alpha \beta}=\frac{1}{2i}[\gamma_{\alpha}, \gamma_{\beta}]$.)~\footnote{
%%%% footnote %%%%
Remark that our $\gamma_5$ definition $\gamma_5\equiv\gamma_x\gamma_y\gamma_z\gamma_t=-\gamma_5^M$
has the opposite sign relative to that in the Minkowski convention 
($\vec{\gamma}^M=i\vec{\gamma}$ and $\gamma_0^{M}=\gamma_t$)
adopted in the particle data group. 
In addition, $q^2$ denoted in this paper, which stands for Euclidean four-momentum
squared, corresponds to the timelike momentum squared as $q_M^2=-q^2<0$
in Minkowski space. }.
%%%%%%%%%%%%%
Here, $M_B$ ($M_b$) denotes the rest mass of the initial (final) state.
Note that although the sign convention of the $f_3$ and $g_3$ form factors is opposite in comparison with that of Ref.~\cite{{Cabibbo:2003cu},{Alavi-Harati:2001xk}}, in our convention both $g_1$ and $g_3$ form factors
are positively defined for neutron beta decay.
%the sign of $f_1$, $f_2$, $g_1$ and $g_3$ for
%neutron beta decay is chosen to be positive in our convention. 
In addition, our adopted normalization of $1/(M_B+M_b)$, instead of $1/M_B$ that adopted in experiments, is theoretically preferable for considering the time-reversal symmetry on the matrix elements~\footnote{
%%%% footnote %%%%
Note that in our convention
the time-reversal process, $b \rightarrow B$, gives
the relation for the first-class form factors 
$f_{1,2}^{\Btob}(q^2)=f_{1,2}^{\btoB}(q^2)$ and 
$g_{1,3}^{\Btob}(q^2)=g_{1,3}^{\btoB}(q^2)$, while
$f_3^{\Btob}(q^2)=-f_3^{\btoB}(q^2)$ and
$g_2^{\Btob}(q^2)=-g_2^{\btoB}(q^2)$
for the first-class form factors~\cite{Weinberg:1958ut}.}.
%%%%%%%%%%%%%

For convenience in later discussion, we consider the scalar form factor 
$f_S(q^2)$ for the vector-current form factors given in Eq.~(\ref{Eq:VcMat}):
\be
f^{\Btob}_S(q^2)\equiv f^{\Btob}_1(q^2)+\frac{q^2}{M_B^2-M_b^2}f^{\Btob}_3(q^2),
\label{Eq:ScalarF}
\ee
which can be defined through the matrix element of the divergence of 
the vector current as $\langle b(p')| \partial_{\alpha}V_{\alpha}(0)|B(p)\rangle 
= (M_b- M_B)f_S(q^2) \bar{u}_{b}(p^{\prime})u_{B}(p)$ (see Appendix A for details), and also introduce a particular linear combination of the 
axial-vector-current form factors given in Eq.~(\ref{Eq:AxMat}) as
\be
\tilde{g}_1^{\Btob}(q^2)\equiv g_1^{\Btob}(q^2)-\frac{M_B-M_b}{M_B+M_b}g_2^{\Btob}(q^2),
\label{Eq:TildeG}
\ee
which is defined in an alternative parametrization of 
${\cal O}^{A}_{\alpha}(q)$ (see Appendix B for details).
Both $f_S$ and ${\tilde g}_1$ form factors are relevant
in lattice calculations~\cite{Guadagnoli:2006gj}.

In the literature, the vector and the axial-vector form factors at zero momentum transfer are called the vector coupling $g_V=f_1(0)$ and 
the axial-vector coupling $g_A=g_1(0)$, respectively. 
According to Weinberg's classification~\cite{Weinberg:1958ut}, the terms $f_3$ and $g_2$ are known 
as the second-class form factors, which are identically zero in the certain symmetric limit (iso-spin symmetry, $U$-spin symmetry or $V$-spin symmetry as 
$SU(2)$ subgroups of the flavor $SU(3)$ symmetry) within
the standard model.
For an example, the second-class form factors 
in $\Delta S=0$ decays such as neutron beta decay 
are prohibited from having nonzero values because of $G$-parity conservation 
in the iso-spin symmetry limit ($m_u=m_d$)~\cite{Cabibbo:2003cu}.
For $\Delta S=1$ decays, the $V$-spin symmetry ($m_d=m_s$) 
plays a similar role instead of the iso-spin symmetry.
Observation of nonzero second-class form factors corresponds to the direct 
signal of flavor $SU(3)$-breaking effects in the hyperon beta decays.
On the other hand, the terms $f_3$ and $g_3$ are suppressed 
in the evaluation of the beta-decay transition amplitude by a factor 
$[m_l/(M_B + M_b)]^2$ where $m_l$ is the charged lepton mass.
Therefore, for the decay $B\rightarrow be\bar{\nu}_e$, their contributions
can be safely ignored. As no accurate experiment has yet been performed 
on muonic hyperon decays, it is hard to access information of 
$f_3$ and $g_3$ form factors in present experiments.

In the exact $SU(3)$ limit, the vector couplings are simply 
given by $SU(3)$ Clebsch-Gordan coefficients as 
$f_1(0)=f_{klm}$, while the axial-vector couplings 
$g_1(0)$ are governed by two parameters $F$ and $D$
as $g_1(0)=Ff_{klm}+Dd_{klm}$~\cite{{Cabibbo:1963yz},{Cabibbo:2003cu}} ($F=0.475(4)$ and $D=0.793(5)$ are quoted in Ref.~\cite{Yamanishi:2007zza} 
for the conventional Cabibbo fit). Here, $d_{klm}$ is the totally 
symmetric tensor of the $SU(3)$ group.
Moreover, the conserved vector current (CVC) hypothesis 
becomes valid in this limit. The vector part of the weak current is 
a conserved current like the electromagnetic current. 
Thus, the value of $f_2(0)$ is described by 
$f_2(0)=(\kappa_p-\kappa_n)f_{klm}-3\kappa_n d_{klm}$. 
where $\kappa_p$ and $\kappa_n$ represent the anomalous 
magnetic moments of the proton and neutron. 
In the case of neutron beta decay,
we get $f^{\ntop}_1(0)=1$, $g^{\ntop}_1(0)=F+D$ and $f^{\ntop}_2(0)=\kappa_{p}-\kappa_{n}$, respectively. 
For the $\XZtoSP$ decay, the exact $SU(3)$ symmetry predicts that $f^{\XtoS}_1(0)$, $g^{\XtoS}_1(0)$ and $f^{\XtoS}_2(0)$
are identical to those of neutron beta decay.

The experimental rate of the hyperon beta decays, 
$B\rightarrow bl\bar{\nu}$, is given by
\be
\Gamma=\frac{G_F^2}{60\pi^3}(M_B-M_b)^5(1-3\delta)
|V_{us}|^2|f^{\Btob}_1(0)|^2\left[1+3
\left|\frac{g^{\Btob}_1(0)}{f^{\Btob}_1(0)}\right|^2
%\left(3-4\delta\frac{g_2(0)}{g_1(0)}\right)
+ \cdot\cdot\cdot
\right],
\ee
where $G_F$ denotes the Fermi constant. The ellipsis can be expressed in terms of a power series in the small quantity $\delta=(M_B-M_b)/(M_B+M_b)$, which is regarded as the size of flavor
$SU(3)$ breaking~\cite{Gaillard:1984ny}. 
The first linear term, which should be given by 
$-4\delta [g_2(0)g_1(0)/f_1(0)^2]_{\Btob}$, 
is safely ignored as small as ${\cal O}(\delta^2)$ since the nonzero value of 
the second-class form factor $g_2$ should be induced 
at first order of the $\delta$ expansion~\cite{Gaillard:1984ny}. 
The absolute value of $g_1(0)/f_1(0)$ can be determined
by measured asymmetries such as electron-neutrino 
correlation~\cite{{Cabibbo:2003cu},{Gaillard:1984ny}}.
Therefore a theoretical estimate of $f_1(0)$ is primarily required for the precise 
determination of $|V_{us}|$. First of all, the value of $f_1(0)$ should 
be equal to the $SU(3)$ Clebsch-Gordan coefficient up to the second 
order in $SU(3)$ symmetry breaking, thanks to the Ademollo-Gatto 
theorem~\cite{Ademollo:1964sr}. 
As the mass splitting among octet baryons is typically 
of the order of 10-15\%, the expected size of the second-order 
corrections is a few percent level. However, either the size, or the sign of 
the second-order corrections are somewhat controversial among various 
theoretical studies at present as summarized in Table~\ref{Tab:Th_estimate_f_1}. 

In the bag-model~\cite{Donoghue:1981uk} and 
quark-model calculations~\cite{{Donoghue:1986th},{Schlumpf:1994fb}}, 
flavor $SU(3)$-breaking effects on $f_1(0)$ are mainly 
accounted for wave-function mismatches 
between strange and non-strange quarks. Both models predict a small 
negative correction. On the other hand, the $1/N_c$ expansion approach 
including $SU(3)$ symmetry breaking up to the second order 
predicts a relatively large and positive correction. Recently, the full
one-loop ${\cal O}(p^4)$ calculation in heavy baryon chiral perturbation 
theory (HBChPT) was completed by Villadoro~\cite{Villadoro:2006nj}. 
However, the author emphasized that the $SU(3)$ version of HBChPT 
does not seem to be of help for the determination of $f_1(0)$. 
This is because a slow convergence of the chiral expansion
is observed. It is also pointed out that a serious convergence problem
is revealed by the inclusion of spin-3/2 decuplet degrees of freedom
into the framework of HBChPT. Subsequently, the complete one-loop order result 
has been checked in a different regularization scheme, covariant 
baryon chiral perturbation theory (CBChPT)~\cite{Lacour:2007wm}.
Both the size and the sign of the second order corrections are found to be
different from results of HBChPT. The authors of Ref.~\cite{Lacour:2007wm}
have estimated partial corrections of ${\cal O}(p^5)$ in HBChPT  
and then reconfirmed that the convergence behavior of $SU(3)$ baryon chiral
perturbation seems to be problematic as pointed out previously
in Ref.~\cite{Villadoro:2006nj}.
Unlike the case of $K_{l3}$ decays, the reliability of the chiral perturbation 
approach is questionable for the hyperon decays. A model independent 
estimate of $f_1(0)$ is highly required to settle both the size, and the sign 
of the second order corrections on $f_1(0)$.

The leading correction to the axial-vector coupling $g_1(0)$ starts 
at first order in symmetry breaking, while flavor $SU(3)$-breaking effects
to $f_1(0)$ are suppressed in first order by the Ademollo-Gatto theorem~\cite{Ademollo:1964sr}. Therefore, sizable breaking corrections, 
which are of the order of 10\% estimated from the mass splitting in the octet 
baryons, are to be expected in the ratio of $g_1(0)/f_1(0)$. 
However, the current experimental precision is not enough
to provide conclusive evidence of the violation of a two-parameter ($F$ and $D$)
fit based on the conventional Cabibbo model to ratios $g_1(0)/f_1(0)$ 
measured in various hyperon decays~\cite{Cabibbo:2003cu}. 
As mentioned earlier, the $\XZtoSP$ beta decay is highly sensitive to 
flavor $SU(3)$ breaking since the ratio $g_1(0)/f_1(0)$ of this particular 
decay should be identical 
to that of neutron beta decay if the flavor $SU(3)$ symmetry is manifest. 
Therefore, flavor $SU(3)$-breaking effects may be easily exposed in the 
$\XZtoSP$ process. Indeed, the center-of-mass correction approach~\cite{Ratcliffe:1998su} and the $1/N_c$ expansion approach~\cite{Flores-Mendieta:1998ii} predict that 
the $[g_1(0)/f_1(0)]_{\XtoS}$ is smaller than 
the $[g_1(0)/f_1(0)]_{\ntop}$ by 8-10\% and 20-30\% respectively.
Such sizable breaking corrections could be distinguishable in experiment. 
However, the first and single experiment done by the KTeV 
collaboration at Fermilab showed no indication of flavor $SU(3)$-breaking effects 
on $g_1(0)/f_1(0)$, measuring 
$[g_1(0)/f_1(0)]_{\XtoS}=1.32\pm^{0.21}_{0.17}$~\cite{Alavi-Harati:2001xk}. 
The KTeV experiment reported no evidence for a nonzero second-class form 
factor $g_2$~\cite{Alavi-Harati:2001xk} within their experimental precision.
The value of $f_2(0)/f_1(0)$ have been also measured 
in the KTeV experiment using the electron energy spectrum. 
Their observed value, $[f_2(0)/f_1(0)]_{\XtoS}=3.8\pm2.3$, 
seems to be consistent with that of neutron beta decay as
$[f_2(0)/f_1(0)]_{\ntop}=\kappa_p-\kappa_n=3.706$. 
Needless to say, its error is too large to discriminate either the exact $SU(3)$ 
value or other theoretical predictions. See Table~\ref{Tab:Th_estimate_f_2}, 
where several theoretical predictions of the value $[f_2(0)]_{\XtoS}$ are compiled. 

In this context, one would tend to conclude that the predictions of 
the exact $SU(3)$ symmetry limit to hold better in the case of hyperons. 
Indeed, it is true that the mass splitting for hyperons is rather small 
compared to mesons. Nevertheless, as we will show from our lattice simulations, 
this is indeed {\it not} the case. 

%
% Models for f_1
%
\begin{table}[htbp]
\caption{
Theoretical uncertainties of $f_1(0)$ for the  $\XZtoSP$ transition process. 
}
\begin{ruledtabular}
\begin{tabular}{llc}
\hline
type of result & $[f_1(0)]_{\XZtoSP}$ & Reference\\ 
\hline
bag model &0.97 & \cite{Donoghue:1981uk} \\
quark model & 0.987 & \cite{Donoghue:1986th}\\
quark model & 0.976 & \cite{Schlumpf:1994fb}\\
$1/N_c$ expansion & 1.12$\pm$0.05 & \cite{Flores-Mendieta:1998ii} \\
full ${\cal O}(p^4)$ HBChPT  & 1.009~\footnote{
The value is obtained  by the iso-spin relation
from that of the $\Xi^- \rightarrow \Sigma^0$ transition process.} 
& \cite{Villadoro:2006nj}\\
full ${\cal O}(p^4)$+partial ${\cal O}(p^5)$ HBChPT & 1.004$\pm$0.026  &  \cite{Lacour:2007wm} \\
full ${\cal O}(p^4)$ CBChPT & 0.944$\pm$0.016 & \cite{Lacour:2007wm} \\
\hline
\end{tabular}
\end{ruledtabular}
\label{Tab:Th_estimate_f_1}
\end{table}
%

%
% Models for f_2
%
\begin{table}[htbp]
\caption{
Theoretical predictions of $f_2(0)$ for the $\XZtoSP$ transition process. 
For evaluations, we use current values of the anomalous magnetic moments of $\Sigma$ and $\Xi$ baryons~\cite{Amsler:2008zz}.
}
\begin{ruledtabular}
\begin{tabular}{  l  l  l  l  c  |}
\hline
type of evaluation (Ref.)  & formula & $[f_2(0)]_{\XZtoSP}$ &$[f_2(0)]_{\XZtoSP}/[f_2(0)]_{\ntop}$ \\
\hline
exact $SU(3)$ case & $\kappa_p-\kappa_n$ & $3.706$ & 1 \\
Cabibbo model~\cite{Cabibbo:2003cu} & 
$\frac{M_{\Xi}+M_{\Sigma}}{2M_N}\left(
\kappa_p-\kappa_n \right)$& $4.958$& 1.338\\
generalized CVC &
$\kappa_{\Sigma^+}-\kappa_{\Xi^0}$ & $2.708$ & 0.731\\
Sirlin's formula~\cite{Sirlin:1979hb} & 
$\frac{M_{\Xi}+M_{\Sigma}}{2M_{\Sigma}}\left(\kappa_{\Sigma^+}+\frac{1}{2}\kappa_{\Sigma^{-}}\right)
-\frac{M_{\Xi}+M_{\Sigma}}{2M_{\Xi}}
\left(\kappa_{\Xi^0}+\frac{1}{2}\kappa_{\Xi^{-}}\right)$&$2.475$ 
& 0.668\\
\hline
experimental value~\cite{Alavi-Harati:2001xk} & N/A & $3.8\pm2.3$~\footnote{
A factor $(M_\Xi+M_\Sigma)/M_{\Xi}$, equal to $\simeq 1.9048$ is
different from definitions of the $f_2$ form factor adopted in Ref.~\cite{Alavi-Harati:2001xk}.} & $1.03\pm0.62$\\
\hline
\end{tabular}
\end{ruledtabular}
\label{Tab:Th_estimate_f_2}
\end{table}
%

%%%%%%%%%%%%%%  SEC 2  %%%%%%%%%%%%%%%%%%%%%%%%
%\clearpage
\section{Simulation details}
\label{Sec:numeric}

\subsection{Lattice set-up}

We have performed a quenched lattice calculation on a $L^3 \times T
=16^3\times 32$ lattice with a renormalization group improved gauge 
action, DBW2 (doubly blocked Wilson in two-dimensional parameter space) 
gauge action~\cite{{Takaishi:1996xj},{de Forcrand:1999bi}}
at $\beta=6/g^2=0.87$. The inverse of lattice spacing is about 1.3 GeV,
set by the $\rho$-meson mass~\cite{Aoki:2002vt}, yielding a physical volume 
of $(2.4\;{\rm fm})^3$. The spatial size 2.4 fm is large enough to accurately 
calculate the axial-vector coupling $g_A=g_1(0)$~\cite{Sasaki:2007gw}, 
which is one of the most sensitive observable to finite 
volume effects~\cite{{Sasaki:2003jh},{Lin:2008uz},{Yamazaki:2008py}}.

The previous quenched DWF studies by the RBC Collaboration 
reported that the residual chiral symmetry breaking of  
DWFs is significantly improved with a moderate size of the fifth-dimension. 
The residual quark mass for $L_s=16$ is measured as small as 
$m_{\rm res}\sim 5 \times 10^{-4}$ in lattice units~\cite{Aoki:2002vt}. 
Although we work with relatively coarse lattice spacing, $a\approx 0.15$ fm, good scaling behaviors of the light hadron spectrum~\cite{Aoki:2002vt}, 
the kaon B-parameter $B_K$~\cite{Aoki:2005ga} and proton decay matrix elements~\cite{Aoki:2006ib} are observed between at $\beta=0.87$ 
($a\approx 0.15$ fm) and 1.04 ($a\approx 0.10$ fm).
Therefore, we may deduce that no large scaling violation is ensured
for other observables as well in our DWF calculations. 
In Table~\ref{tab:old_simulation_results}, some basic physics results are
compiled from Ref.~\cite{Aoki:2002vt}.

In this study, DWF quark propagators were generated with three lighter quark 
masses $m_{ud}=$ 0.04, 0.05 and 0.06 for up and down quarks~\footnote{
In this paper, we restrict ourselves to considering the iso-spin symmetric
case as $m_{ud}=m_u=m_d$} and 
with two heavier quark masses $m_s=$0.08 and 0.10 for the strange quark 
with $L_s=16$ and $M_5=1.8$. We then take 5 different combinations between
the up (down) quark and the strange quark as $(m_{ud}, m_s)$=(0.04, 0.08),
(0.05, 0.08), (0.06, 0.08), (0.04, 0.10) and (0.05, 0.10), which yield different
$SU(3)$-breaking patterns characterized by 
$\delta=(M_{B}-M_{b})/(M_{B}+M_{b})$ for the $\Btob$ process
in the range of 0.009 to 0.028. 
Our results are analyzed on 377 
configurations. Preliminary results were first reported in Ref.~\cite{Sasaki:2006jp}~\footnote{
%%%%%%% footnote %%%%%%
In our actual simulations, the time-reversal process, $\Sigma^+\rightarrow \Xi^0$,
was actually utilized. This gives us some initial confusion in data analysis of Ref.~\cite{Sasaki:2006jp}. We now update all of results in this paper.
}. 
%%%%%%%%%%%%%%%%%%

As mentioned earlier, the previous study of neutron beta decay 
with the same simulation parameters successfully yields a value of 
$g_1(0)/f_1(0)$ as $1.212\pm0.027$, which just underestimates the 
experimental one by less than 5\%~\cite{Sasaki:2003jh}.
This success encourages us to study flavor $SU(3)$-breaking effects in
the hyperon beta decays through a comparison between neutron beta decay and 
$\Xi^0$ beta decay.
 
\subsection{Mass spectra and dispersion relation}
In order to compute baryon masses or beta-decay matrix elements, 
we use the following baryon interpolating operator
\be
(\eta^{S}_{X})_{ ijk}(t,{\bf p})=\sum_{\bf x} e^{-i{\bf p}\cdot{\bf x}}
\varepsilon_{abc}[q^{T}_{a, i}({\bf y}_1,t) C\gamma_5 q_{b, j}({\bf y}_2,t)]
q_{c, k}({\bf y}_3,t) \times \phi({\bf y}_1-{\bf x})\phi({\bf y}_2-{\bf x})
\phi({\bf y}_3-{\bf x}),
\ee
where $C$ is the charge conjugation matrix defined as 
$C=\gamma_t \gamma_y$ and
the index $X\in \{ B, b\}$ distinguishes between the initial ($B$) 
and final ($b$) states. 
The superscript $T$ denotes transpose and the indices $abc$ and $ijk$ label color and flavor, respectively. The superscript $S$ of the interpolating operator $\eta$ specifies the smearing for the quark propagators. 
In this study, we use two types of source: local source as
$\phi({\bf x}_i -{\bf x})=\delta({\bf x}_i - {\bf x})$ and Gaussian smeared source.
Here we take ${\bf x}_1={\bf x}_2={\bf x}_3={\bf 0}$ in our calculation. 
As for the Gaussian smeared source, we apply the gauge-invariant Gaussian 
smearing~\cite{{Gusken:1989qx},{Alexandrou:1992ti}} with $N=30$, $\omega=4.35$.
Details of our choice of smearing parameters are described 
in Ref.~\cite{Berruto:2005hg}.

We construct two types of the two-point function for the baryon states. 
One interpolating operator at the source location is constructed 
from Gaussian smeared quark fields, 
while the other interpolating operator at the sink location is either constructed from 
local quark fields (denoted LG) or Gaussian smeared ones (denoted GG):
\be
C_X^{SG}(t-t_{\rm src}, {\bf p})=\frac{1}{4}{\rm Tr}\left\{
{\cal P_+}
\langle 
{\eta}_{X}^{S}(t, {\bf p})
 \overline{{\eta}_{X}^{G}}(t_{\rm src},-{\bf p})
\right\}
\ee
with $S=L$ or $G$. The projection operator ${\cal P}_+=\frac{1+\gamma_t}{2}$
can eliminate contributions from the opposite-parity state for 
$|{\bf p}|=0$~\cite{{Sasaki:2001nf}, {Sasaki:2005ug}}. 
It is rather expensive to make the Gaussian smeared interpolating operator 
projected onto a specific finite momentum at the source location ($t_{\rm src}$).
However, it is sufficient to project only the sink operator onto the desired momentum
by virtue of momentum conservation. Thus, the quark fields at the source
location are not projected onto any specific momentum in this calculation.
For the momentum at the sink location ($t_{\rm sink}$), 
we take all possible permutations of the three momentum ${\bf p}$ including both 
positive and negative directions in this study.

All hadron masses are computed by 
using the LG-type correlators. We use the conventional interpolating 
operators, $\bar{u}\gamma_5 d$ ($\bar{u}\gamma_5 s$) for the $\pi$ ($K$)
state, $\varepsilon_{abc}(u^T_a C\gamma_5d_b)u_c$ for the nucleon,
$\varepsilon_{abc}(u^T_a C\gamma_5s_b)u_c$ for the $\Sigma$ state and
$\varepsilon_{abc}(s^T_a C\gamma_5u_b)s_c$ for the $\Xi$ state.
All fitted values, which are obtained from the conventional single exponential 
fit for baryons ($N$, $\Sigma$ and $\Xi$) and 
the conventional cosh fit for mesons ($\pi$ and $K$), are summarized in Tables
~\ref{Tab:mass_spect1} and \ref{Tab:mass_spect2}. 
Our simulated values of the pion mass range from 0.54 GeV to 0.67 GeV. 

The evaluation of the squared four-momentum transfer $q^2$ 
requires precise knowledge of the baryon energies 
$E_X({\bf p})$ ($X=N$, $\Sigma$, $\Xi$) with finite momentum.
This can be achieved by an estimation of the energy with the help 
of the dispersion relation and the measured rest mass that can be 
most precisely measured. As we reported in Ref.~\cite{Sasaki:2007gw}, 
the relativistic dispersion relation 
\be
E_X({\bf p}) = \sqrt{{\bf p}^2 + M_X^2},
\ee
where ${\bf p}=(p_x, p_y, p_z)$ with continuum-like momenta $p_i=\frac{2\pi}{L}n_i$ ($n_i=0, 1, 2, \cdot\cdot\cdot, (L-1)$),
is indeed fairly consistent with the energies computed at least at
the four lowest nonzero momenta: $(1,0,0)$, $(1,1,0)$, $(1,1,1)$ and
$(2,0,0)$ in our simulations. It implies that our simulations do not much
suffer from large ${\cal O}(a^2)$ errors even at finite $q^2$.
We utilize such estimated energies instead of actually measured values 
in our whole analysis~\footnote{
%%%%%%%%% footnote %%%%%%%%%%%
There is no differences in the final results
between using the dispersion relation, and the fitted energies 
at non-zero momenta within statistical errors, while the former 
statistical errors are slightly smaller than the latter.
%%%%%%%%%%%%%%%%%%%%%%%%%
}. 

%
% Table. 3
% 
\begin{table}[htbp]
\caption{
The residual mass $m_{\rm res}$, inverse lattice spacing ($a_{\rho}^{-1}$,
set by the $\rho$ meson mass), the renormalization factor of the axial-vector
current ($Z_A$), the pion decay constant ($F_{\pi}$) and the Kaon decay constant ($F_K$).
Those values are taken from Ref.~\cite{Aoki:2002vt}, where
simulations are performed on a $16^3\times 32$ volume.}
\begin{ruledtabular}
\begin{tabular}{lccccccc}
\hline
Gauge action ($\beta$) & $M_5$ & $L_s$ & $m_{\rm res}$ & $a_{\rho}^{-1}$ [GeV]& $Z_A(m_f=-m_{\rm res})$
& $F_{\pi}$ [MeV] & $F_{K}$ [MeV] \\
\hline
DBW2 (0.87) & 1.8 & 16 & 5.69(26)$\times 10^{-4}$ & 1.31(4) & 0.77759(45) & 91.2(5.2) & 104.2(3.8)\\
\hline
\end{tabular}
\end{ruledtabular}
\label{tab:old_simulation_results}
\end{table}
%

%
% Table. 4
%
\begin{table}[htbp]
\caption{Mass spectrum of non-strange hadrons (pion and nucleon) in lattice units.}
\begin{ruledtabular}
\begin{tabular}{ccc}
\hline
$m_{ud}$ & $M_\pi$ & $M_N$ \\
\hline
0.06 & 0.5050(8)  & 1.0821(42) \\
0.05 & 0.4617(9)  & 1.0358(46) \\
0.04 & 0.4148(9)  & 0.9869(50) \\
\end{tabular}
\end{ruledtabular}
\label{Tab:mass_spect1}
\end{table}
%

%
% Table. 4
%
\begin{table}[htbp]
\caption{Mass spectrum of strange hadrons (Kaon, $\Sigma$ and $\Xi$) in lattice units.}
\begin{ruledtabular}
\begin{tabular}{ccccc}
\hline
$m_{s}$ & $m_{ud}$ & $M_K$ & $M_{\Sigma}$  & $M_{\Xi}$\\
\hline
0.08 & 0.06 & 0.5455(7)  & 1.1161(41) & 1.1375(39) \\
          & 0.05 & 0.5257(8)  & 1.0895(43) & 1.1210(39) \\
          & 0.04 & 0.5055(8)  & 1.0626(46) & 1.1039(40) \\
0.10 & 0.05 & 0.5652(8)  & 1.1201(46) & 1.1741(40) \\
          & 0.04 & 0.5462(8)  & 1.0941(50) & 1.1577(41) \\
\hline          
\end{tabular}
\end{ruledtabular}
\label{Tab:mass_spect2}
\end{table}
%

%%%%%%%%%%%%%%%%%%%%%%%%%%%%%%%%%%%%%%%%
%\clearpage
\subsection{Three-point correlation functions}
\label{Sec:numeric_3ptfunc}

We next define the finite-momentum three-point functions 
for the relevant components of either the local vector current
 (${\cal J}^{V}_{\alpha}$) or
the local axial current (${\cal J}_{\alpha}^{A}$) with the interpolating
operators $\eta_{B}$ and $\eta_{b}$ for the $B$ and $b$ states:
\be
\langle \eta_b (t', {\bf p '}){\cal J}^{\Gamma}_{\alpha}(t, {\bf q}) 
\overline{\eta}_B(0, -{\bf p})\rangle
=
{\cal G}_{\alpha}^{\Gamma, \Btob}(p, p')\times f(t, t', E_{B}({\bf p}), E_{b}({\bf p}')) + \cdot\cdot\cdot,
\ee
where the initial ($B$) and final ($b$) states carry fixed momenta
${\bf p}$ and ${\bf p}'$ respectively and then the current operator
has a three-dimensional momentum transfer ${\bf q}={\bf p}-{\bf p}'$.
Here, Dirac indices have been suppressed. The ellipsis denotes excited state 
contributions which can be ignored in the case of $t'-t\gg 1$ and $t \gg 1$.
The ground state contribution of the three-point correlation function is 
described by two parts. The first part, ${\cal G}_{\alpha}^{\Gamma}(p,p')$, 
is defined as
%
% eq.
%
\be
{\cal G}^{\Gamma, \Btob}_{\alpha}(p,p')=
(-i\gamma\cdot p'+M_{b}){\cal O}_{\alpha}^{\Gamma, \Btob}(q)(-i\gamma\cdot p+M_{B}),
\ee
where ${\cal O}_{\alpha}^{\Gamma}(q)$ corresponds to either Eq.~(\ref{Eq:VcMat}) or Eq.~(\ref{Eq:AxMat}), and the factor $f(t, t', E_{B}({\bf p}), E_{b}({\bf p}'))$ collects all the kinematical factors, normalization of states, and time 
dependence of the correlation function.
The trace of ${\cal G}_{\alpha}^{\Gamma}(p,p')$ with some appropriate 
projection operator ${\cal P}$ for specific combinations of $\Gamma$ 
and $\alpha$ yields some linear combination of form factors 
in each $\Gamma$ channel.
On the other hand, all time dependences of the factor 
$f(t, t', E_{B}({\bf p}), E_{b}({\bf p}'))$
can be eliminated by the appropriate ratio of three- and two-point functions~\cite{Hagler:2003jd}
\be
{\cal R}^{\Btob}(t)=\frac{C^{{\cal P},\Btob}_{\Gamma, \alpha}(t, {\bf p}', {\bf p})}
{C_b^{GG}(t_{\rm sink}-t_{\rm src}, {\bf p}')}
\left[
\frac{C_B^{LG}(t_{\rm sink}-t, {\bf p})C_b^{GG}(t-t_{\rm src}, {\bf p}')C_b^{LG}(t_{\rm sink}-t_{\rm src}, {\bf p}')}
{C_b^{LG}(t_{\rm sink}-t, {\bf p}')C_B^{GG}(t-t_{\rm src}, {\bf p})C_B^{LG}(t_{\rm sink}-t_{\rm src}, {\bf p})} 
\right]^{\frac{1}{2}},
\label{Eq:LHPCratio}
\ee
where 
\be
C_{\Gamma, \alpha}^{{\cal P}, \Btob}(t, {\bf p}', {\bf p})=\frac{1}{4}{\rm Tr}
\left\{
{\cal P}
\langle {\eta}^{G}_{b}(t_{\rm sink}, {\bf p}')
{\cal J}^{\Gamma}_{\alpha}(t, {\bf q})
 \overline{{\eta}^{G}_{B}}(t_{\rm src},-{\bf p})
\rangle
\right\},
\ee
which are calculated by the sequential source method described 
in Ref.~\cite{Sasaki:2003jh}.

In this study, we consider the hyperon decay process 
$B({\bf p}) \rightarrow b({\bf 0})$ at the rest flame of the final ($b$) state
(${\bf p}'={\bf 0}$), which leads to ${\bf q}={\bf p}$. 
Therefore the squared four-momentum transfer is given by 
$q^2=2M_b(E_B({\bf p})-M_B)-(M_B-M_b)^2$.
The energies of the initial and final baryon state is
simply abbreviated as $E_B$ and $E_b$, hereafter.
In this kinematics, ${\cal G}_{\alpha}^{\Gamma}(p, p')$ is represented by a simple
notation ${\cal G}_{\alpha}^{\Gamma}(p)$.
Then, the ratio~(\ref{Eq:LHPCratio}) gives the asymptotic form as a
function of the current operator insertion time $t$,
\be
{\cal R}^{\Btob}(t)
\rightarrow \frac{1}{4}{\rm Tr}\left\{
{\cal P} {\cal G}_{\Gamma, \alpha}^{\Btob}(q) 
\right\} \times \frac{1}{\sqrt{2M_b^2E_B(E_B+M_B)}}
\label{Eq:Ratio2}
\ee
in the limit when the Euclidean time separation between all operators is
large, $t_{\rm sink}\gg t\gg t_{\rm src}$ with fixed $t_{\rm src}$ and $t_{\rm sink}$. 

We choose particular combinations of the projection operator ${\cal P}$ and 
the current operator ${\cal J}^{\Gamma}_{\alpha}$ ($\Gamma=V$ or $A$).
Two types of the projection operator, ${\cal P}^t={\cal P}_{+}\gamma_{t}$
and ${\cal P}_5^{z}={\cal P}_{+}\gamma_5\gamma_{z}$
are considered in this study.
The latter operator implies that the $z$-direction is chosen 
as the polarized direction.
We then obtain some linear combination of desired form factors from
the following projected correlation functions:
%
% eq.
%
\bea
\frac{1}{4}\tr \{{\cal P}^{t}{\cal G}^{V, \Btob}_{t}(q) \}&=& M_b(E_B + M_B) \left[
f^{\Btob}_1(q^2)-\frac{E_B - M_B}{M_B + M_b} f^{\Btob}_2(q^2) - \frac{E_B -M_b}{M_B+M_b} f^{\Btob}_3(q^2) 
\right],
\label{Eq:3pt_vec_time}
\\
\frac{1}{4}\tr \{{\cal P}^{t}{\cal G}^{V, \Btob}_{i}(q) \}&=& -iq_i M_b
\left[
f^{\Btob}_1(q^2)-\frac{E_B-M_b}{M_B+M_b}f^{\Btob}_2(q^2)-\frac{E_B+M_B}{M_B+M_b}f^{\Btob}_3(q^2)
\right],
\label{Eq:3pt_vec_spatial}
\\
\frac{1}{4}\tr\{ {\cal P}^{z}_5{\cal G}^{V, \Btob}_{i}(q) \}&=& -i\varepsilon_{ijz}q_j M_b
\left[
f^{\Btob}_1(q^2)+f^{\Btob}_2(q^2)
\right]
\label{Eq:3pt_vec_tensor}
\eea
for the vector currents ${\cal J}^{V}_{t}$ and ${\cal J}^{V}_{i}$ $(i=x,y,z)$.
Similarly, we get
%
% eq.
%
\bea
\frac{1}{4}\tr\{ {\cal P}_5^{z}{\cal G}^{A, \Btob}_{t}(q) \}&=& iq_zM_b
\left[
g^{\Btob}_1(q^2)-\frac{E_B+M_B}{M_B+M_b}g^{\Btob}_2(q^2)
-\frac{E_B-M_b}{M_B+M_b}g^{\Btob}_3(q^2)
\right],
\label{Eq:3pt_axial_time}
\\
\frac{1}{4}\tr\{ {\cal P}_5^{z}{\cal G}^{A, \Btob}_{i}(q) \}&=& M_b
\left[(E_B+M_B)
\left(g^{\Btob}_1(q^2)
-\frac{M_B-M_b}{M_B+M_b}g^{\Btob}_2(q^2) \right) \delta_{iz}\right.
 \nonumber\\
&&
\left.
-\frac{q_i q_z}{M_B+M_b} \left(
g^{\Btob}_2(q^2)+g^{\Btob}_3(q^2)
\right)
\right]
\label{Eq:3pt_axial_spatial}
\eea
for the axial-vector currents ${\cal J}^{A}_{t}$ and ${\cal J}^{A}_{i}$ $(i=x, y, z)$. 
In this calculation, we use the four nonzero three-momentum transfer ${\bf q}=\frac{2\pi}{L}{\bf n}$ (${\bf n}^2=1$, 2, 3, 4).
All possible permutations of the lattice momentum including both positive 
and negative directions are properly taken into account.
All three-point correlation functions are calculated with a source-sink separation of 10
in lattice units, which is the same in the previous DWF calculations of 
the axial-vector coupling $g_A$~\cite{Sasaki:2003jh} and
the weak matrix elements of the nucleon~\cite{Sasaki:2007gw}.

Here, it is worth noting that the longitudinal momentum ($q_z$) 
dependence explicitly appears in Eq.~(\ref{Eq:3pt_axial_spatial})
due to our choice of the polarized direction.
This fact provides two kinematical constraints on 
determination of the three-point functions in our calculation.
First, there are two types of kinematics, $q_z\neq 0$ and $q_z=0$ 
in the longitudinal component ($i=z$) of Eq.~(\ref{Eq:3pt_axial_spatial}), except 
for the case of ${\bf n}^2=3$ where $q_z$ is always nonzero. 
Secondly, the transverse components ($i=x$ or $y$)
of Eq.~(\ref{Eq:3pt_axial_spatial}) are prevented from vanishing by the kinematics 
only if ${\bf n}^2=2$ and 3, where two components of the momentum 
including the polarized  direction ($z$-direction) are nonzero. 

\begin{figure}[hbtp]
\bc
\includegraphics[width=.60\textwidth,clip]{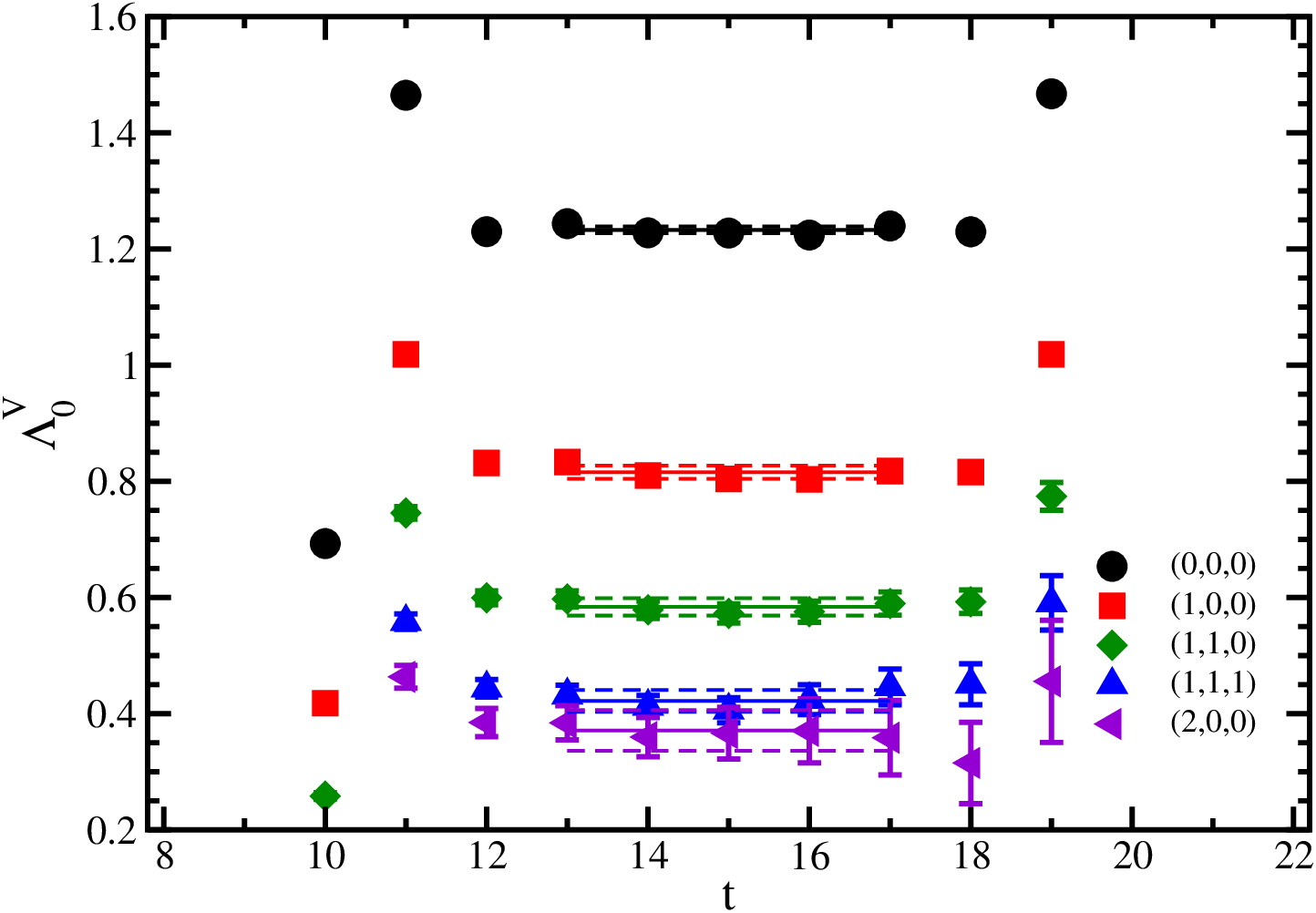}
\includegraphics[width=.60\textwidth,clip]{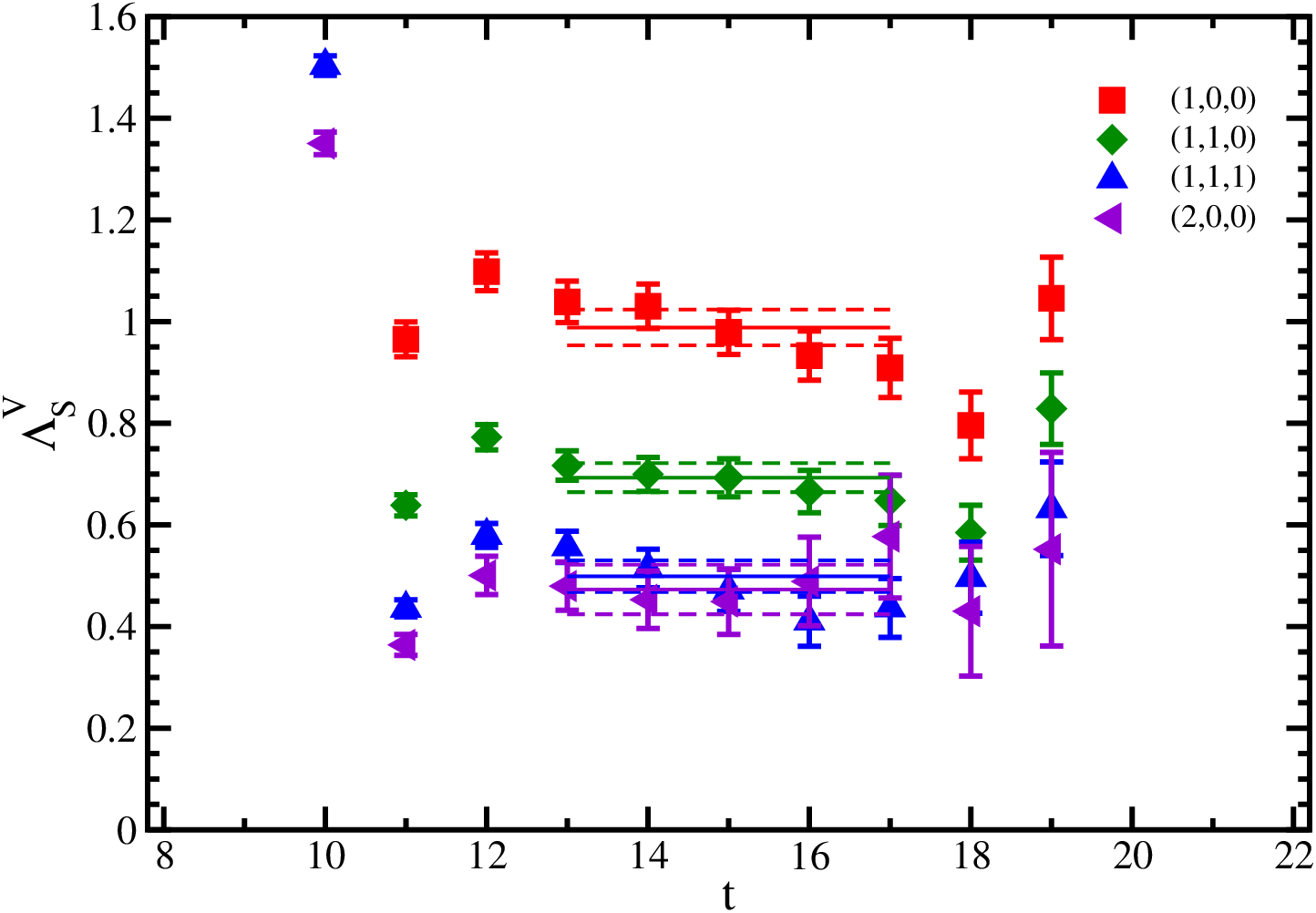}
\includegraphics[width=.60\textwidth,clip]{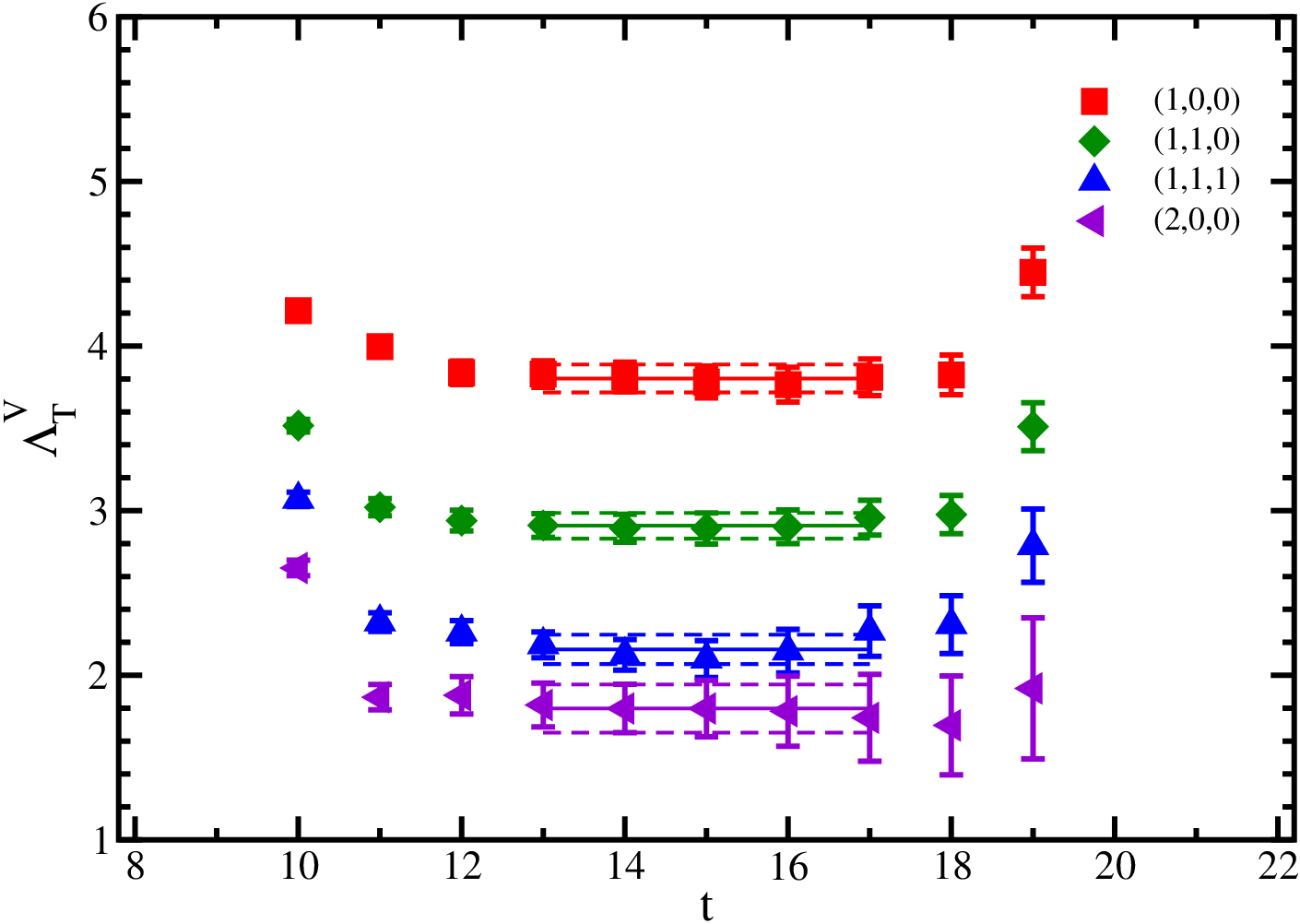}
\ec
\caption{
Relevant ratios of three- and two-point functions, ${\Lambda}^{V}_{0}$ (top),
${\Lambda}^{V}_{S}$ (middle) and ${\Lambda}^{V}_{T}$ (bottom), for all possible 
three-momentum transfer ${\bf q}$ as a function of the current insertion time slice 
at $(m_{ud}, m_s)=(0.04, 0.08)$.}
\label{FIG:Lambda^V}
\end{figure}

First of all, in Fig.~\ref{FIG:Lambda^V}, we plot the dimensionless projected correlators of the vector part  
%
% eq.
%
\bea
{\Lambda}^{V, \Btob}_{0}&=&\frac{\frac{1}{4}\tr\{ {\cal P}^t {\cal G}^{V, \Btob}_{t}(q)\}}
{M_b(E_B+M_B)},
\label{Eq:Lambda_V_0}
\\
{\Lambda}^{V, \Btob}_{S}&=&-\frac{1}{3}\sum_{i=x,y,z} \frac{\frac{1}{4}\tr\{{\cal P}^t {\cal G}^{V,\Btob}_{i}(q)\}}{iq_iM_b}, 
\label{Eq:Lambda_V_S}
\\
{\Lambda}^{V, \Btob}_{T}&=&-\frac{1}{2}\left(
\frac{\frac{1}{4}\tr\{ {\cal P}^{z}_5 {\cal G}^{V,\Btob}_{x}(q)\}}{iq_yM_b} 
-\frac{\frac{1}{4}\tr\{ {\cal P}^{z}_5 {\cal G}^{V,\Btob}_{y}(q)\}}{iq_xM_b} 
\right)
\label{Eq:Lambda_V_T}
\eea
as a function of the current insertion time slice $t$ for
the $\XZtoSP$ process at $(m_{ud}, m_{s})=(0.04, 0.08)$
as typical examples. Good plateaus for all squared three-momentum transfer are
observed in the middle region between the source and sink points. The quoted errors
are estimated by a single elimination jackknife method. The lines plotted in 
each figure represent the average value (solid lines) and their 1 standard 
deviations (dashed lines) in the time-slice range $13\le t \le 17$.

\begin{figure}[htbp]
\bc
\includegraphics[width=.60\textwidth,clip]{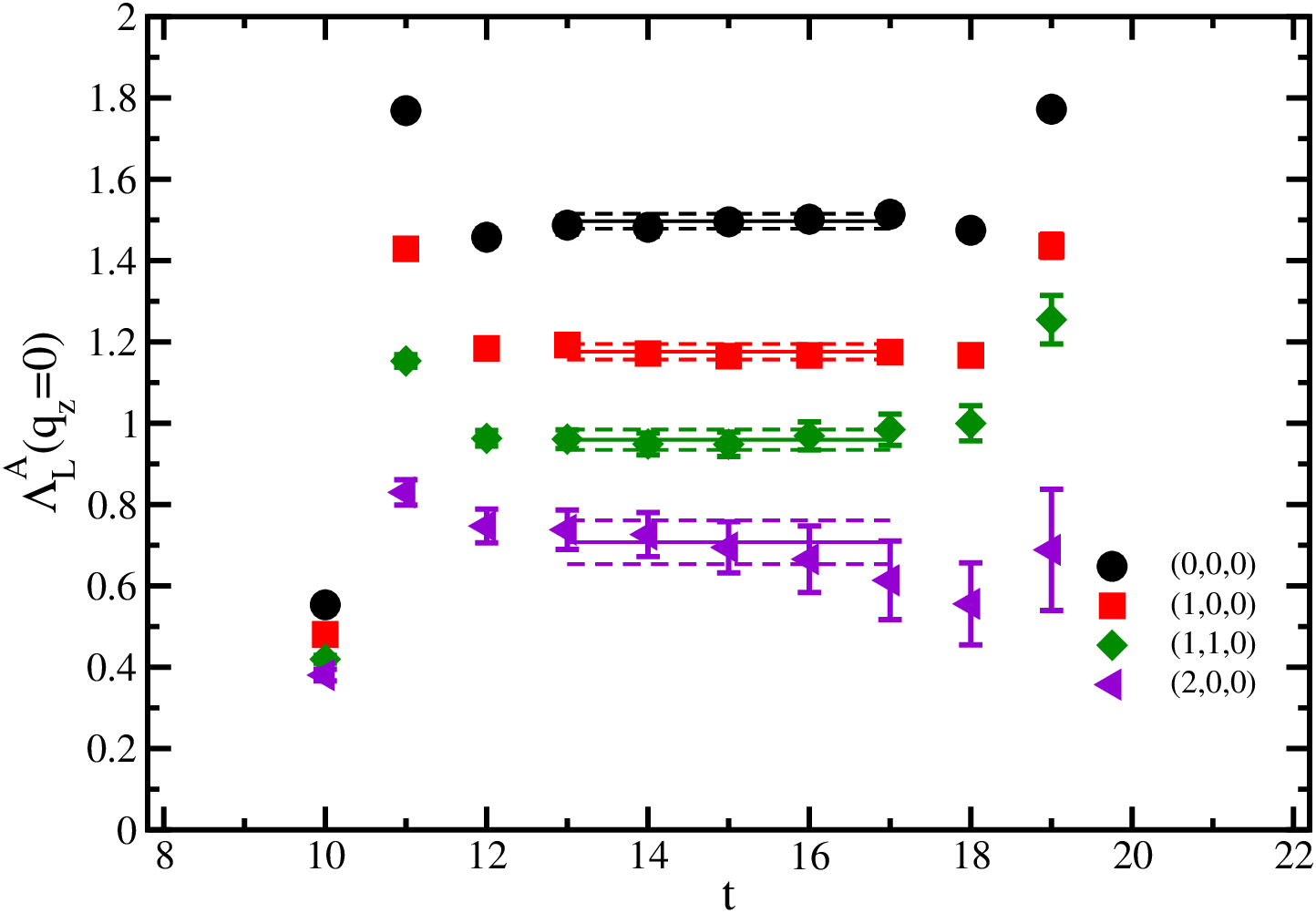}
\includegraphics[width=.60\textwidth,clip]{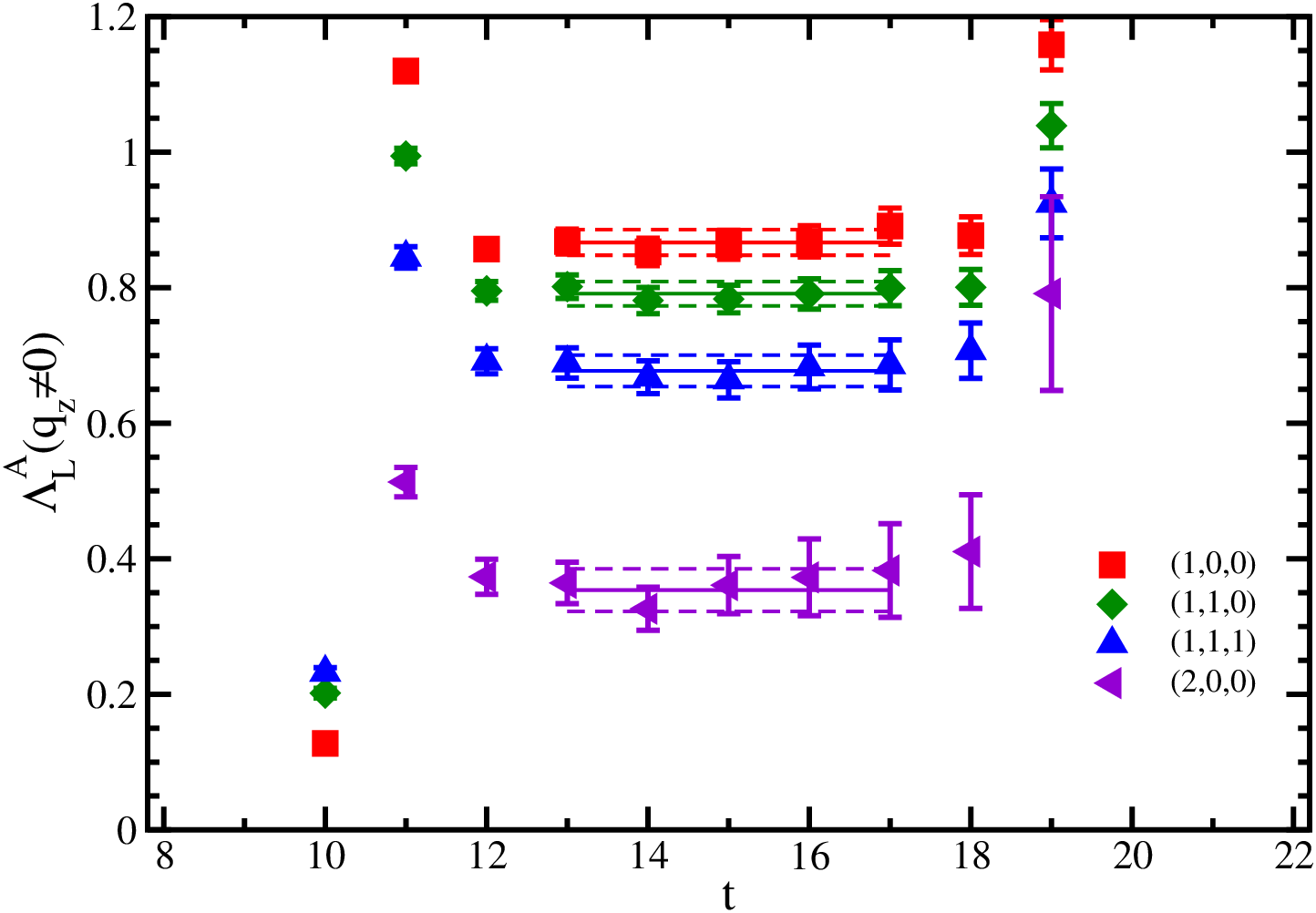}
\ec
\caption{
Relevant ratios of three- and two-point functions,
${\Lambda}^{A}_{L}(q_z=0)$ (top) and ${\Lambda}^{A}_{L}(q_z\neq 0)$ (bottom), 
for all possible three-momentum transfer ${\bf q}$ as a function of the current insertion 
time slice at $(m_{ud}, m_s)=(0.04, 0.08)$.}
\label{FIG:Lambda^A1}
\end{figure}

Similarly, we also define the dimensionless projected correlators of the 
axial-vector part as
%
% eq.
%
\bea
{\Lambda}^{A,\Btob}_{L}&=&\frac{\frac{1}{4}\tr\{{\cal P}_5^z {\cal G}^{A,\Btob}_{z}(q)\}}
{M_b(E_B+M_B)},
\label{Eq:Lambda_A_L}
\\
{\Lambda}^{A,\Btob}_{T}&=&-
\frac{1}{2}\left(
\frac{\frac{1}{4}\tr\{ {\cal P}_5^z {\cal G}^{A,\Btob}_{x}(q)\}}{q_z q_x} 
+\frac{\frac{1}{4}\tr\{ {\cal P}_5^z {\cal G}^{A,\Btob}_{y}(q)\}}{q_z q_y} 
\right),
\label{Eq:Lambda_V_T}
\\
{\Lambda}^{A,\Btob}_{0}&=&\frac{\frac{1}{4}\tr\{ {\cal P}_5^z {\cal G}^{A,\Btob}_{t}(q)\}}{iq_zM_b},
\label{Eq:Lambda_A_0}
\eea
which are also evaluated from the plateau of the ratio (\ref{Eq:LHPCratio}). 
Fig.~\ref{FIG:Lambda^A1} is plotted for $\Lambda^{A}_L$, which
explicitly depends on the longitudinal momentum $q_z$
because of the chosen direction of the polarization. 
Two figures represent two types of kinematics, $q_z\neq 0$ and $q_z=0$. 
Solid and dashed lines are defined as in Fig.~\ref{FIG:Lambda^V}.
Good plateaus for all squared three momentum transfer ${\bf q}^2$ 
are observed, similar to what we observe in the vector channel. 

\begin{figure}[htbp]
\bc
\includegraphics[width=.60\textwidth,clip]{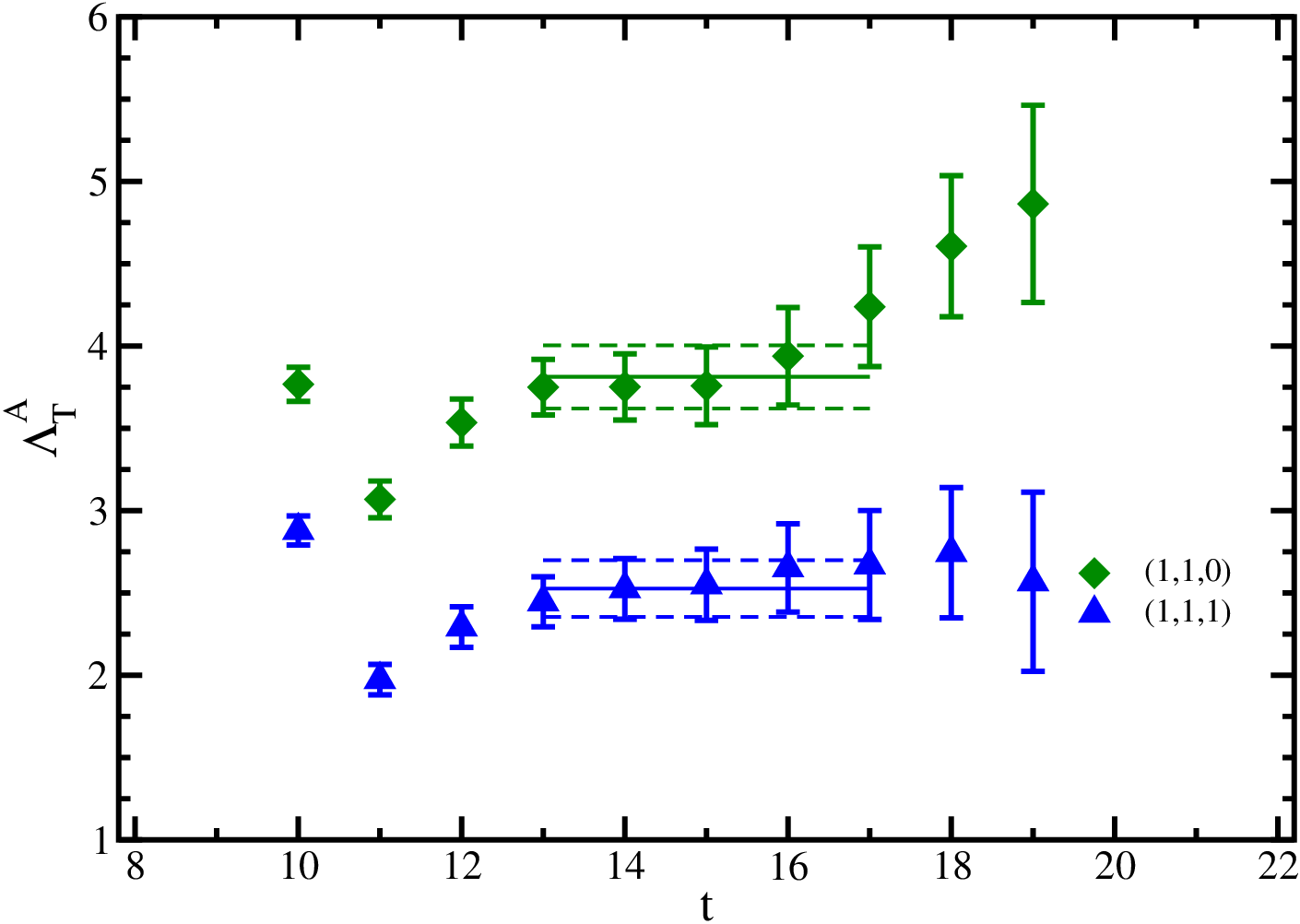}
\includegraphics[width=.60\textwidth,clip]{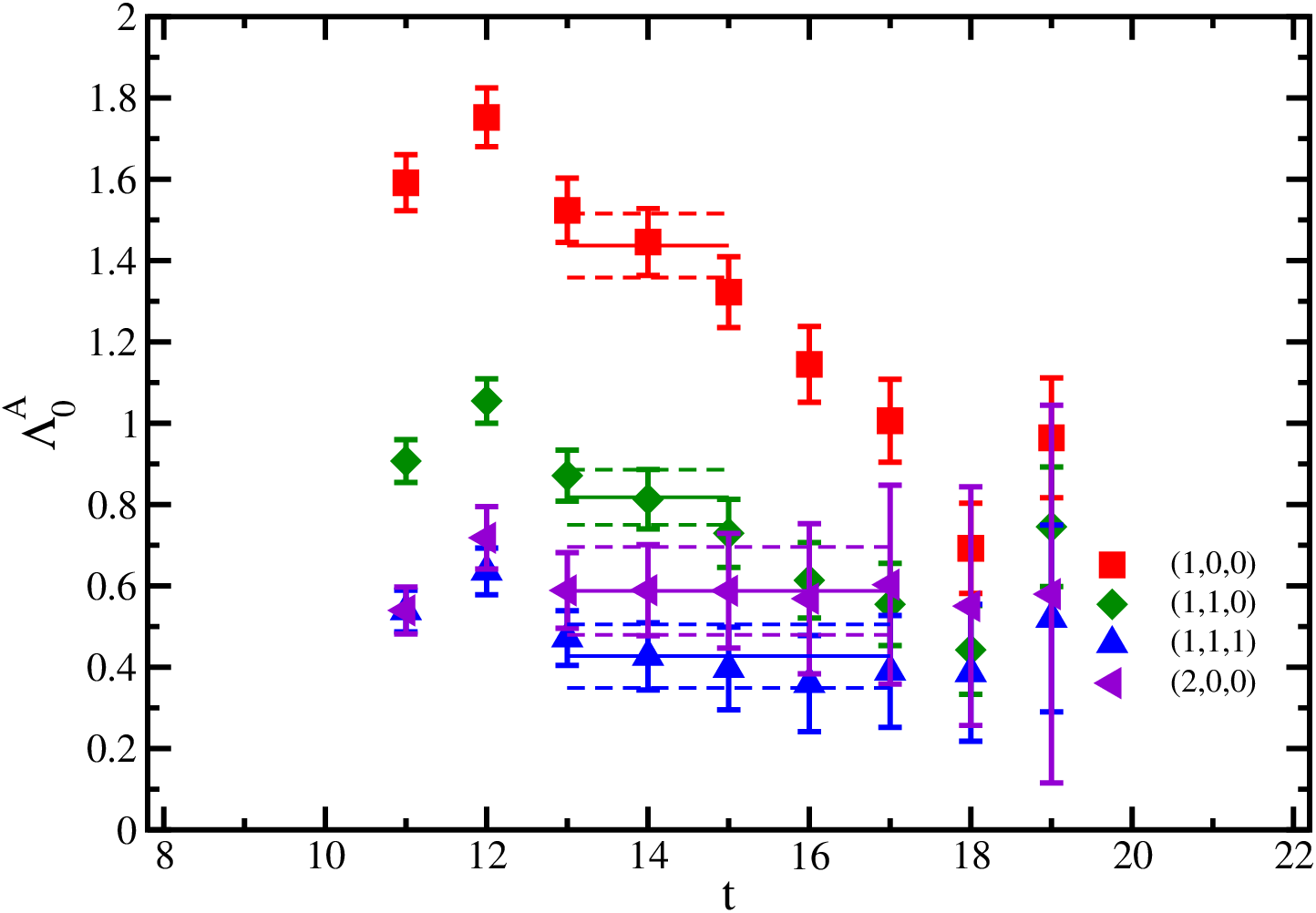}
\ec
\caption{
Relevant ratios of three- and two-point functions,
${\Lambda}^{A}_{T}$ (top) and ${\Lambda}^{A}_{0}$ (bottom), 
for all possible three-momentum transfer ${\bf q}$ as a function of the current insertion 
time slice at $(m_{ud}, m_s)=(0.04, 0.08)$.}
\label{FIG:Lambda^A2}
\end{figure}

The remaining two ratios $\Lambda^{A}_T$ and $\Lambda^{A}_0$ are shown in
Fig.~\ref{FIG:Lambda^A2}. The upper figure is for $\Lambda^{A}_T$, which
is accessible only for ${\bf q}^2=2$ and 3 in units of $(2\pi/L)^2$, where two 
components of the momentum including the polarized direction ($z$-direction)
are allowed to be nonzero. Again, we observe reasonable good plateaus
in the time-slice range $13\le t \le 17$. However, in the lower figure, 
the ratio $\Lambda^{A}_0$ at two lower ${\bf q}^2$ doesn't exhibit
a clear plateau, while the reasonable plateau can be observed at two 
higher ${\bf q}^2$ similar to other ratios.
We count a short shoulder plateau in the time-slice range $13\le t \le 15$
to take the average value of $\Lambda^{A}_0$ at two smaller ${\bf q}^2$.
It is worth mentioning that $g_1(q^2)$ and $g_3(q^2)$ are mainly determined
by either $\Lambda^{A}_L$ or $\Lambda^{A}_T$, since contributions of $\Lambda^{A}_0$ in Eqs.~(\ref{Eq:AxvFormFacts_g1}) and (\ref{Eq:AxvFormFacts_g3}) 
%which will be defined in Sec.~\ref{Sec:other_FF_2ndclass}, 
are numerically much smaller than others.
The precise determination of the second-class form factor $g_2(q^2)$ may be 
affected by the poor plateau observed in $\Lambda^{A}_0$ at two lower ${\bf q}$.
However, the subtraction procedure for the second-class form factors $f_3(q^2)$ and $g_2(q^2)$, which will be described in the proceeding section (Sec.~\ref{Sec:other_FF_2ndclass}), 
may reduce the systematic error stemming from above mentioned issue.

%

%\clearpage

%
%

%%%%%%%%%%%%%%%%%%%%%%%%%%%%%%%%%%%%%%%%%%%%%%%%
\subsection{Renormalization}

In general, lattice operators receive finite renormalizations relative
to their continuum counterparts since the exact symmetries
of the continuum are usually realized only in the continuum
limit $a\rightarrow0$. Fortunately, the well-preserved chiral
and flavor symmetries of 
DWFs~\cite{{Kaplan:1992bt},{Shamir:1993zy},{Furman:1995ky}} make
this task much easier than in the more conventional fermions.
In this study, we use the vector and axial-vector local currents, which
shares a common renormalization: $Z_V=Z_A$, up to higher order
discretization errors, ${\cal O}(a^2)$~\cite{Blum:2001sr}. 
Therefore, we first focus on the vector renormalization.

The vector form factors, especially in the precise determination of $f_1(0)$, 
require some independent estimation of $Z_V$, the renormalization of the
quark bilinear vector currents,
%
% eq
%
\be
[\bar{q}_f\gamma_{\alpha} q_{f'}]^{\rm ren}
=Z^{\bar{f}f'}_V[\bar{q}_f\gamma_{\alpha} q_{f'}]^{\rm lattice}
\label{Eq:RenormCur}
\ee
where a subscript $f$ denotes the flavor index. In this study, we need two vector renormalizations, 
$Z_V^{\bar{u}d}$ and $Z_V^{\bar{u}s}$, for neutron beta decay and its $SU(3)$ 
counterpart, the $\XZtoSP$ transition process. The former can be 
evaluated by the inverse of the forward limit of the $\ntop$ vector matrix element because of $\lim_{q^2 \rightarrow 0}\langle p|[{\bar u}\gamma_0d]^{\rm ren}|n\rangle =1$ 
in the present calculation under the exact iso-spin symmetry
($m_u=m_d$). For the latter, this prescription is not directly applicable because of the 
presence of the flavor $SU(3)$ breaking. 
However, we may calculate $Z_V^{\bar{u}s}=Z_V^{\bar{u}s}(m_{ud}, m_s)$ 
for $m_{ud}\neq m_s$ through the following relation:
\be
Z_{V}^{\bar{u}s}(m_l, m_h)=
\sqrt{Z_V^{\bar{u}s}(m_l, m_l)Z_V^{\bar{u}s}(m_h, m_h)}
\label{Eq:ZvRenorm}
\ee
where $m_l$ and $m_h$ ($m_l < m_h$) are simulated quark masses for either 
up (down) quark or strange quark. 
$Z_V^{\bar{u}s}(m_l, m_l)$ and $Z_V^{\bar{u}s}(m_h, m_h)$ correspond to 
the case of degenerate quark masses ($m_{ud}=m_s$). 
Therefore, for an evaluation of those two vector renormalizations,
we can utilize the relation, $\lim_{q^2 \rightarrow 0}\langle \Sigma^+|[{\bar u}\gamma_0s]^{\rm ren}|\Xi^0\rangle =1$,
which is valid in the exact $SU(3)$ symmetry limit. 

In the case of the flavor current ($f=f'$ in Eq.(\ref{Eq:RenormCur})), we had already observed that the relation $Z_V=Z_A$ is well 
satisfied in the chiral limit, up to higher order discretization errors 
${\cal O}(a^2)$ and neglecting explicit chiral symmetry breaking
due to the moderate size of the fifth-dimensional extent $L_s$. 
This good chiral property of DWFs is known to be maintained
even for the heavy-light vector and axial-vector currents~\cite{Yamada:2004ri}, 
which correspond to the extreme case of the flavor changing current ($f\neq f'$) 
in Eq.(\ref{Eq:RenormCur}). Therefore, in this study, we use the common renormalization given in Eq. (\ref{Eq:ZvRenorm})
for both vector and axial-vector local currents.

%%%%%%%%%%%%%%%%%%%%%%%%%%%%%%%%%%%%%%%%%
%\clearpage

\section{Determination of $f_1(0)$}
\label{Sec:f_1}

\subsection{Scalar form factor $f_S(q^2)$ at $q^2=q_{\rm max}^2$}

\begin{figure}[htbp]
\bc
\includegraphics[width=.60\textwidth,clip]{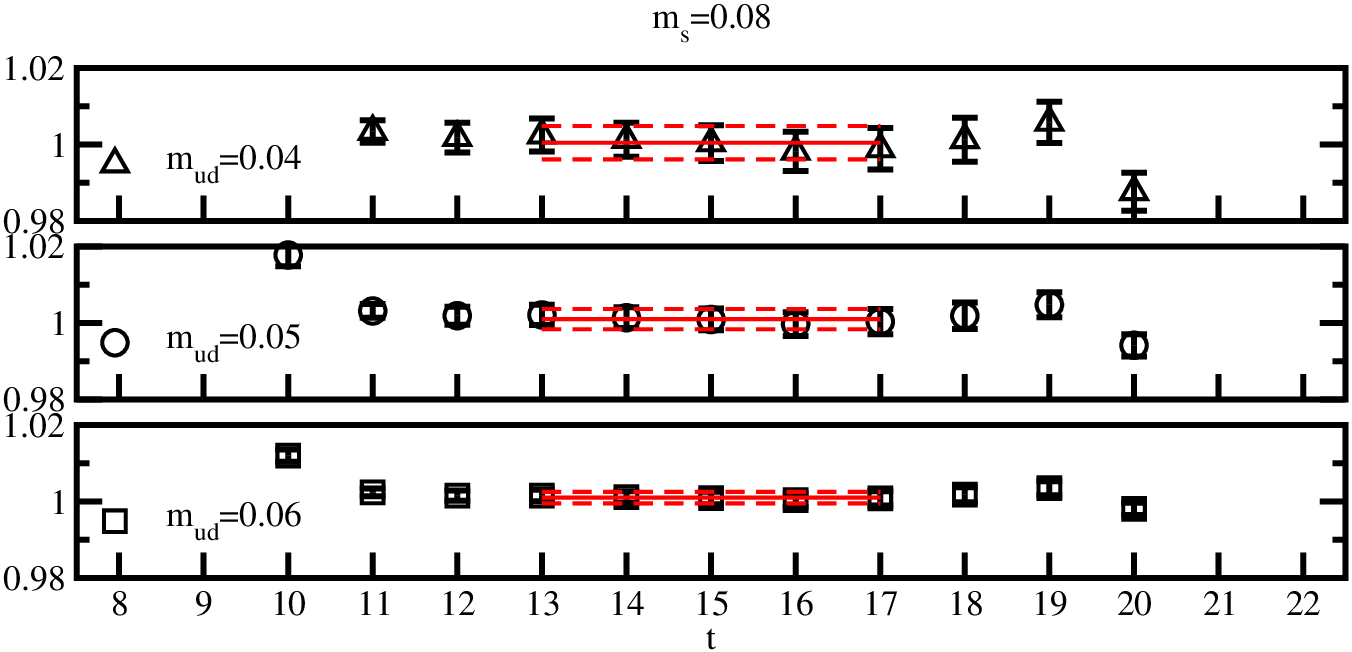}
\includegraphics[width=.60\textwidth,clip]{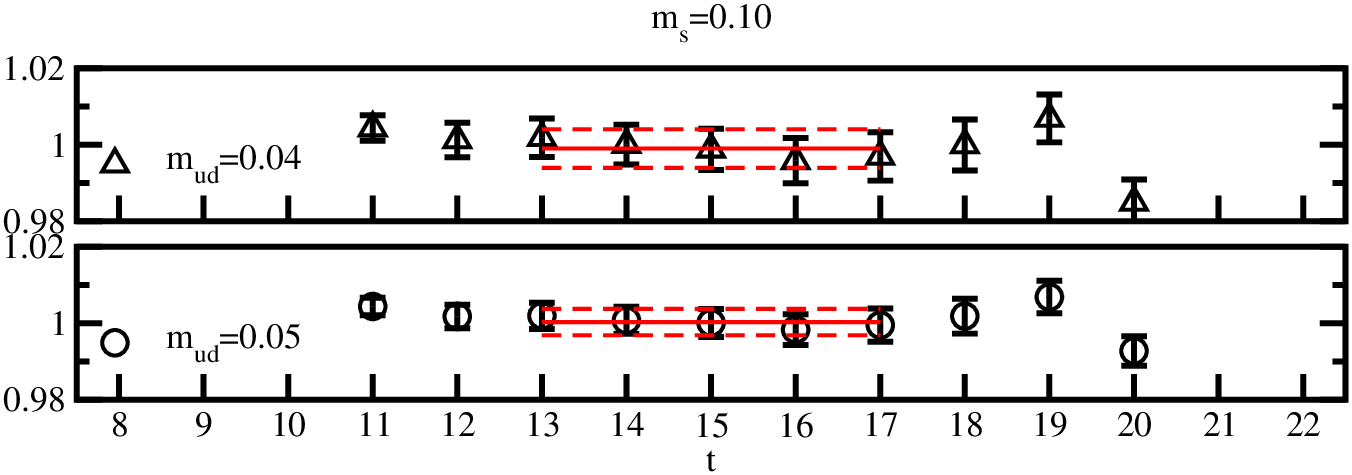}
\caption{$f_S^{\rm ren}(q^2_{\rm max})$ as a function of the current insertion time slice.
A source-sink location of three-point functions is set at $[t_{\rm src}, t_{\rm sink}]=[10,20]$.
The lines represent the average value (solid lines) and their 1 standard deviations (dashed
lines) in the time-slice range $13 \le t \le 17$.
}
\label{FIG:plateau_F0max}
\ec
\end{figure}

In the vector channel, only the time component of the vector current, namely the three-point correlation function $\frac{1}{4}\tr\{ {\cal P}^{t}{\cal G}^{V, \Btob}_{t}(q) \}$ is prevented from vanishing at zero three
momentum transfer $|{\bf q}|=0$ by the kinematics~\cite{Sasaki:2003jh}.
This non-vanishing correlator gives the scalar form factor at specific four-momentum
transfer as
\be
f_S^{\Btob}(q_{\rm max}^2)=
\Lambda^{V, \Btob}_0({\bf q}={\bf 0}),
\label{Eq:scalar_ff}
\ee
where $q_{\rm max}^2=-(M_B-M_b)^2$~\footnote{
%%%%%% footnote %%%%%%%%%%
Here, we should comment that we did not use the double ratio method, 
which is originally proposed in Ref.~\cite{Hashimoto:1999yp} for $B$ meson
decays and also adopted in Ref.~\cite{Guadagnoli:2006gj} 
for the hyperon beta decays.}.
%%%%%%%%%%%%%%%%%%%%% 
In Fig.~\ref{FIG:plateau_F0max}, we plot the renormalized $f_S(q^2_{\rm max})$ 
as a function of the current insertion time slice.
Good plateaus are observed in the middle region between the source and sink points.
The lines represent the average value (solid lines) and their 1 standard deviations (dashed lines) in the time-slice range $13 \le t \le 17$. 
We stress that the statistical accuracy 
is less than about 0.5\% even in the worst case ($m_{ud}, m_{s}$)=(0.04, 0.10).
The obtained values of the renormalized $f_S(q^2_{\rm max})$
as well as the bare one and its renormalization factor $Z_V$
are summarized in Table~\ref{Tab:f0_qmax}.
There is a tendency that the error of $f_S(q_{\rm max}^2)$
increases as $m_s$ deviates from $m_{ud}$, which was also
observed in the scalar form factor of $K_{l3}$ decays~\cite{Dawson:2006qc}.

We should note that the renormalized $f_S(q_{\rm max}^2)$ is exactly equal to
unity in the flavor $SU(3)$ symmetric limit, where $f_S(q_{\rm max}^2)$ 
becomes $f_1(0)$. Thus, the deviation from unity in 
$f_S(q_{\rm max}^2)$ is attributed to three types of the $SU(3)$ breaking effect:
1) the recoil correction ($q^2_{\rm max}\neq 0$) stemming from
the mass difference of $B$ and $b$ states, 
2) the presence of the second-class form factor $f_3(q^2)$ and
3) the deviation from unity in the renormalized $f_1(0)$. 
Taking the limit of zero four-momentum transfer of $f_S(q^2)$ can separate
the third effect from the others, since the scalar form factor 
at $q^2=0$, $f_S(0)$, is identical to $f_1(0)$. 
Indeed, to measure the third one is our main target. 
%
%	plateaus of f0_qmax
%
%

%\clearpage
\subsection{Interpolation to zero four-momentum squared}

\begin{figure}[htbp]
\bc
\includegraphics[width=.47\textwidth,clip]{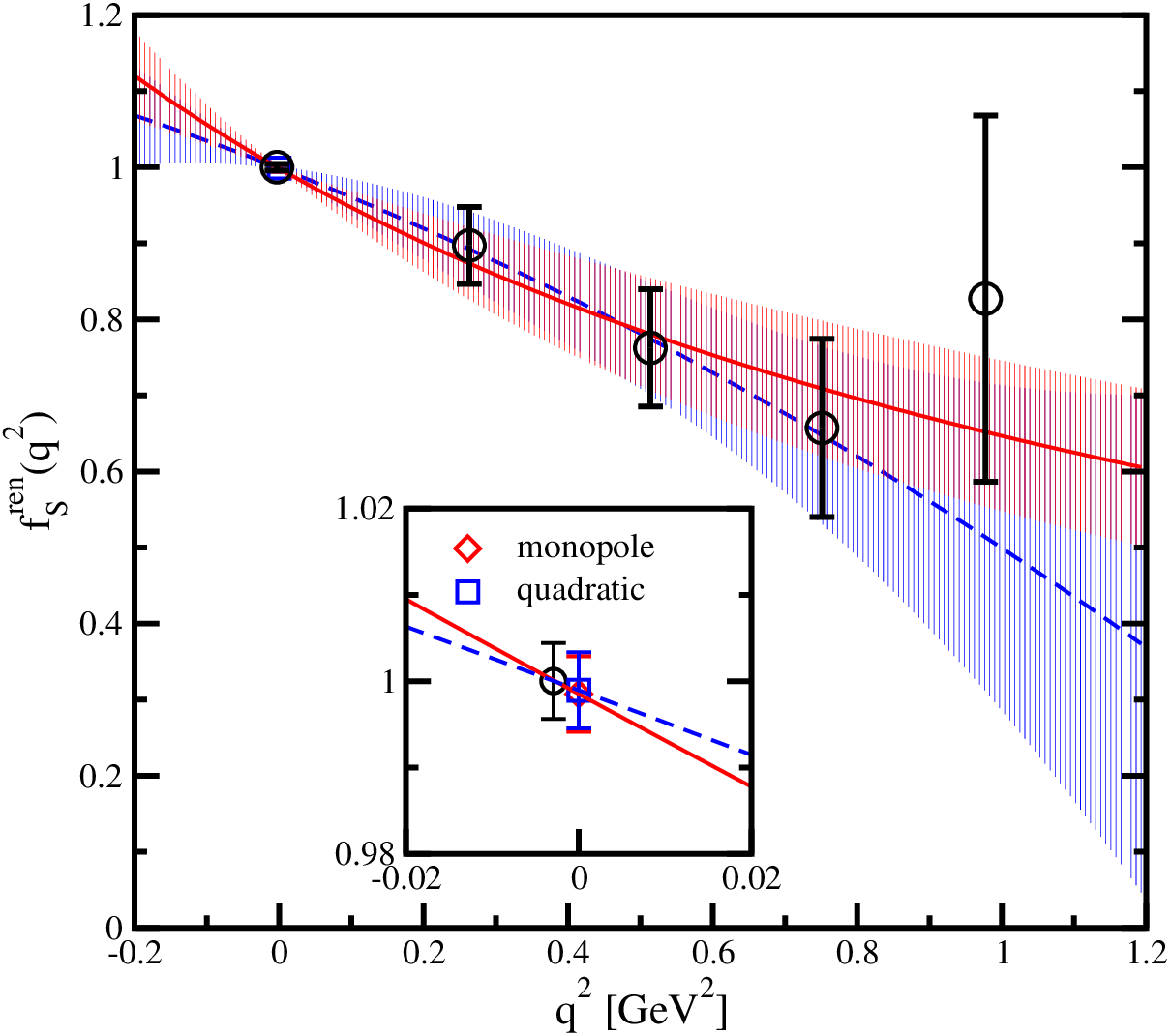}
\caption{Interpolation of $f_S(q^2)$ to $q^2=0$ for $(m_{ud}, m_s)=(0.04, 0.08)$.
Open circles are $f_S(q^2)$ at the simulated $q^2$. The solid (dashed) curve is the fitting result
by using the monopole (quadratic) interpolation form, while the open diamond (square) represents the interpolated value to $q^2=0$. 
}
\label{FIG:Qextra_Fs_408}
\ec
\end{figure}

The scalar form factor $f_S(q^2)$ at $q^2 >0$ is calculable  
with nonzero spatial momentum transfer 
($|{\bf q}|\neq 0$)~\footnote{Strictly speaking, 
it is true only if the spatial momentum transfer ${\bf q}$ satisfies the condition 
${\bf q}^2<(M_B^2-M_b^2)^2/(4M_b^2)$.
Our accessible finite momentum is much larger than $(M_B^2-M_b^2)^2/(4M_b^2)$ in this calculation.}.
We can make the $q^2$ interpolation of $f_S(q^2)$ to $q^2=0$ together with the precisely
measured value of $f_S(q^2)$ at $q^2=q^2_{\rm max}<0$. 
First of all, we calculate the following combinations of two projected correlators (\ref{Eq:Lambda_V_0}) and (\ref{Eq:Lambda_V_S}) with nonzero three 
momentum transfer ($|{\bf q}|\neq 0$):
\be
f_S^{\Btob}(q^2)=\frac{E_B - M_b}{M_B-M_b}{\Lambda^{V, \Btob}_0}-
\frac{E_B-M_B}{M_B-M_b}{\Lambda^{V, \Btob}_S}
\ee
and then study the $q^2$ dependence of the scalar form factor.
In Fig.~\ref{FIG:Qextra_Fs_408}, we plot the renormalized $f_S(q^2)$
as a function of four-momentum squared $q^2$ in physical units
for $(m_{ud}, m_s)=(0.04, 0.08)$ as a typical example.
Either theoretically or phenomenologically, the $q^2$ dependence
of $f_S(q^2)$ is not known due to the lack of knowledge of the
second-class form factor $f_3$. Our measured $f_S(q^2)$ up to at 
least $q^2<1.0$ GeV$^2$ exhibit a monotonic decrease
with increasing $q^2$. This observation is barely consistent 
with an expectation that $f_S(q^2)$ is dominated by $f_1(q^2)$, which is 
supposed to be the dipole form at low $q^2$.

In practice, the lack of the precise knowledge about the $q^2$ dependence
of $f_S(q^2)$ is not a serious issue to determine $f_S(0)$ reliably. 
The simulated value of $q_{\rm max}^2$
is not far from $q^2=0$~\footnote{
%%%%%%%% footnote %%%%%%%%
Even the physical value of 
$q^{2}_{\rm max}$ is relatively close to $q^2=0$ in the case of the
hyperon beta decays}.
%%%%%%%%%%%%%%%%%%%%%
Therefore, $f_S(0)$ can be determined by a 
very short interpolation from $q^2_{\rm max}$,
where we have very accurate data of $f_S(q^2_{\rm max})$. 
This allows us to expect that the choice of the interpolation form does 
not affect the interpolated value $f_S(0)$ significantly.
To demonstrate it, we test 
the monopole form
\be
f_S(q^2)=\frac{f_S(0)}{1+\lambda_S^{(1)}q^2}
\ee
and the quadratic form
\be
f_S(q^2)=f_S(0)(1+\lambda_S^{(1)}q^2 + \lambda_S^{(2)} q^4)
\ee
for the $q^2$ interpolation of $f_S(q^2)$ to $q^2=0$. 
Indeed, both functional forms are adopted to evaluate $f_S(0)$ in a published work 
of the $\Sigma^- \rightarrow n$ transition process~\cite{Guadagnoli:2006gj}.
As shown in Fig.~\ref{FIG:Qextra_Fs_408}, two determinations
to evaluate $f_S(0)=f_1(0)$ from measured points are indeed consistent 
with each other. All obtained values of $f_S(0)=f_1(0)$ from both the monopole  
and quadratic form fits are summarized in Table~\ref{Tab:f0andf1}. 
Although the quadratic fit achieves the slightly smaller value of 
$\chi^2/{\rm dof}\sim 0.04$ than that of the monopole fit ($\sim 0.23$), 
the highest $q^2$ point, which is not included in our fits, rather agrees 
with the monopole fit. We, therefore, do not use the results from the 
quadratic fit in the following discussion.

\begin{figure}[htbp]
\bc
\includegraphics[width=.47\textwidth,clip]{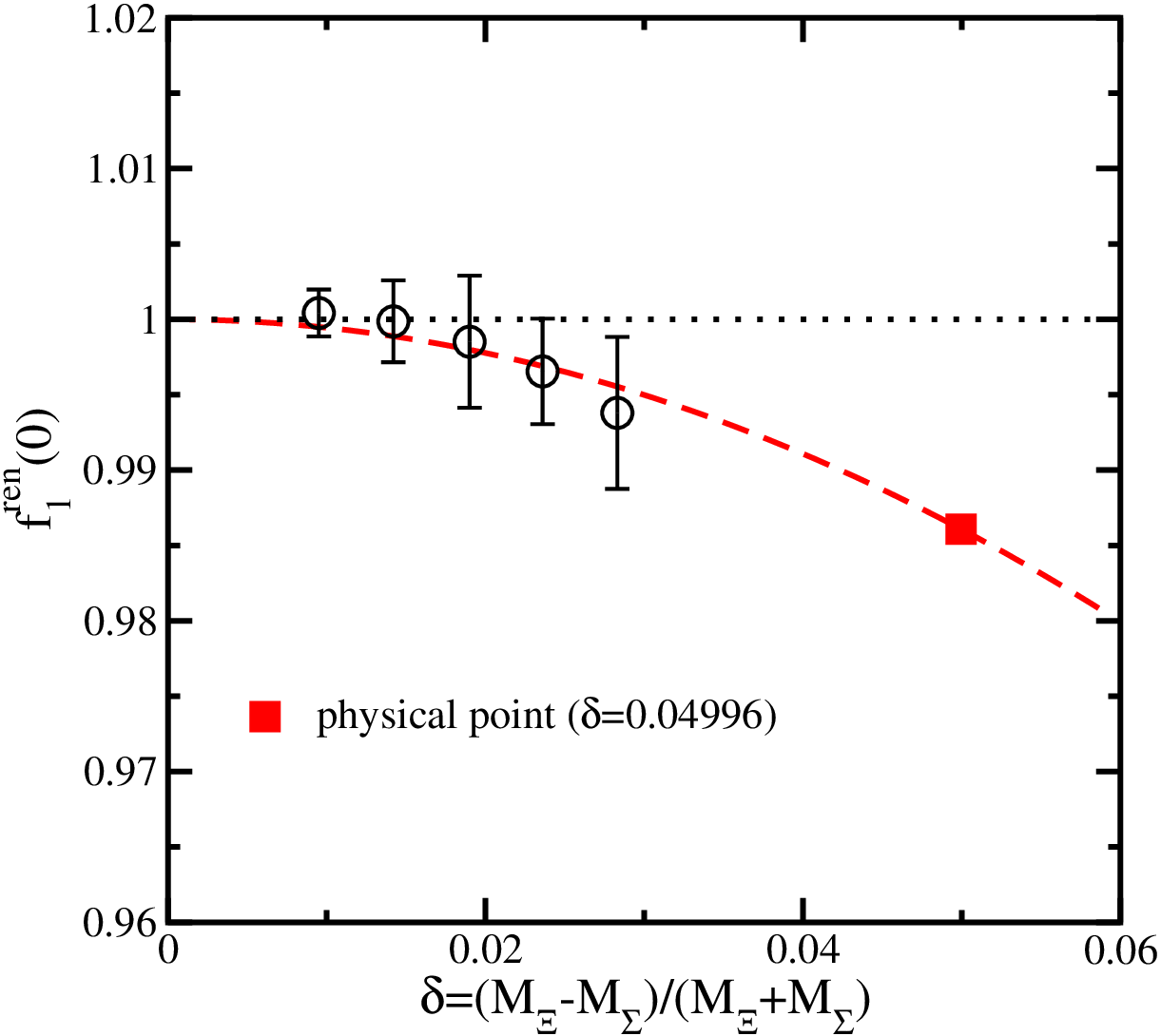}
\caption{The deviation from unity on $f_1(0)$
increases as the $SU(3)$ breaking $\delta$ increases. 
All measured values of $f_1(0)$ exhibit quadratic scaling with respect to the $SU(3)$-breaking parameter, which is suggested by the Ademollo-Gatto theorem. 
The dashed (quadratic) curve is a guide to the eye.
The physical point is represented by a filled square. 
}
\label{FIG:2ndOrd_on_f1}
\ec
\end{figure}

Although it is hard to make a firm conclusion within the current 
statistical uncertainty, our measured values of $f_1(0)$ at the 
simulated points seem to receive small negative corrections 
of the $SU(3)$ breaking.
We then introduce the parameter of flavor $SU(3)$ breaking, which is 
characterized by the measured mass difference between the 
$\Xi$ and $\Sigma$ states at the simulated points as 
$\delta=(M_{\Xi}-M_{\Sigma})/(M_{\Xi}+M_{\Sigma})$.
Our observed $SU(3)$-breaking effect on $f_1(0)$, which corresponds to the
deviation from unity, exhibits the quadratic dependence of this $SU(3)$-breaking
parameter $\delta$ in consistent with the Ademollo-Gatto theorem as shown in 
Fig.~\ref{FIG:2ndOrd_on_f1}. 
Therefore, our results indicate that a sign of the 
second-order correction of the $SU(3)$ breaking on $f_1(0)$ is likely negative.

%
%	Q2-extrapolation for f0
%
%

%
%	quadratic dependence of delta_mass
%
%

%\clearpage
%%%%%%%%%%%%%%%%%%%%%%%%%%%%%%%%%%%%%%%%%%
\subsection{Chiral extrapolation of $f_1(0)$}

\begin{figure}[htbp]
\bc
\includegraphics[width=.47\textwidth,clip]{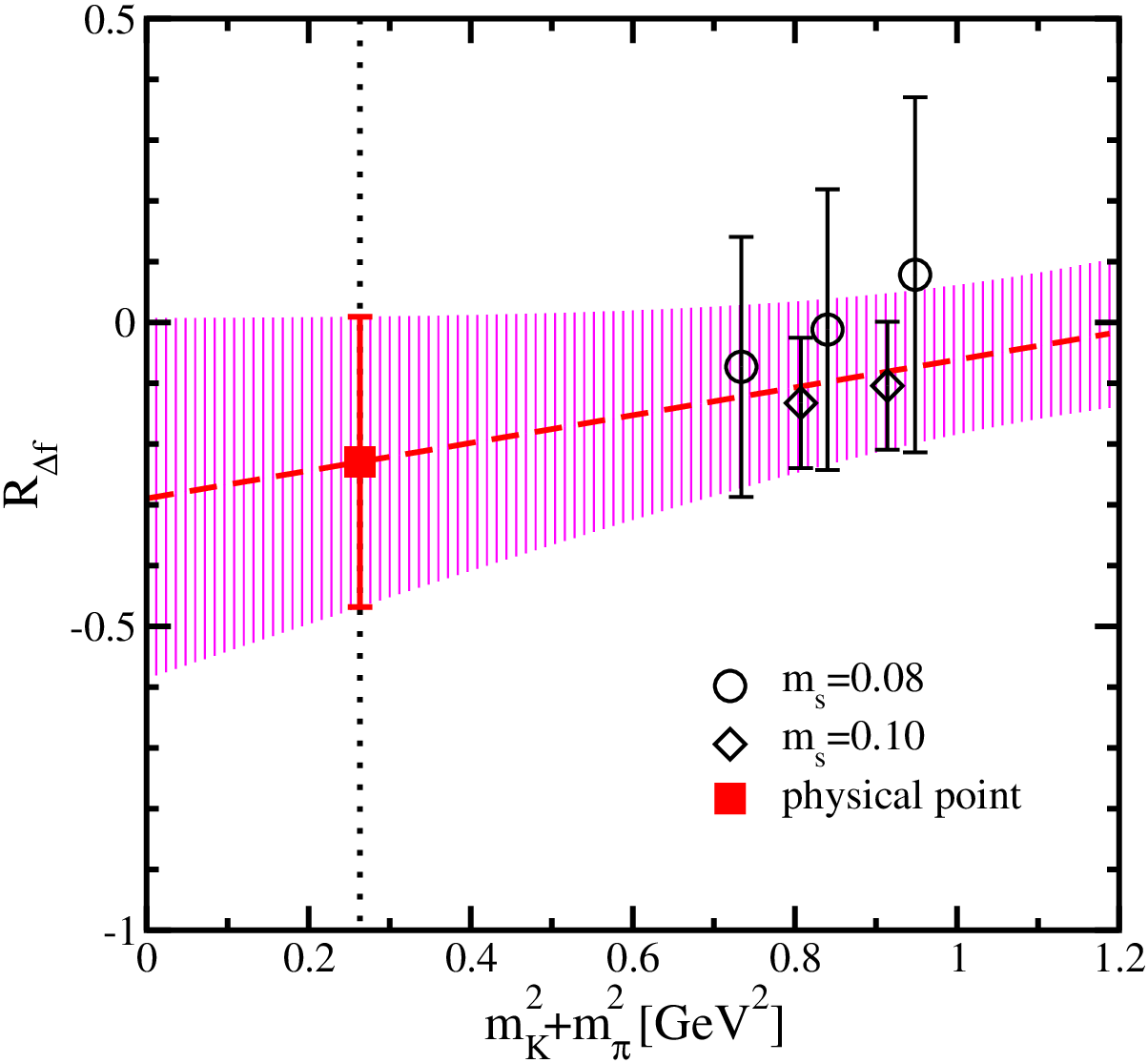}
\caption{Chiral extrapolation of $R_{\Delta f}$.  Open circles (diamonds)
are calculated with $m_s$=0.08 (0.10). 
The extrapolated point at the physical point is 
represented by a filled square. %{\bf [TEMP: $\chi^2/dof=0.58(1.32)/3$]}
}
\label{FIG:ChExt_on_f1}
\ec
\end{figure}

In order to estimate $f_1(0)$ at the physical point, we perform the chiral extrapolation of 
$f_1(0)$. As described earlier, $f_1(0)$ can be parameterized as
\be
f_1(0)=1+\Delta f , 
\ee
where $\Delta f$ represents all $SU(3)$ breaking corrections on $f_1(0)$.
We then introduce the following ratio:
%
% eq.
%
\be
R_{\Delta f}(M_K, M_\pi)=\frac{\Delta f}{(M_K^2 - M_{\pi}^2)^2} ,
\ee
where  the leading symmetry-breaking correction, which is predicted by the  Ademollo-Gatto theorem, is explicitly factorized out~\cite{Guadagnoli:2006gj}.
As reported in Ref. \cite{Guadagnoli:2006gj}, the remaining dependence of the $SU(3)$ 
breaking is hardly observed within the statistical errors. To examine the quark mass dependence, we plot $R_{\Delta f}$, which is evaluated by the monopole form for the $q^2$ interpolation, as a function of $M_K^2 + M_\pi^2$ 
in Fig.~\ref{FIG:ChExt_on_f1}. 
There is no appreciable dependence of simulated quark masses within 
the statistical errors. This observation is consistent with what we observe in Fig.\ref{FIG:2ndOrd_on_f1}, where
all measured values $f_1(0)$ at different simulated quark masses
exhibit quadratic scaling with respect to the $SU(3)$-breaking parameter $\delta$. 
We then consider a linear fit in terms of $M_K^2 + M_\pi^2$:
\be
R_{\Delta f}(M_K, M_\pi) = A_0 + A_1 \cdot (M_K^2+M_\pi^2).
\label{Eq:ChExt_linear}
\ee
A dashed line in Fig.~\ref{FIG:ChExt_on_f1} corresponds to 
the chiral extrapolation using the linear form~(\ref{Eq:ChExt_linear}). 
We obtain the extrapolated value of $R_{\Delta f}$ at the physical point as
%
% eq.
%
\be
R_{\Delta f}(M_{K}^{\rm phys}, M_{\pi}^{\rm phys})=-0.22(24)
\mbox{ in (GeV)}^{-4}
\ee
by employing result from the monopole form for the $q^2$ interpolation.
We finally quote 
%
% eq.
%
\be
\left[f^{\rm ren}_1(0)\right]_{\XtoS}= 0.987(19)
\ee
at the physical point. By combining with a single estimate of $|V_{us}f^{\XtoS}_1(0)|=0.216(33)$
from the KTeV experiment~\cite{Alavi-Harati:2001xk}, we obtain
\be
|V_{us}|=0.219(27)_{\rm exp}(4)_{\rm theory},
\ee
which is consistent with the value obtained from $K_{l3}$ decays
and the CKM unitary predicted 
value~\cite{{Boyle:2007qe},{Becirevic:2004ya}, {Dawson:2006qc}}.

%
%	Phys-extrapolation for f1
%
%

%
%	f0_qmax
%
\begin{table}[htbp]
\caption{Results for $Z_V^{\rm latt}(m_{ud}, m_{s})$, $f^{\rm latt}_S(q_{\rm max}^2)$ and 
$f^{\rm ren}_S(q_{\rm max}^2)$, which are evaluated in the region of 
$[t_{\rm min}, t_{\rm max}]=[13,17]$. 
}
\begin{ruledtabular}
\begin{tabular}{c c |  l  l  l}
\hline
$m_s$  & $m_{ud}$  
& $Z_V^{\rm latt}(m_{ud}, m_{s})$ 
& $f^{\rm latt}_S(q_{\rm max}^2)$
& $f^{\rm ren}_S(q_{\rm max}^2)$
\\
\hline
 0.08 
 &0.06 
 &  0.8144(7)
 & 1.2286(24)
 &  1.0010(15) \\
 &0.05 
 &  0.8128(8)
 &  1.2310(37)
 &  1.0010(27) \\
 &0.04 
 &  0.8112(10)
 &  1.2327(58)
 &  1.0005(44) \\
 0.10
 &0.05 
 &  0.8168(7)
 &  1.2240(46)
 &  1.0003(35) \\
 &0.04 
 &  0.8152(10)
 &  1.2248(65)
 &  0.9990(51)  \\
\hline
\end{tabular}
\end{ruledtabular}
\label{Tab:f0_qmax}
\end{table}
%

%
%	q^2 extrapolation f_s
%
\begin{table}[htbp]
\caption{Results for 
$\left[f^{\rm ren}_S(0)\right]_{\XtoS}=\left[f^{\rm ren}_1(0)\right]_{\XtoS}$, 
which are evaluated by the $q^2$ extrapolation with the monopole and 
quadratic functional form.
}
\begin{ruledtabular}
\begin{tabular}{c c |  l  l}
\hline
& &$\left[f^{\rm ren}_1(0)\right]_{\XtoS}$ &\\
$m_s$  & $m_{ud}$ 
& monopole
& quadratic
\\
\hline
 0.08 
 &0.06 
 &  1.0004(16)  &  1.0006(16) \\
 &0.05 
 &  0.9986(27) & 1.0001(28) \\
 &0.04 
 &  0.9985(44) & 0.9989(44) \\
 0.10
 &0.05 
 &  0.9965(35) & 0.9972(35) \\
 &0.04 
 &  0.9938(50) & 0.9947(51) \\
  \hline
---
&phys. point
& 0.9868(191)
& 0.9897(192)
\\
\hline
\end{tabular}
\end{ruledtabular}
\label{Tab:f0andf1}
\end{table}
%

%%%%%%%%%%%%%%%%%%%%%%%%%%%%%%%%%%%%%%%%%%%
%%%%%%%%%%%%%%%%%%%%%%%%%%%%%%%%%%%%%%%%%%%
%\clearpage
\section{Determination of $g_1(0)/f_1(0)$}
\label{Sec:ratio_g_1_ov_f_1}
\subsection{Ratio $\tilde{g}_1(q^2)/f_S(q^2)$ at $q^2=q^2_{\rm max}$}

\begin{figure}[htbp]
\bc
\includegraphics[width=.60\textwidth,clip]{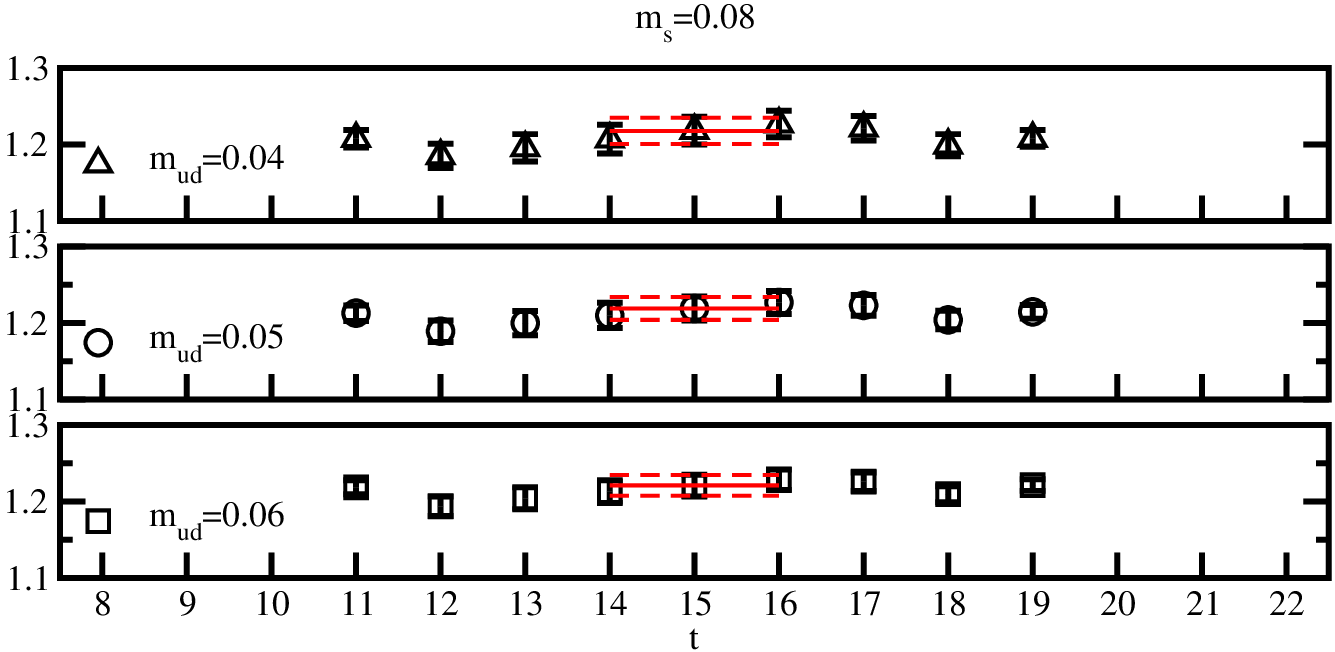}
\includegraphics[width=.60\textwidth,clip]{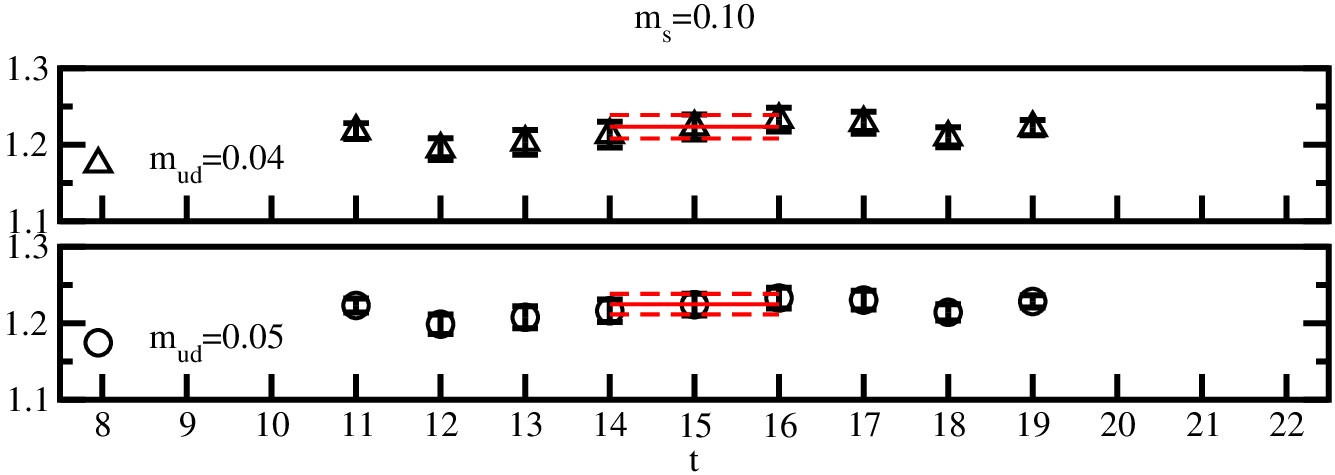}
\caption{$\tilde{g}_1(q^2_{\rm max})/f_S^{\rm ren}(q^2_{\rm max})$ as a function of the current insertion time slice.
The lines represent the average value (solid lines) and their 1 standard 
deviations (dashed lines) in the time-slice range $14 \le t \le 16$.
}
\label{FIG:plateau_G0F0max}
\ec
\end{figure}

Let us consider the ratio of $\Lambda_L^{A, \Btob}$ and 
$\Lambda_0^{V, \Btob}$,
which are both accessible at zero three-momentum transfer $|{\bf q}|=0$
in the axial-vector and vector channels respectively. 
From this ratio at $|{\bf q}|=0$, we can evaluate the value 
of $\tilde{g}_1(q^2)/f_S(q^2)$ at $q^2=q^2_{\rm max}$,
\be
\frac{\tilde{g}_1^{\Btob}(q^2_{\rm max})}{f_S^{\Btob}(q^2_{\rm max})}=
\frac{\Lambda_L^{A, \Btob}({\bf q}={\bf 0})}{\Lambda_0^{V, \Btob}({\bf q}={\bf 0})},
\ee
where $q_{\rm max}^2=-(M_B-M_b)^2$. We note that
this ratio is exactly equal to $g_1(0)/f_1(0)$ 
in the flavor $SU(3)$ symmetric limit, which corresponds to that of $\ntop$.
In Fig.~\ref{FIG:plateau_G0F0max}, we plot the ratio as a function
of the current insertion time slice. Good plateau behaviors 
of $\tilde{g}_1(q^2_{\rm max})/f_S(q^2_{\rm max})$ 
are observed in the middle of region between the source and sink points.
Results of $\tilde{g}_1(q^2_{\rm max})/f_S(q^2_{\rm max})$, which are
summarized in Table~\ref{Tab:g0f0_qmax_and_Dratio}, are averaged over 
appropriate time slice range $14\le t \le 16$.

\begin{figure}[htbp]
\bc
\includegraphics[width=.60\textwidth,clip]{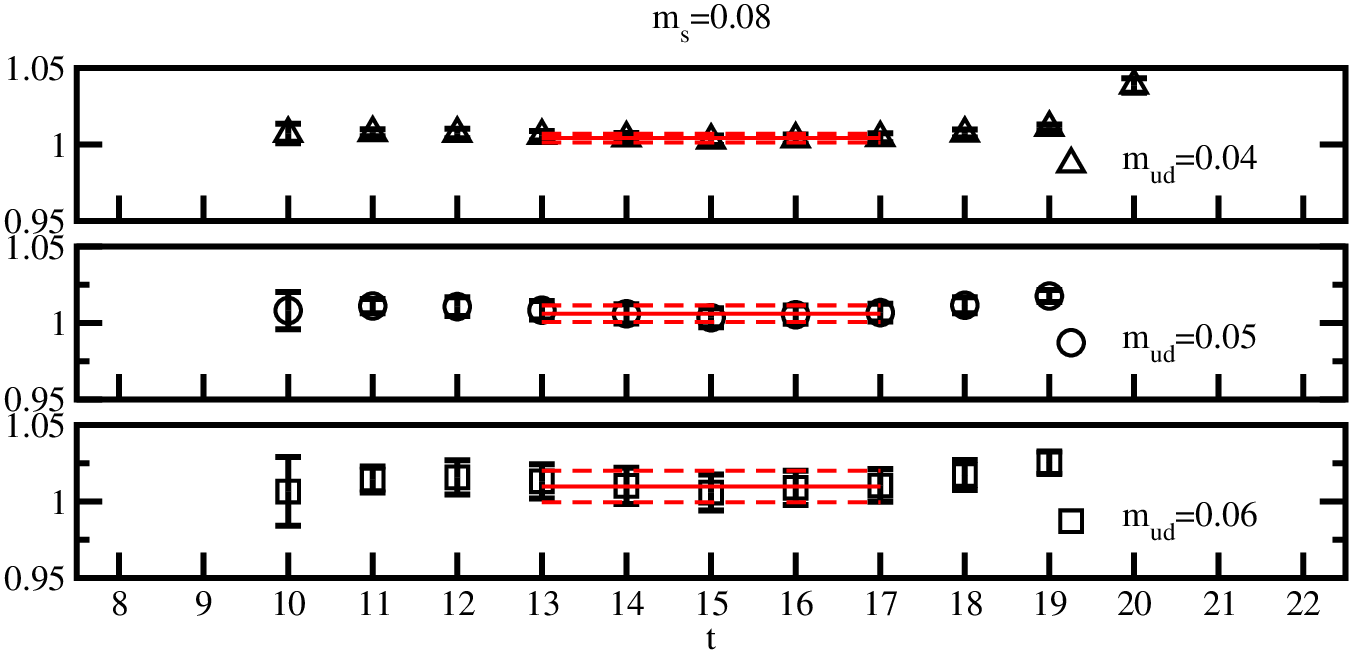}
\includegraphics[width=.60\textwidth,clip]{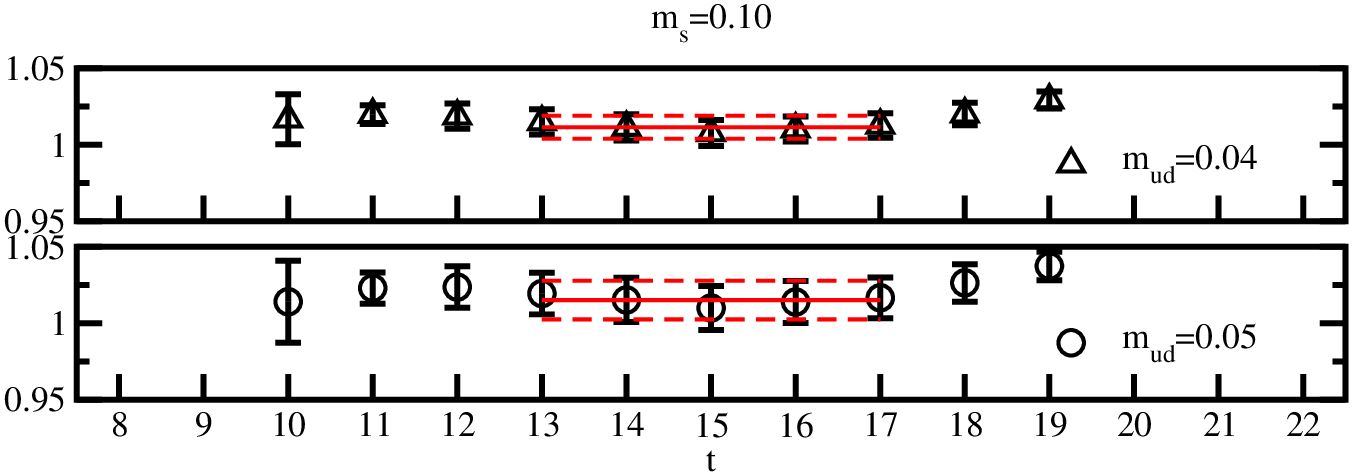}
\caption{The double ratio $R_{D}$ as a function of the current insertion time slice.
A source-sink location of three-point functions is set at 
$[t_{\rm src}, t_{\rm sink}]=[10,20]$.
The lines represent the average value (solid lines) and their one 
standard deviations (dashed lines) in the time-slice range $13 \le t \le 17$.
}
\label{FIG:plateau_DoubleR}
\ec
\end{figure}

In our exploratory study~\cite{Sasaki:2006jp}, we proposed 
the following double ratio:
\be
R_{D}=\frac{\Lambda^{A, \XtoS}_L({\bf q}={\bf 0}) \cdot \Lambda^{V, \ntop}_0({\bf q}={\bf 0})}{\Lambda^{V, \XtoS}_0({\bf q}={\bf 0}) \cdot \Lambda^{A, \ntop}_L({\bf q}={\bf 0})}
=\left.\left(\frac{g_1^{\XtoS}(q^2_{\rm max})-\delta g_2^{\XtoS}(q^2_{\rm max})}
{f_1^{\XtoS}(q^2_{\rm max})-\delta f_3^{\XtoS}(q^2_{\rm max})}\right)
\right/\left(\frac{g_1^{\ntop}(0)}{f_1^{\ntop}(0)}\right),
\ee
where $\delta=\frac{M_{\Xi}-M_{\Sigma}}{M_{\Xi}+M_{\Sigma}}$.
Since this double ratio is exactly equal to unity in the flavor 
$SU(3)$ symmetric limit, the deviation form unity exposes 
flavor $SU(3)$-breaking effects in 
the $\XtoS$ decay. As shown in Fig.~\ref{FIG:plateau_DoubleR}, the double ratio exhibits good plateau behavior slightly above unity in the time slice range $13\le t \le 17$. The deviation from unity becomes large as increasing the size of 
the flavor $SU(3)$-breaking, which is characterized by the size of $\delta$ as listed in Table~\ref{Tab:g0f0_qmax_and_Dratio}

The observed deviation indeed contains three types 
of the $SU(3)$ breaking effect, similar to what we explained for 
$f_S(q^2_{\rm max})$. Here, we note that $g_1(0)/f_1(0)$ receives the first-order corrections since the axial-vector form factors are not protected by the Ademollo-Gatto theorem. Therefore, we expect that the flavor $SU(3)$-breaking observed in the double ratio could be dominated by 
the leading symmetry-breaking correction on $[g_1(0)/f_1(0)]_{\XtoS}$. The reasons as follows:
1) The $q^2$ dependence of form factors
at $q^2=q^2_{\rm max}$, which is proportional to $\delta^2$, can be involved in the second-order corrections as 
$f_1^{\XtoS}(q^2_{\rm max})=f_1^{\XtoS}(0)+{\cal O}(\delta^2)$ and
$g_1^{\XtoS}(q^2_{\rm max})=g_1^{\XtoS}(0)+{\cal O}(\delta^2)$.
2) The nonzero value of the second-class form factors $f_3$ and $g_2$
starts from the first-order corrections. These contributions in $\tilde{g_1}$ 
and $f_S$ are involved in the second-order corrections as well.
As a result, the double ratio is expressed by 
\be
R_{D} =\frac{[g_1(0)/f_1(0)]_{\XtoS}}{[g_1(0)/f_1(0)]_{\ntop}}+{\cal O}(\delta^2),
\ee
where the first term is responsible for the leading first-order correction.
% is supposed to receive the first-order corrections.
The deviation from unity observed in the double ratio may be able to 
exhibit the size of the leading $SU(3)$-breaking correction on 
$g_1(0)/f_1(0)$ for small $\delta$~\footnote{
It should be reminded that the value of $\delta$ for $\XtoS$ 
at the physical point is 0.04996, which is indeed small.
Here, we evaluate $\delta$ with $M_\Sigma=1193.2$ MeV and $M_\Xi=1318.7$ MeV, which are the isospin-averaged masses.}.
As listed in Table~\ref{Tab:g0f0_qmax_and_Dratio}, the observed size of flavor $SU(3)$ breaking effects is indeed comparable of the size of the leading order 
${\cal O}(\delta)$~\cite{Sasaki:2006jp} and glows linearly with 
increasing the parameter $\delta$.

We simply perform the linear fit in two mass combinations $M_K^2+M_{\pi}^2$
and $M_K^2-M_{\pi}^2$ on $R_{D}$ as
\be
R_D = B_0 + B_1 \cdot (M_K^2+M_\pi^2) + B_2 \cdot (M_K^2-M_{\pi}^2).
\ee
We then obtain $R_{D}=1.022(31)$ at the physical point. A sign of the observed corrections to $[g_1(0)/f_1(0)]_{\XtoS}$ seems to be opposite to model predictions from the center-of-mass correction approach~\cite{Ratcliffe:1998su} and the $1/N_c$ expansion approach~\cite{Flores-Mendieta:1998ii}. 
However we recall that the observed corrections less than a few percents
are too small to justify neglect of the second-order corrections in our analysis
since the natural size of the flavor $SU(3)$ breaking is around 10\%~\footnote{
%%%%% footnote %%%%%%%
In the quantitative sense, the size of $2\delta$ is more relevant than the size of 
$\delta$ for quoting the real size of the flavor $SU(3)$ breaking.}.
%%%%%%%%%%%%%%%%%
To make a firm conclusion, we have to removes systematic uncertainties
induced by neglecting both the recoil corrections and the presence of 
the second-class form factors.

%
%	plateaus of g0f0
%
%

%
%	plateaus of DoubleRatio
%
%

%
%	Table for g0/fs at q_max^2
%
\begin{table}[htbp]
\caption{
Results for $\tilde{g}_1(q^2_{\rm max})/f_S(q^2_{\rm max})$, which are evaluated 
in the region of $[t_{\rm min}, t_{\rm max}]=[14,16]$. 
The double ration $R_{D}$ and the $SU(3)$ breaking parameter $\delta$
are also listed.
}
\begin{ruledtabular}
\begin{tabular}{cc|cll}
\hline
$m_s$  & $m_{ud}$  
& $\tilde{g}_1(q_{\rm max}^2)/f_S(q_{\rm max}^2)$ & $R_{D}$
& $\delta=\frac{M_{\Xi}-M_{\Sigma}}{M_{\Xi}+M_{\Sigma}}$
\\
\hline
 0.08 
 & 0.06 
 & 1.221(13)
 & 1.0042(28) 
 & 0.0095(3)
 \\
 & 0.05 
 &  1.219(15)
 &  1.0061(55)
 &  0.0142(5)
\\
 &0.04 
 &  1.218(17)
 &  1.0099(103)
 &  0.0190(7)
 \\
  0.10
 &0.05 
 &  1.225(13)
 &  1.0114(75)
 &  0.0236(7)
\\
 &0.04 
 &  1.223(15)
 &  1.0151(126)
 &  0.0283(10)
\\
\hline
& phys. point
& ---
& 1.0217(309) 
& 0.04996
\\
\hline
\end{tabular}
\end{ruledtabular}
\label{Tab:g0f0_qmax_and_Dratio}
\end{table}
%

%
%	Table for g0(0)/fs(0)
%
\begin{table}[htbp]
\caption{
Results for $\tilde{g}_1(0)/f_1(0)$, which are evaluated 
by the $q^2$ extrapolation with the monopole and quadratic functional
form.
}
\begin{ruledtabular}
\begin{tabular}{cc|cc}
\hline
&&$\tilde{g}_1(0)/f_1(0)$ &\\
$m_s$  & $m_{ud}$  
& monopole
& quadratic
\\
\hline
 0.08 
 &0.06 
 & 1.220(13)
 &1.220(13) \\
 &0.05 
 &  1.217(15)
 &  1.217(15) \\
 &0.04 
 &  1.215(17)
 &  1.215(17)\\
 0.10
 &0.05 
 &  1.222(13)
 &  1.222(13)\\
 &0.04 
 &  1.220(15)
 &  1.220(15)\\
\hline
---
& phys. point
& 1.205(27)
& 1.206(27)\\
\hline
\end{tabular}
\end{ruledtabular}
\label{Tab:g0f0_q0}
\end{table}
%

%\clearpage
\subsection{Results for $\tilde{g}_1(0)/f_1(0)$ and $g_1(0)/f_1(0)$}

\begin{figure}[htbp]
\bc
\includegraphics[width=.47\textwidth,clip]{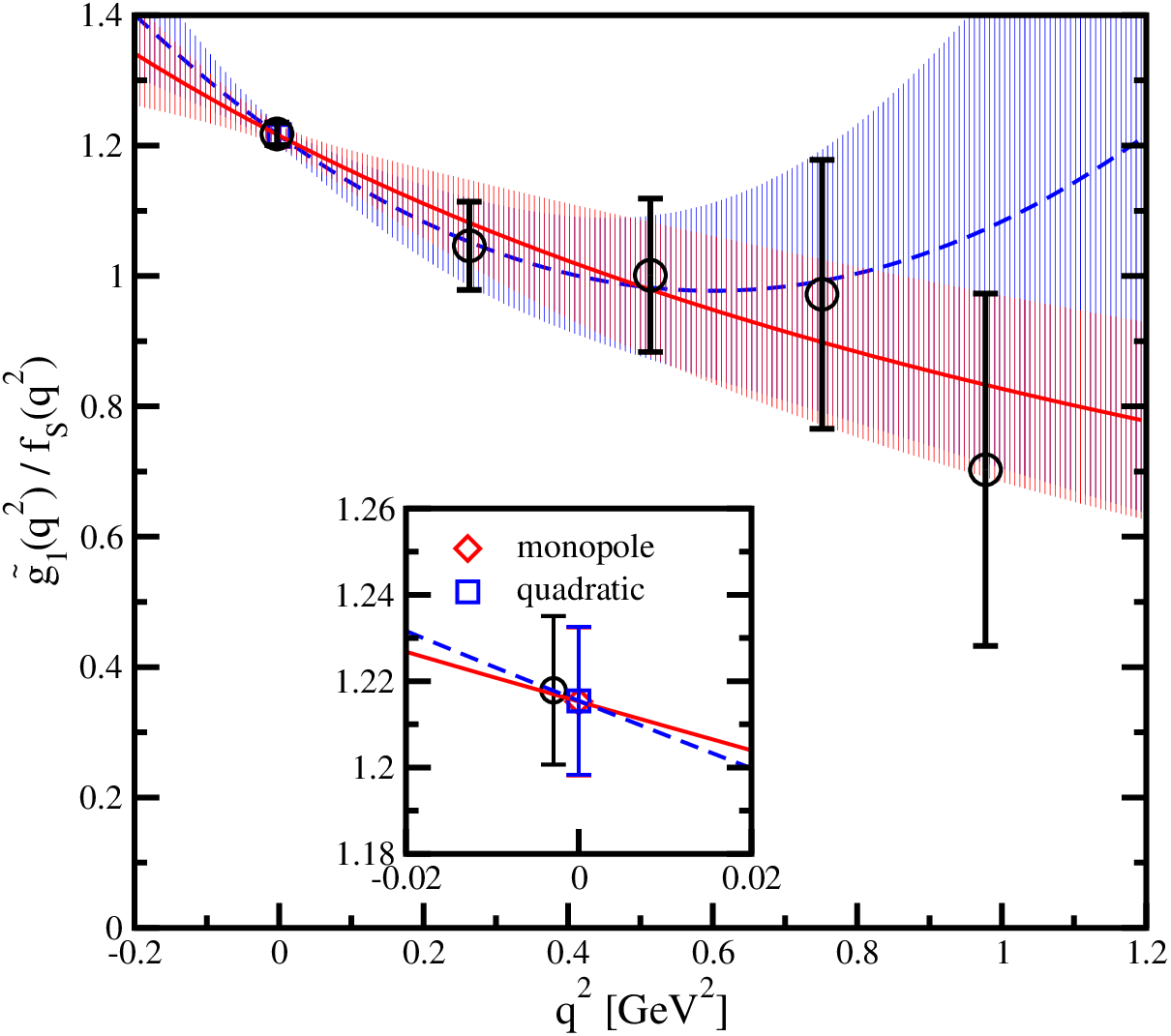}
\caption{Interpolation of $\tilde{g}_1(q^2)/f_S(q^2)$ to $q^2=0$ for $(m_{ud}, m_s)=(0.04, 0.08)$.
Open circles are $\tilde{g}_1(q^2)/f_S(q^2)$ at the simulated $q^2$. 
The solid (dashed) curve is the fitting result
by using the monopole (quadratic) interpolation form, while the open diamond (square) represents the interpolated value to $q^2=0$. }
\label{FIG:Qextra_G0F0_408}
\ec
\end{figure}

The recoil corrections are removed by taking the limit of considered form factors 
to $q^2=0$. In similar to the case of $f_S(q^2)$, the value of $\tilde{g}_1(0)/f_S(0)$ can be evaluated by the $q^2$ interpolation of $\tilde{g}_1(q^2)/f_S(q^2)$ to $q^2=0$. The form factor $\tilde{g}_1^{\Btob}(q^2)$ at $q^2>0$ can be calculated by the projected correlator (\ref{Eq:Lambda_A_L})
with zero longitudinal momentum ($q_z=0$) but nonzero transverse momentum 
($q_x$ or $q_y\neq 0$)~\footnote{In the case of ${\bf n}^2=3$, we use Eq.~(\ref{Eq:Lambda_L_qz=0}) for evaluate $\Lambda_L^{A, \Btob}(q_z=0)$.}:
\be
\tilde{g}_1^{\Btob}(q^2) = \Lambda_L^{A, \Btob}(q_z=0).
\label{Eq:g1tilde_ff}
\ee
One can calculate the ratio of $\tilde{g}_1(q^2)/f_S(q^2)$ with 
Eqs.~(\ref{Eq:scalar_ff}) and (\ref{Eq:g1tilde_ff}).
In Fig.~\ref{FIG:Qextra_G0F0_408}, we plot the ratio 
of $\tilde{g}_1(q^2)/f_S(q^2)$ as a function of $q^2$. 
We consider two types of the interpolation form, the monopole 
and quadratic forms, the same as in the case of $f_S(q^2)$. 
The highest $q^2$ data point is not included in our fits.
All obtained values of $\tilde{g}_1(0)/f_S(0)=\tilde{g}_1(0)/f_1(0)$
from both the monopole and quadratic form fits are summarized
in Table~\ref{Tab:g0f0_q0}.
Again we observe that two determinations to evaluate 
$\tilde{g}_1(0)/f_S(0)$ from measured points 
are fairly consistent with each other. 
Therefore, we do not use the results from the quadratic fit in the 
following discussion, the same in the case of $f_S(0)$.

Next, we perform a linear fit in two mass combinations $M_K^2+M_\pi^2$ and $M_K^2-M_{\pi}^2$ for the values of $\tilde{g}_1(0)/f_1(0)$ and then obtain $[\tilde{g}_1(0)/f_1(0)]_{\XtoS}=1.205(27)$
at the physical point. It may be compared with its $SU(3)$-symmetric
value of 1.191(49), which corresponds to the chiral extrapolated value 
of $[g_1(0)/f_1(0)]_{\ntop}$ to the physical point by using the simple 
linear fitting form in terms of $M_{\pi}^2$. These results give
$[\tilde{g}_1(0)/f_1(0)]_{\XtoS}=1.016(31)\times [g_1(0)/f_1(0)]_{\ntop}$.

After the $q^2$ dependence is taken into account, the deviation from unity is
now reduced by 0.6\% from the double ratio $R_D$.
If the conventional assumption $g_2(0)=0$ 
is adopted here similar to usual experimental analyses, the flavor $SU(3)$-breaking 
found in $g_1(0)/f_1(0)$ tends to be tiny. Although it does not conflict with 
the Cabibbo theory, the following alternative interpretation 
is still {\it not} ruled out. It is possible that a relatively small first-order correction 
on $g_1(0)$ is accidently canceled out in the form $\tilde{g}_1=g_1-\delta g_2$ by an opposite contribution stemming from the {\it large and positive value} of 
$g_2(0)$ such as $g_2(0)/g_1(0)\sim 1$. As we will discuss in next
section, {\it this is indeed the case}. 
Therefore, we have to subtract the contribution of the second-class 
form factor properly in order to estimate the true size of the first-order 
correction on $g_1(0)/f_1(0)$. 

A complete analysis requires information of the $g_2$ form factor.  
The individual form factors in Eqs.~(\ref{Eq:VcMat}) and (\ref{Eq:AxMat}) 
can be determined at finite $|{\bf q}|$.
Then, the value of the ratio of $g_2(q^2)/g_1(q^2)$ at zero 
momentum transfer are obtained by an appropriate $q^2$ extrapolation. 
We finally obtain $g_2(0)/g_1(0)=0.677(177)$ at the physical point.
See the next section for details. 

After the subtraction of the $g_2(0)$ contribution, we obtain $[g_1(0)/f_1(0)]_{\XtoS}=1.248(29)$ at the physical point. 
It implies that the relatively large and positive value of $g_2(0)/g_1(0)$ 
has induced a shift of the value of $\tilde{g}_1(0)/f_1(0)$ toward 
the exact $SU(3)$-symmetric value. Finally, we obtain
\be
\left[\frac{g_1(0)}{f_1(0)}\right]_{\XtoS}=1.051(35)\times \left[\frac{g_1(0)}{f_1(0)}\right]_{\ntop}
\ee
at the physical point. The deviation from unity is increased by 3.5\% in comparison with the unsubtracted case. Although the size of the $SU(3)$-breaking corrections on $g_1(0)/f_1(0)$ is still relatively smaller than the expected size evaluated from the mass splitting among the octet baryons ($\sim 10 \%$), the similar size of the flavor $SU(3)$ breaking in the $\Sigma^+\rightarrow n$ decay was also reported in 
Ref.~\cite{Guadagnoli:2006gj}. 

\begin{figure}[htbp]
\includegraphics[width=0.6\textwidth,clip]{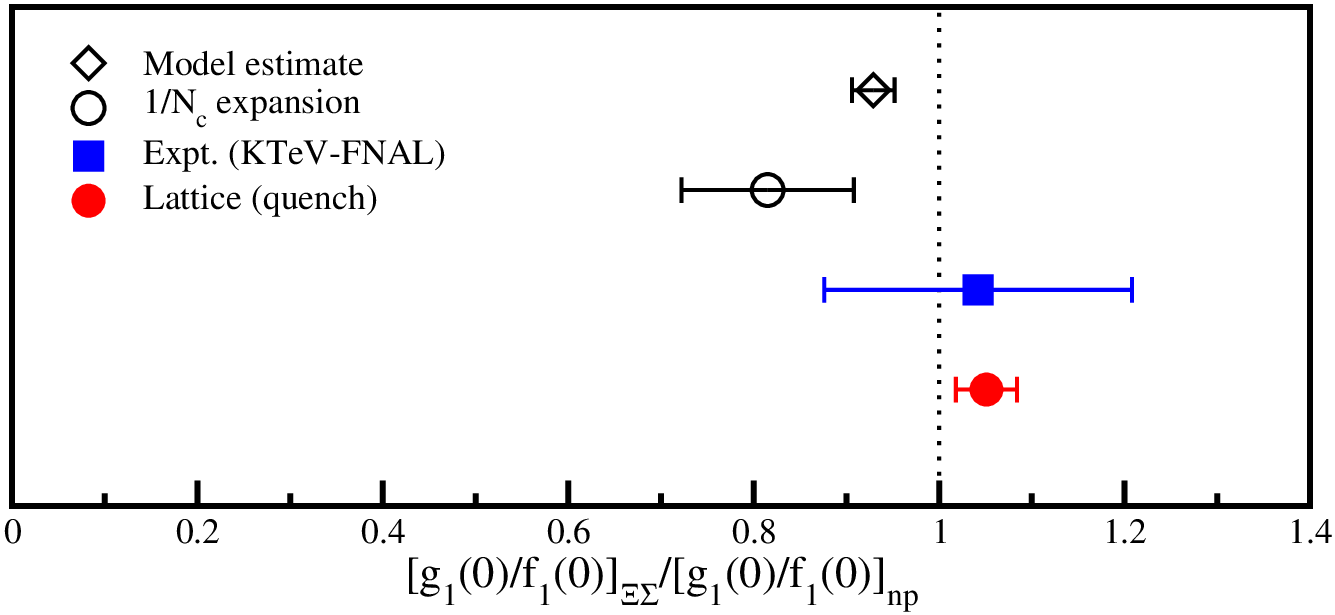}
\caption{Comparison among model predictions, experimental data (KTeV-FNAL)
and our lattice result for the ratio of $[g_1(0)/f_1(0)]_{\XtoS}$ and
its $SU(3)$ counterpart $[g_1(0)/f_1(0)]_{\ntop}$.}
\label{FIG:g1f1ratioComp}
\end{figure}

In Fig.~\ref{FIG:g1f1ratioComp}, we summarize our result and the experimental values combined with predictions from the center-of-mass correction approach~\cite{Ratcliffe:1998su} and the $1/N_c$ expansion approach~\cite{Flores-Mendieta:1998ii}.  
Although the experimental data is not yet sufficiently precise to determine 
either the size, or the sign, of the $SU(3)$-breaking corrections, our result 
suggests that the symmetry-breaking correction is {\it likely small but positive}. 
It is worth mentioning that the sign of our observed corrections is opposite 
to the model predictions, but in agreement with that of the 
$\Sigma^+\rightarrow n$ decay measured 
in quenched lattice QCD~\cite{Guadagnoli:2006gj}.

%
%	Q2-extrapolation for g0/fs
%
%

%%%%%%%%%%%%%%%%%%%%%%%%%%%%%%%%%%%%%%%%%%%
%%%%%%%%%%%%%%%%%%%%%%%%%%%%%%%%%%%%%%%%%%%
%\clearpage
\section{Results for other form factors}
\label{Sec:other_FF}
\subsection{Computational Method}

The kinematics of $|{\bf q}|^2=0$ allows only a particular 
combination of the projection operator (${\cal P}$) and the Lorentz index of 
the currents ($\alpha$) in either vector or axial-vector channels~\cite{Sasaki:2003jh}. 
However, in the case if spatial momentum transfer $\bf q$ is nonzero, all three-point 
correlation functions defined in Sec.~\ref{Sec:numeric_3ptfunc} 
are calculable. Therefore, three form factors at finite $|{\bf q}|$ can be obtained individually by solving simultaneous linear equations. For the vector channel, the simultaneous linear equations are given by
%
% eq.
%
\be
\left(
\begin{array}{ccc}
1 & -\frac{E_B -M_B}{M_B+M_b} & -\frac{E_B-M_b}{M_B+M_b} \\
1 & -\frac{E_B -M_b}{M_B+M_b} & -\frac{E_B+M_B}{M_B+M_b} \\
1 & 1 & 0
\end{array}
\right)
\left(
\begin{array}{c}
f^{\Btob}_1 (q^2)\\
f^{\Btob}_2 (q^2)\\
f^{\Btob}_3 (q^2)
\end{array}
\right)
=
\left(
\begin{array}{c}
{\Lambda}^{V,\Btob}_0 \\
{\Lambda}^{V,\Btob}_S \\
{\Lambda}^{V,\Btob}_{T}
\end{array}
\right). 
\ee
One gets each form factor by inverting the 
above equations~\footnote{We found an error in 
the third term of Eq. (20) in Ref.~\cite{Guadagnoli:2006gj}.
Note that we formulate both of  Eqs.~(\ref{Eq:VecFormFacts_f1})-(\ref{Eq:VecFormFacts_f3})
and Eqs.~(\ref{Eq:AxvFormFacts_g1})-(\ref{Eq:AxvFormFacts_g3}) in a fully covariant way, 
while only a single component of the Dirac indices is considered 
in Ref.~\cite{Guadagnoli:2006gj}.} as
\bea
f_1^{\Btob}(q^2) &=& \frac{M_B+M_b}{2M_b} 
\left[
{\Lambda}^{V,\Btob}_0 
- \frac{E_B-M_b}{E_B+M_B} {\Lambda}^{V,\Btob}_S 
- \frac{M_B^2+M_b^2-2E_BM_b}{(M_B+M_b)(E_B+M_B)} {\Lambda}^{V,\Btob}_T
\right] ,
\label{Eq:VecFormFacts_f1}
\\
f_2^{\Btob}(q^2) &=& \frac{M_B+M_b}{2M_b}
\left[-{\Lambda}^{V,\Btob}_0 
+ \frac{E_B-M_b}{E_B+M_B} {\Lambda}^{V,\Btob}_S 
+ \frac{M_B+M_b}{E_B+M_B} {\Lambda}^{V,\Btob}_T
\right] ,
\label{Eq:VecFormFacts_f2}
\\
f_3^{\Btob}(q^2) &=& \frac{M_B+M_b}{2M_b}
\left[{\Lambda}^{V,\Btob}_0 
- \frac{E_B+M_b}{E_B+M_B} {\Lambda}^{V,\Btob}_S 
- \frac{M_B-M_b}{E_B+M_B} {\Lambda}^{V,\Btob}_T
\right] .
\label{Eq:VecFormFacts_f3}
\eea

Similarly, three-point correlation functions of the axial-vector part are
also described by the following simultaneous linear equations,
\be
\left(
\begin{array}{ccc}
1 & -\frac{M_B-M_b}{M_B+M_b}& 0\\
0 & \frac{M_b}{M_B+M_b} & \frac{M_b}{M_B+M_b}\\
1 & -\frac{E_B+M_B}{M_B+M_b} & -\frac{E_B-M_b}{M_B+M_b}
\end{array}
\right)
\left(
\begin{array}{c}
g_1^{\Btob}(q^2) \\
g_2^{\Btob}(q^2) \\
g_3^{\Btob}(q^2)
\end{array}
\right)
\\
=
\left(
\begin{array}{c}
{\Lambda}^{A, \Btob}_{L}(q_z=0) \\
{\Lambda}^{A, \Btob}_{T} \\
{\Lambda}^{A, \Btob}_{0}
\end{array}
\right)
\ee
and then each individual form factor is given by inverting the above equation as
\bea
g_1^{\Btob}(q^2)&=&
\frac{M_B+M_b}{2M_b}
\left[
{\Lambda}^{A, \Btob}_{L}(q_z=0)
-\frac{M_B-M_b}{M_B+M_b}\left\{
{\Lambda}^{A, \Btob}_{0}
+\frac{E_B-M_b}{M_b}{\Lambda}^{A, \Btob}_{T}
\right\}
\right],
\label{Eq:AxvFormFacts_g1}
\\
g_2^{\Btob}(q^2)&=&
\frac{M_B+M_b}{2M_b}\left[
{\Lambda}^{A, \Btob}_{L}(q_z=0)-{\Lambda}^{A, \Btob}_{0}
-\frac{E_B-M_b}{M_b} {\Lambda}^{A, {\Btob}}_{T}\right],
\label{Eq:AxvFormFacts_g2}
\\
g_3^{\Btob}(q^2)&=&
\frac{M_B+M_b}{2M_b}\left[
-{\Lambda}^{A, \Btob}_{L}(q_z=0)+{\Lambda}^{A, \Btob}_{0}
+\frac{E_B+M_b}{M_b}
{\Lambda}^{A, \Btob}_{T}\right].
\label{Eq:AxvFormFacts_g3}
\eea
Here we remark that as described previously, we utilize all possible 
permutations of the lattice momentum including 
both positive and negative directions and adopt four nonzero values of three-momentum transfer 
${\bf q}=\frac{2\pi}{L}{\bf n}$ (${\bf n}^2=1$, 2, 3, 4).
Here, it should be reminded that the $z$-direction is chosen as the polarized direction 
in this study. This fact makes  the analysis of the axial-vector channel 
more complex than the vector channel. Indeed, the longitudinal momentum 
($q_z$) dependence explicitly enters in $\Lambda^{A}_{L}$. 
Accordingly, there are two types of kinematics, $q_z\neq 0$ and $q_z=0$ 
in the three-momentum transfer, except for the case of ${\bf n}^2=3$ where 
$q_z$ is always nonzero. In other words, ${\Lambda}^{A, \Btob}_{L}(q_z=0)$ 
at ${\bf n}^2=3$ can not be calculated directly. However, instead, we may 
evaluate ${\Lambda}^{A, \Btob}_{L}(q_z=0)$ at ${\bf n}^2=3$
from the longitudinal correlator ${\Lambda}^{A, \Btob}_{L}$ 
and the transverse correlator ${\Lambda}^{A, \Btob}_{T}$
by using a relation
%
% eq.
%
\be
{\Lambda}^{A, \Btob}_L(q_{z}=0)={\Lambda}^{A, \Btob}_L(q_{z}\neq 0) +\frac{q_z^2}{M_b(E_B+M_B)}
{\Lambda}^{A, \Btob}_T,
\label{Eq:Lambda_L_qz=0}
\ee
which is easily read off from Eq.~(\ref{Eq:3pt_axial_spatial}).
Note that $\Lambda_L^A(q_z\neq 0)$ are always calculable at finite $|{\bf q}|$.

%\clearpage
\subsection{Second-class form factors: $f_3(q^2)$ and $g_2(q^2)$}
\label{Sec:other_FF_2ndclass}
\subsubsection{Subtraction method}

\begin{figure}[htbp]
\bc
\includegraphics[width=.47\textwidth,clip]{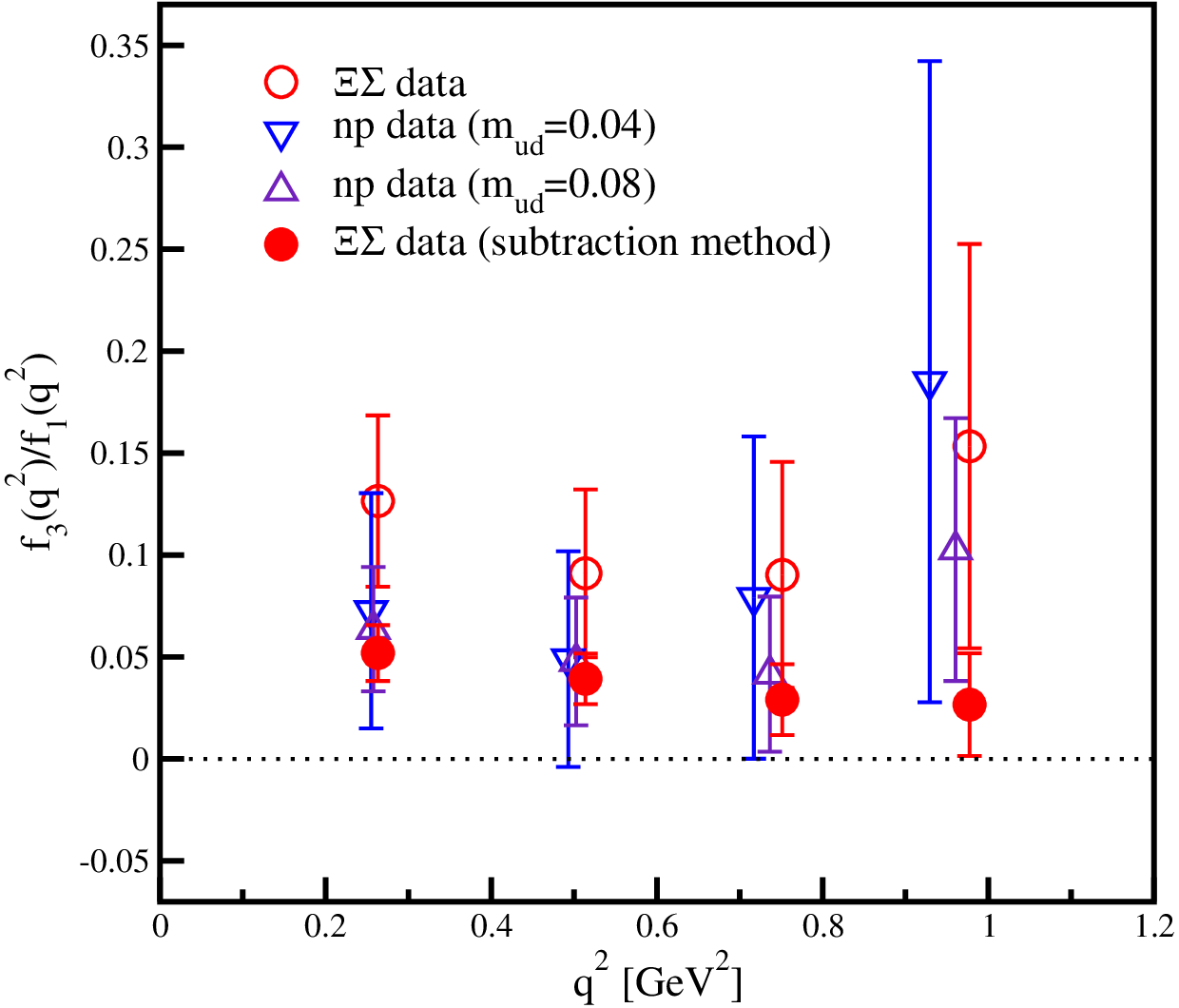}
\includegraphics[width=.47\textwidth,clip]{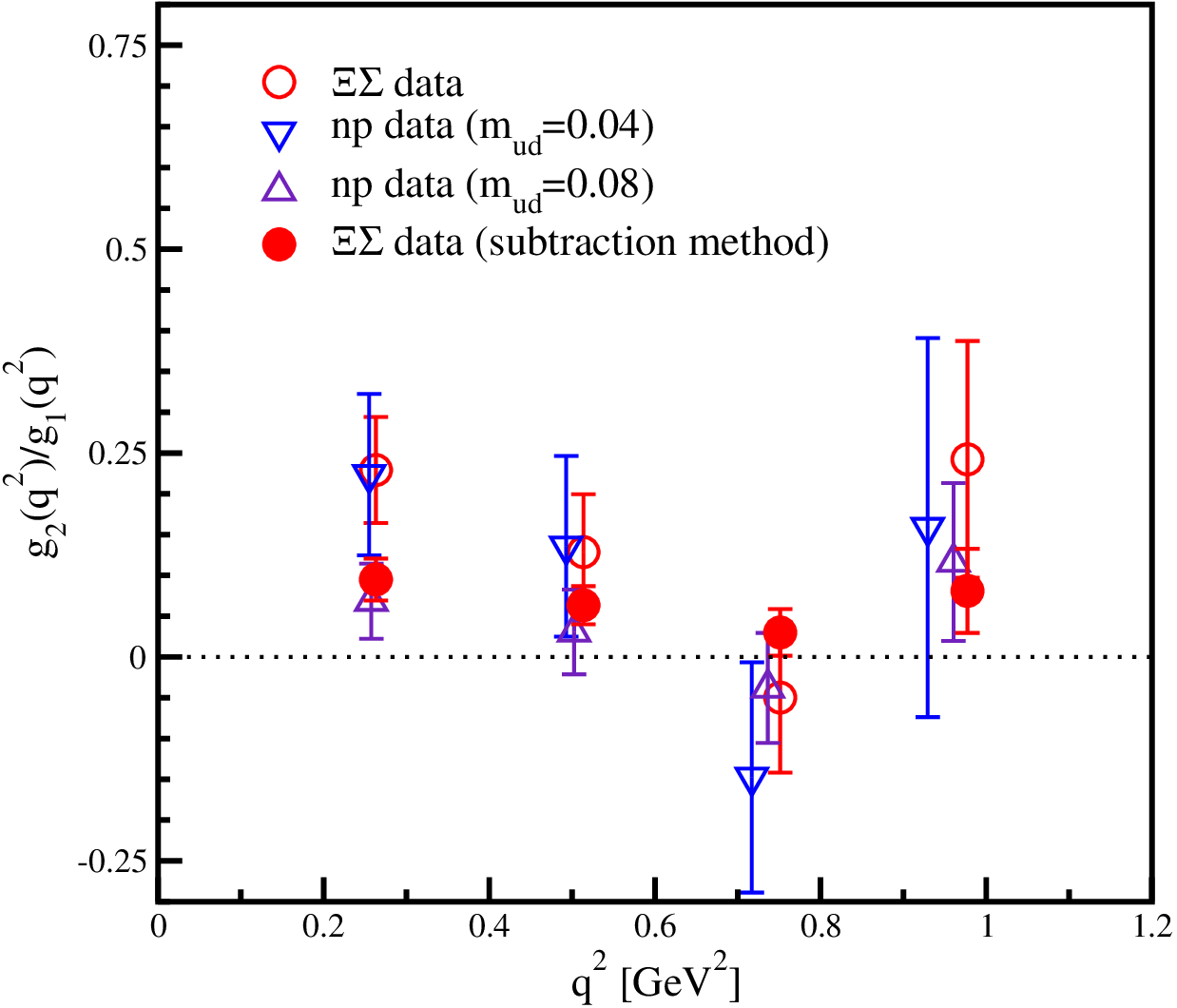}
\caption{
%Subtraction
The subtraction method can expose non-zero values of 
the second-class form factors, which corresponds to 
the $SU(3)$-breaking effect, with better statistical precision.
}
\label{FIG:2ndClass_comp}
\ec
\end{figure}

Figure~\ref{FIG:2ndClass_comp} shows the ratios of $f_3(q^2)/f_1(q^2)$
(left panel) and $g_2(q^2)/g_1(q^2)$ (right panel) as a function of 
four-momentum squared $q^2$
for $(m_{ud}, m_s)=(0.04, 0.08)$. 
Open circles represent the values measured for $\XtoS$ at the simulated $q^2$.
Although we observe non-negligible values of the second-class
form factors, it is still questionable whether nonzero signals
correspond to a pure effect from the flavor $SU(3)$ breaking. 
Indeed, the same analysis for the case of the exact $SU(3)$ limit ($\ntop$), 
yields comparable values of the second-class form factors to those of $\XtoS$. 
Lower and upper triangle symbols correspond to the cases of 
$\ntop$ with $m_{ud}=0.04$ and 0.08.
The lighter $m_{ud}$ calculations yield central values closer to 
those of $\XtoS$ with larger statistical uncertainties, while results from 
the heavier $m_{ud}$ calculations are marginally consistent with zero values 
for both $f_3$ and $g_2$ form factors within 1-2 standard deviation. 
Although it seems that observed nonzero values of the 
second-class form factors suffer much from large statistical fluctuations, 
we are also concerned about another possibility. 

For the case of the $f_3$ 
form factor, a nonvanishing contribution even in the exact $SU(3)$ limit 
stems from a subtle difference of $\Lambda_S^V$ and $\Lambda_0^V$ 
correlators, which is possibly due to the discretization error~\footnote{
%%%%%% footnote %%%%%%
We observe that this difference becomes very small
at a finer lattice spacing, $a\approx 0.11$ fm, in dynamical $N_f=2+1$ 
flavor DWF calculations~\cite{Progress}.
%%%%%%%%%%%%%%%%%
}. 
Although there is no such clear correspondence
in the case of the $g_2$ form factor, the discretization error may 
equally cause a systematic uncertainty on the determination 
of the $g_2$ form factors as well.

To control both statistical fluctuations and systematic uncertainties, 
we subtract the measured values of the second-class form factors in the $SU(3)$ limit calculation from those of $\XtoS$ as
%
% eq
%
\bea
\left[f^{\rm sub}_3(q^2)\right]_{\Xi\rightarrow \Sigma}
&=&
\left[f_3(q^2)
\right]_{\XtoS}
-
\frac{M_\Xi+M_\Sigma}{2M_N}\left[
f_3(q^2)
\right]_{\ntop},
\label{Eq:Sub_f3}
\\
\left[g^{\rm sub}_2(q^2)\right]_{\XtoS}
&=&
\left[g_2(q^2)
\right]_{\XtoS}
-
\frac{M_\Xi+M_\Sigma}{2M_N}\left[
g_2(q^2)
\right]_{\ntop},
\label{Eq:Sub_g2}
\eea
where a factor $(M_{\Xi}+M_{\Sigma})/(2M_N)$ is accounted 
for the mass difference between the $\XtoS$ and $\ntop$ decays, which is determined by simulated masses~\footnote{
%%%%%%%% footnote %%%%%%%%%%%%%%%
The subtraction procedure is not unique. Alternatively, we 
may consider a simple difference of measured second-class form factors between
$\XtoS$ and $\ntop$ processes as $[f_3^{\rm sub}(q^2)]_{\XtoS}=[f_3^{\rm sub}(q^2)]_{\XtoS}-[f_3^{\rm sub}(q^2)]_{\ntop}$. 
The leading-order behavior of the flavor $SU(3)$ breaking 
in the second-class form factors is not changed by a variation of 
the subtraction term. The resulting difference appears only 
in higher-order corrections. Indeed, analyses with two types 
of the subtraction term yield fairly consistent results 
of $f_3(0)$ and $g_2(0)$ at the physical point.}.
%%%%%%%%%%%%%%%%%%%%%%%%%%%%%% 

Although the $q^2$ value for the $\XtoS$ transition differs from that 
of the $\ntop$ transitions at the same three-momentum transfer, 
above subtraction is simply performed at every given three-momentum ${\bf q}$.
There are choices to set a reference value of the subtraction, since the 
second-class form factors for the $\ntop$ transition are supposed to 
vanish with any value of $m_{ud}$. However, as mentioned above, 
the lighter $m_{ud}$ calculations provide larger statistical uncertainties 
on the second-class form factors than those of $\XtoS$. 
In Eqs.~(\ref{Eq:Sub_f3}) and (\ref{Eq:Sub_g2}),
we adopt the single reference values of $[f_3(q^2)]_{\ntop}$ and $[g_2(q^2)]_{\ntop}$ evaluated at $m_{ud}=0.06$. Our chosen value 
of $m_{ud}$ corresponds to the heaviest $m_{ud}$ mass in all 
combinations of $(m_{ud}, m_{s})$ for the $\XtoS$ calculation 
in this study. 

Results from the subtraction method are also plotted 
in Fig.~\ref{FIG:2ndClass_comp} 
as filled circle symbols. It is clearly observed that statistical 
errors are significantly reduced after such subtraction
due to a strong correlation between those two
form factors,  while center values are slightly shifted to zero. 
Non-vanishing signals of both $f_3$ and $g_2$ form factors 
turn out to be more pronounced.  The subtraction method can expose the real $SU(3)$-breaking effect with better statistical precision. 
          
%
% FIG for bare and subtracted 2ndClass 
%
          
\subsubsection{Extrapolation to zero four-momentum squared}

\begin{figure}[htbp]
\bc
\includegraphics[width=.47\textwidth,clip]{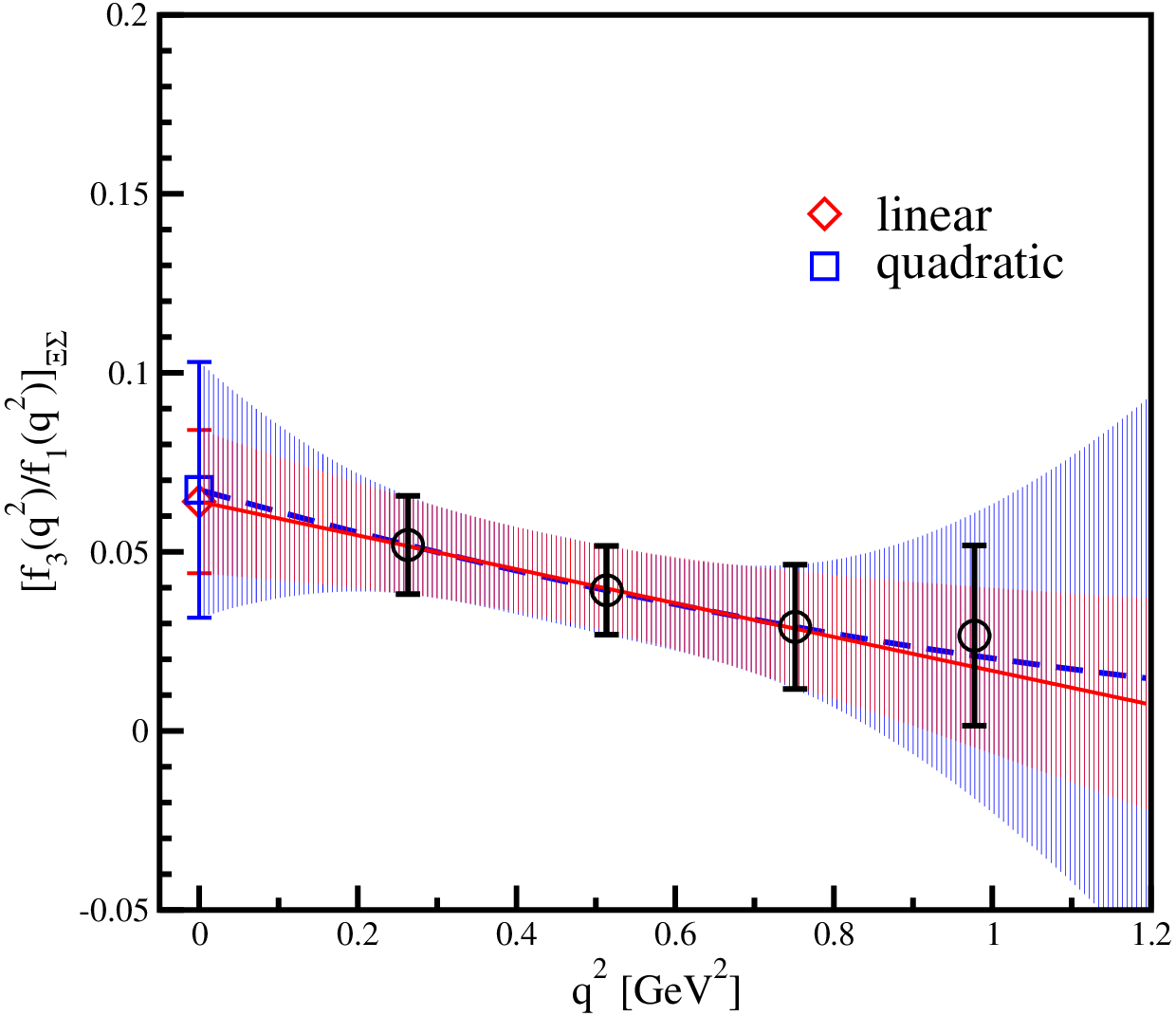}
\includegraphics[width=.47\textwidth,clip]{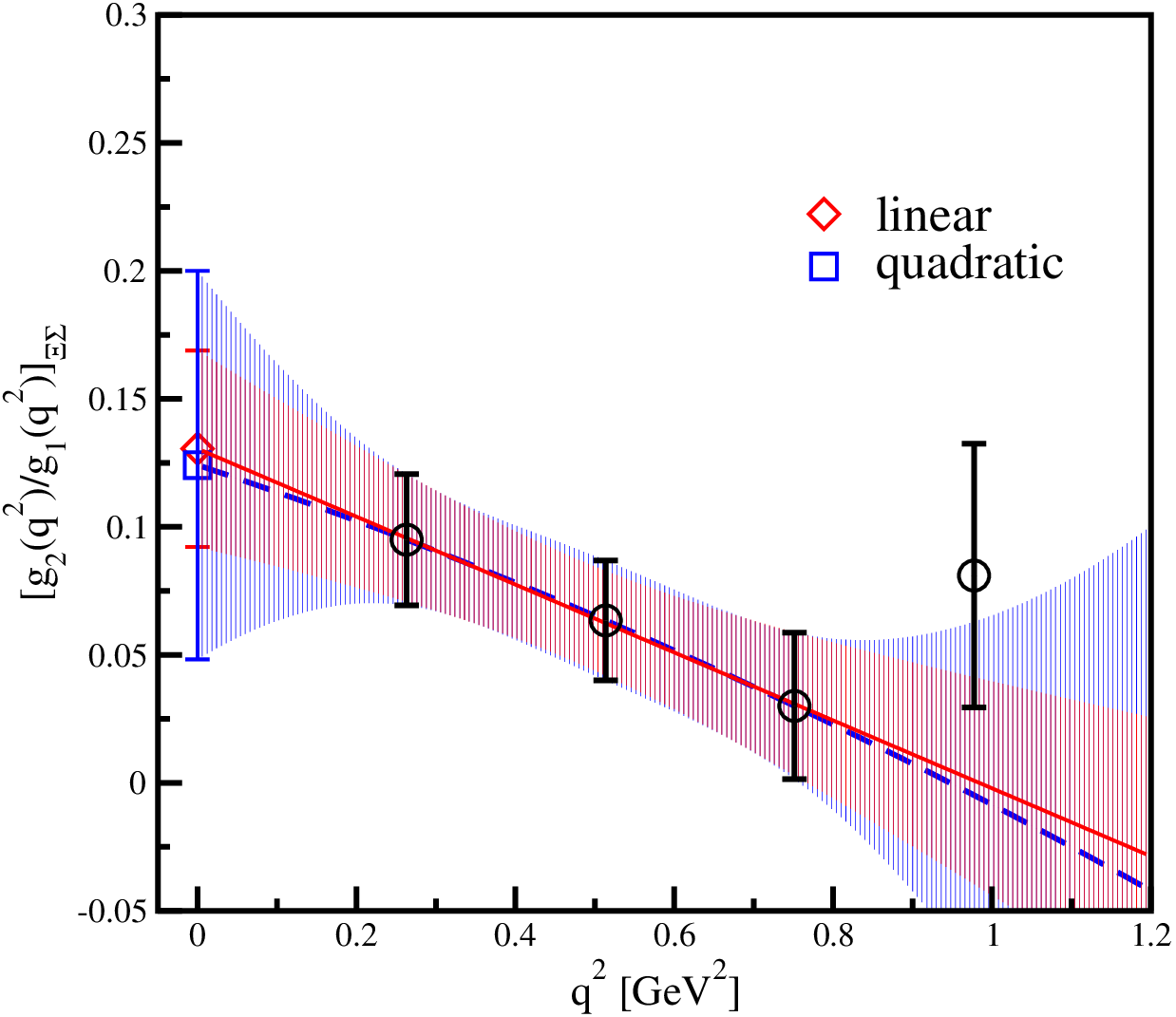}
\caption{The ratios $f_3(q^2)/f_1(q^2)$ (left panel)
and $g_2(q^2)/g_1(q^2)$ (right panel) as functions
of $q^2$ as in the case of $(m_{ud}, m_s)=(0.04, 0.08)$.
The solid and dashed lines represent linear and quadratic fits for three lowest $q^2$ data, respectively.}
\label{FIG:2ndClass_Qdep}
\ec
\end{figure}

We next examine the $q^2$ dependence of the ratios $f^{\rm sub}_3(q^2)/f_1(q^2)$ and $g^{\rm sub}_2(q^2)/g_1(q^2)$. In Fig.~\ref{FIG:2ndClass_Qdep}, we show the case of $(m_{ud}, m_s)=(0.04, 0.08)$ as typical examples. 
One can easily observe that both ratios do not yield any strong $q^2$ dependence
at least in our simulated range of 0.25 to 0.93 ${\rm GeV}^2$. It is worth mentioning that there is no theoretical prediction of the $q^2$ dependence 
of the second-class form factors.
Therefore, we simply adopt the linear or quadratic extrapolation with respect to $q^2$ to estimate $f_3(0)/f_1(0)$ and $g_2(0)/g_1(0)$. All evaluated 
values of $f_3(0)/f_1(0)$ and $g_2(0)/g_1(0)$ from two functional 
forms are summarized in Table~\ref{Tab:2ndClassF}. 
The extrapolated values from both determinations agree well with each other 
within their errors. As a result, the extrapolated values are not
significantly affected by the specific fitting form adopted for describing 
the $q^2$ dependence of form factors.

% 
%	Q2-extrapolation for f3f1 and g2g1
%

\subsubsection{Chiral extrapolation}
\label{Sec:other_FF_2ndclass_chiral}

\begin{figure}[htbp]
\bc
\includegraphics[width=.47\textwidth,clip]{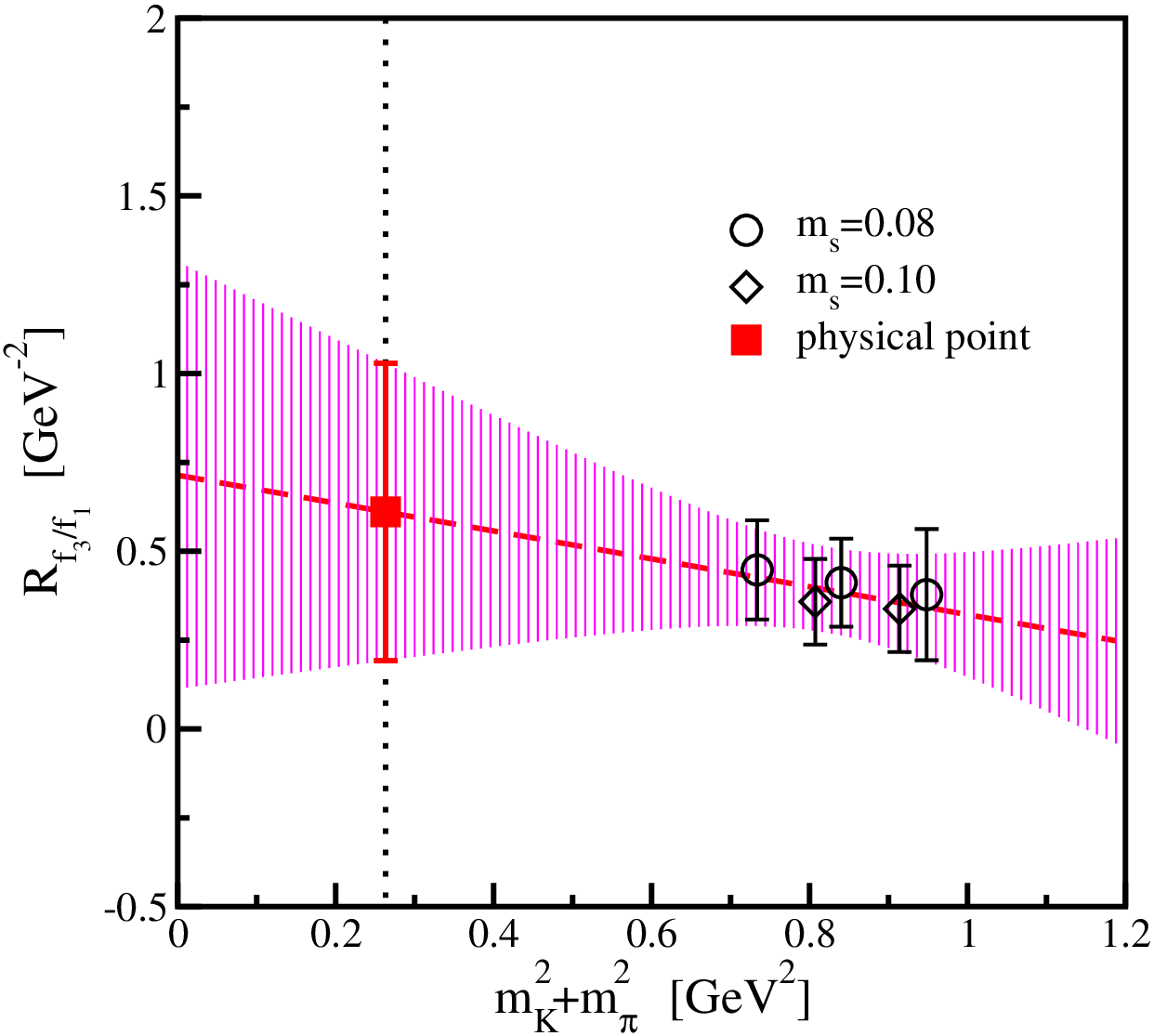}
\includegraphics[width=.47\textwidth,clip]{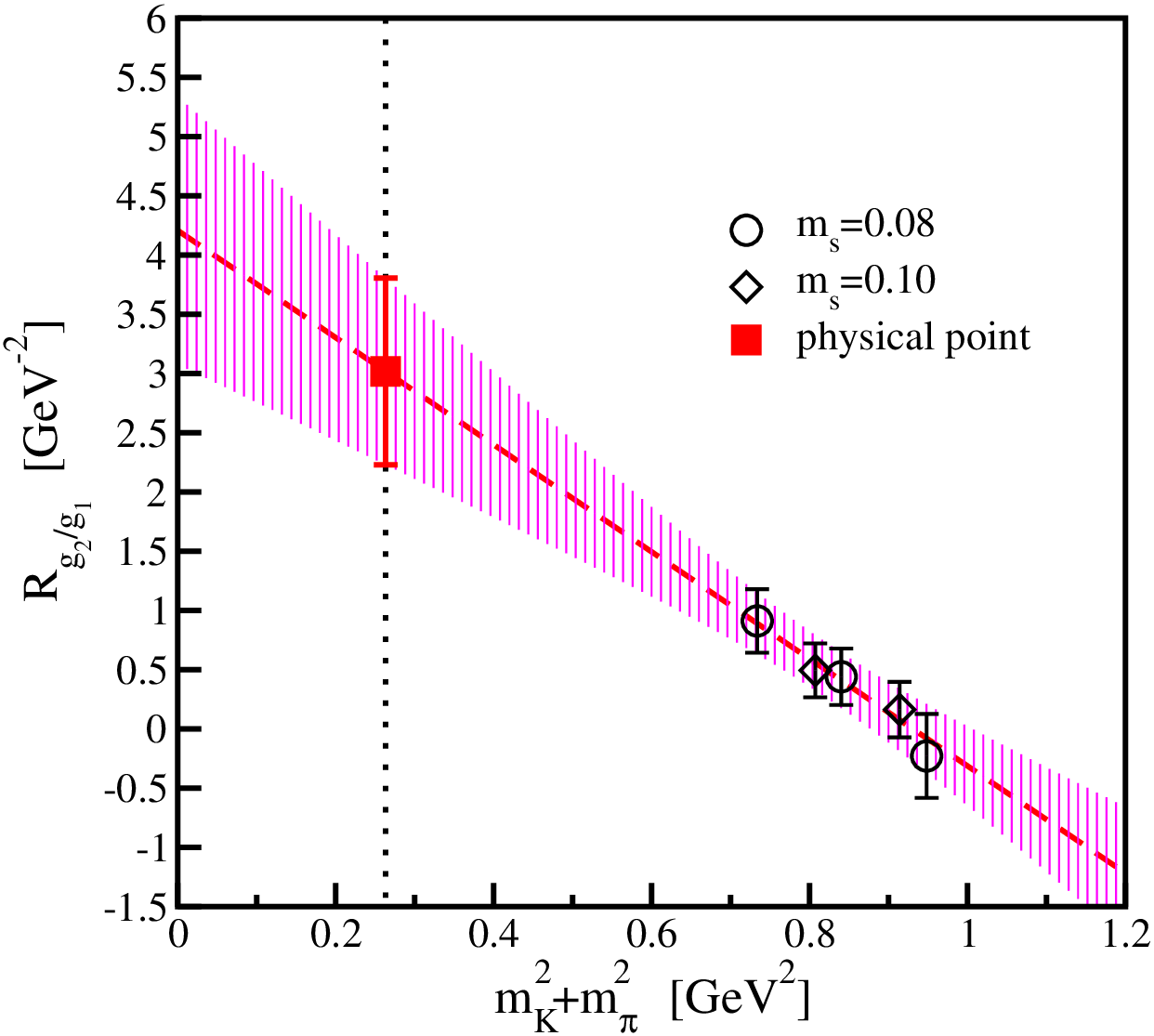}
\caption{Chiral extrapolation of $R_{f_3/f_1}$ (left panel) and $R_{g_2/g_1}$
(right panel). Symbols are defined as in Fig.~\ref{FIG:2ndOrd_on_f1}.}
\label{FIG:2ndClass_phy}
\ec
\end{figure}

In this subsection we show the results of $f_3(0)/f_1(0)$ and 
$g_2(0)/g_1(0)$ obtained by two types of the chiral extrapolation.
In the first method, which we call method-A, we take into account
the fact that the second-class form factors $f_3$ and 
$g_2$ vanish in the exact $SU(3)$ symmetry limit and nonzero values are induced 
by the first-order corrections. We then introduce the following ratios for
both $f_3(0)/f_1(0)$ and $g_2(0)/g_1(0)$:
\bea
R_{f_3/f_1}(M_{K}, M_{\pi}) &=&\frac{f_3(0)/f_1(0)}{M_{K}^2-M_{\pi}^2}, 
\\
R_{g_2/g_1}(M_{K}, M_{\pi})&=&\frac{g_2(0)/g_1(0)}{M_{K}^2-M_{\pi}^2},
\eea
where the leading first-order corrections of the flavor $SU(3)$ breaking
are explicitly factorized out. In Fig.~\ref{FIG:2ndClass_phy}, we
show the quark mass dependences of $R_{f_3/f_1}$ and $R_{g_2/g_1}$
as functions of $M_K^2 + M_\pi^2$. The ratio $R_{f_3/f_1}$ reveals
the mild dependence of $M_K^2 + M_\pi^2$, while rather
strong dependence of $M_K^2 + M_\pi^2$ appears in the ratio $R_{g_2/g_1}$.
In either cases, obtained data is well described by the linear fitting form~\footnote{
%%%%% footnote %%%%%%%%%%%
If we consider inclusion of either
$(M_{K}^2+M_{\pi}^2)^2$ or $M_{K}^2-M_{\pi}^2$
terms into Eq.~(\ref{Eq:ChET_Atype}), our limited data 
points can not sufficiently determine either coefficients, 
which receive more than 100\% error. The resulting 
values at the physical point are consistent with the results
obtained from the simple linear fit (\ref{Eq:ChET_Atype})
within the statistical errors. 
}:
%%%%%%%%%%%%%%%%%%%%%
%
%
%
\be
R_{f_3/f_1}(M_{K}, M_{\pi})\;{\rm or}\;R_{g_2/g_1}(M_{K}, M_{\pi})=A_0 + A_1 \cdot (M_{K}^2+M_{\pi}^2).
\label{Eq:ChET_Atype}
\ee
Here, the values of $f_3(0)$ and $g_2(0)$ given by the simplest linear fit in $q^2$
are used for calculating $f_3(0)/f_1(0)$ and $g_2(0)/g_1(0)$.
We then get the extrapolated values of $R_{f_3/f_1}$ and $R_{g_2/g_1}$
at the physical meson masses as
%
% eq.
%
\bea
R_{f_3/f_1}(M_{K}^{\rm phys}, M_{\pi}^{\rm phys})&=&0.61(42)
\mbox{ in (GeV)}^{-2}, \\
R_{g_2/g_1}(M_{K}^{\rm phys}, M_{\pi}^{\rm phys})&=&3.02(89)
\mbox{ in (GeV)}^{-2}, 
\eea
which finally provide the values $f_3(0)/f_1(0)=0.137(94)$ and $g_2(0)/g_1(0)=0.677(177)$. 

In an alternative method indicated by method-B, we may perform a linear fit 
in two mass combinations $M_{K}^2+M_{\pi}^2$ and $M_{K}^2-M_{\pi}^2$
\be
f_3(0)/f_1(0)\;{\rm or}\;g_2(0)/g_1(0)=B_0 + B_1 \cdot (M_{K}^2+M_{\pi}^2) + B_2 \cdot (M_{K}^2 - M_{\pi}^2).
\label{Eq:ChET_Btype}
\ee
We then obtain $f_3(0)/f_1(0)=0.147(60)$ and $g_2(0)/g_1(0)=0.450(110)$.
Both method-A and B provide consistent results. Although the errors in the latter approach are relatively smaller than that of the former, 
the former leads to a much smaller value of $\chi^2/{\rm dof}$ than the latter.
Therefore, we quote the values obtained from the method-A for 
our final values at the physical point:
\bea
\left[\frac{f_3(0)}{f_1(0)}\right]_{\XtoS}&=&0.137 \pm 0.094, \\
\left[\frac{g_2(0)}{g_1(0)}\right]_{\XtoS}&=&0.677 \pm 0.177,
\eea
which show firm evidence for nonzero second-class form factors 
in the $\XZtoSP$ beta decay. It should be reminded that the KTeV experiment
reported no evidence for a nonzero second-class form factor $g_2$~\cite{Alavi-Harati:2001xk}, measuring $g_2(0)/f_1(0)=-0.89\pm1.05$, which corresponds to 
$g_2(0)/g_1(0)\simeq-0.73\pm 0.89$~\footnote{
%%%%%%% footnote %%%%%%
A factor $(M_{\Xi}+M_{\Sigma})/M_{\Xi}$, equal to $\simeq 1.9048$
is different from definitions of $f_3$ and $g_2$ form factors adopted 
in Ref.\cite{Alavi-Harati:2001xk}.}. 
%%%%%%%%%%%%%%%%%%

%
% FIG for 2ndClass 
%

%
% 2nd-Class form factors
%
\begin{table}[htbp]
\caption{Results for 
$\left[f_3(0)/f_1(0)\right]_{\XtoS}$
and
$\left[g_2(0)/g_1(0)\right]_{\XtoS}$, both of 
which are evaluated by the $q^2$ extrapolation
with the linear and quadratic functional form.
}
\begin{ruledtabular}
\begin{tabular}{cc| l l | l l}
\hline
&
&$\left[f_3(0)/f_1(0)\right]_{\XtoS}$ &
&\phantom{$-$}$\left[g_2(0)/g_1(0)\right]_{\XtoS}$ &
\\
$m_s$  & $m_{ud}$ 
& linear & quadratic
& \phantom{$-$}linear & quadratic
\\
\hline
 0.08 
 &0.06 
 & 0.028(13)
 & 0.011(22)
 & $-$0.017(26)
 & 0.027(52)\\
 &0.05 
 &  0.045(13)
 &  0.037(22)
 &  \phantom{$-$}0.048(26)
 &  0.069(49) \\
 &0.04 
 &  0.064(20)
 &  0.067(36)
 &  \phantom{$-$}0.131(38)
 &  0.124(76)\\
 0.10
 &0.05 
 &  0.062(22)
 &  0.040(35)
 &  \phantom{$-$}0.030(43)
 &  0.088(83)\\
 &0.04 
 &  0.078(26)
 &  0.066(42)
 &  \phantom{$-$}0.107(49)
 &  0.143(92)\\
  \hline
(From method-A)
 &  phys. point 
 & 0.137(94)
 & 0.241(173)
 & \phantom{$-$}0.677(177)
 & 0.414(372)\\
(From method-B)
 &  phys. point
 & 0.147(60)
 & 0.193(110)
 & \phantom{$-$}0.450(114)
 & 0.344(232)\\
\hline
\end{tabular}
\end{ruledtabular}
\label{Tab:2ndClassF}
\end{table}
%

%%%%%%%%%%%%%%%%%%%%%%%%%%%%%%%%%%%%%
%\clearpage
\subsection{Weak magnetism form factor $f_2(q^2)$}

\begin{figure}[htbp]
\bc
\includegraphics[width=.47\textwidth,clip]{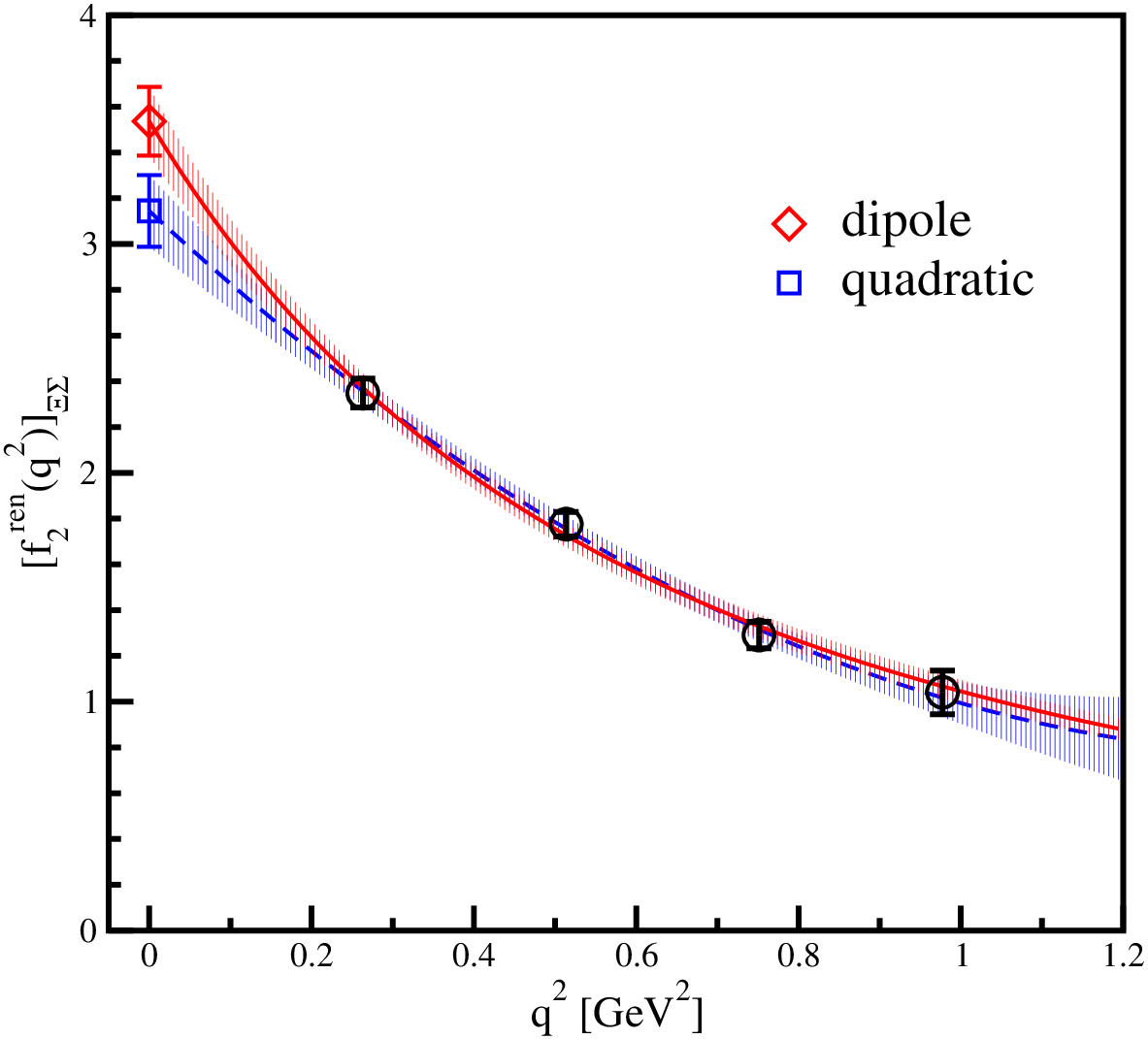}
\includegraphics[width=.47\textwidth,clip]{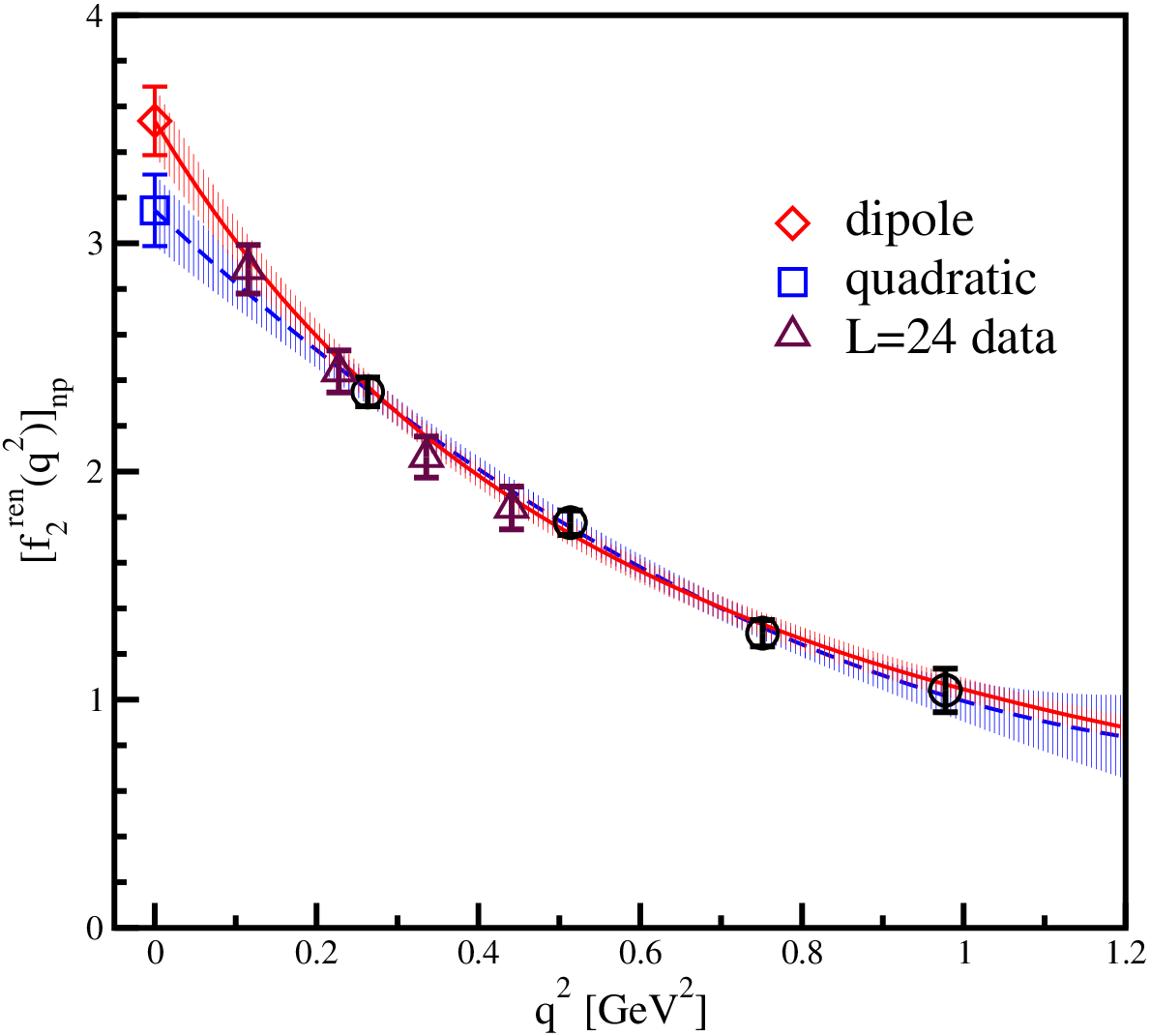}
\caption{The renormalized weak-magnetism form factor $f_2(q^2)$ as functions of four-momentum squared $q^2$ for $\XZtoSP$ (left panel) 
and $\ntop$ (right panel) at $(m_{ud}, m_s)=(0.04, 0.08)$.
For the $q^2$ extrapolation, we apply two types of fitting form. 
The solid (dashed curve) is the fitting result by using the dipole (quadratic) form, 
while the open diamond (square) represents the extrapolated value to $q^2=0$.
The highest $q^2$ data point is not included in our fits.
In the right panel, results from the larger volume ($L=24$)~\cite{Sasaki:2007gw} 
are also included as open triangles for a comparison.
}
\label{FIG:f2_Qdep}
\ec
\end{figure}

In Fig.~\ref{FIG:f2_Qdep}, we show the weak magnetism form factors 
$f_2^{\Btob}(q^2)$ for $\XtoS$ (left panel)
and $n\rightarrow p$ (right panel) at $(m_{ud}, m_s)=(0.04, 0.08)$
as functions of four-momentum squared $q^2$. The form factors plotted here
are scaled by the renormalization factor $Z_V$ determined in Sec~\ref{Sec:f_1}.
To determine $f_2^{\rm ren}(0)$,
we adopt two functional forms for the $q^2$ dependence of $f_2^{\rm ren}(q^2)$:
the dipole form
\bea
f^{\rm ren}_2(q^2) = \frac{f^{\rm ren}_2(0)}{(1 + \lambda^{(1)}_2q^2)^2},
\eea
and the quadratic form
\be
f^{\rm ren}_2(q^2) = f^{\rm ren}_2(0)(1 + \lambda^{(1)}_2 q^2 + \lambda^{(2)}_2 q^4).
\ee
The former functional form is phenomenologically adopted in 
the nucleon's electromagnetic form factors, which are related 
to the weak nucleon form factors under 
the CVC hypothesis on the weak processes.

Figure~\ref{FIG:f2_Qdep} shows that both functional forms
are equally fitted to data of $f_2^{\rm ren}(q^2)$ in either case of 
$\XtoS$ or $\ntop$ decay processes. 
However, there appears to be a sensitivity of the choice 
of the fitting form in extrapolated values at $q^2=0$. This is simply because
our simulated $q^2$ points are not close enough to $q^2=0$. 
In Ref.~\cite{Sasaki:2007gw},
we have studied the $q^2$ dependence of the weak nucleon form factors 
at low $q^2$ down to about 0.1 ${\rm GeV}^2$, which is accessible with 
the spatial extent $L=24$. Indeed, in our previous study, the weak 
magnetism form factor for $\ntop$ is observed to be well described 
by the dipole form~\cite{Sasaki:2007gw}. 
To make this point clear, in the right panel of Fig.~\ref{FIG:f2_Qdep},
we also include results from the larger volume ($L=24$) for a comparison. The steep
raising behavior of $L=24$ data as $q^2$ decreases clearly favors the dipole form.
We, therefore, do not use the results from the quadratic fit in the following discussion.
All extrapolated values with both determinations are summarized 
in Table~\ref{Tab:renorm_f2}.
As for the $\ntop$ transition, the extrapolated values of $f_2(0)$ by using
the dipole fit are summarized in Table~\ref{Tab:nbdecay}, together 
with other relevant quantities.

We also perform the chiral extrapolation of $f_2^{\rm ren}(0)$ through
a linear fit in two mass combinations $M_K^2+M_{\pi}^2$ and $M_K^2-M_{\pi}^2$,
the same as Eq.(\ref{Eq:ChET_Btype}). We obtain $[f_2^{\rm ren}(0)]_{\XtoS}
= 3.30 \pm 0.24$ by employing result from the dipole form for the $q^2$ extrapolation.
Finally, we compare the ratio $[f_2(0)/f_1(0)]_{\XtoS}$ 
to its $SU(3)$ counterpart $[f_2(0)/f_1(0)]_{\ntop}$
and then observe an order of 16\% breaking effect as
\be
\left[\frac{f_2(0)}{f_1(0)}\right]_{\XtoS}=1.16(11) \times \left[\frac{f_2(0)}{f_1(0)}\right]_{\ntop},
\label{Eq:f_2}
\ee
which implies the violation of the exact $SU(3)$ relation.
It is worth mentioning that $[f_2(0)/f_1(0)]_{\XtoS}$
tends to be bigger than $[f_2(0)/f_1(0)]_{\ntop}$, while the 
$f_2(0)/f_1(0)$ values evaluated from both 
the generalized CVC hypothesis~\cite{Yamanishi:2007zza} 
and Sirlin's formula~\cite{Sirlin:1979hb} yield opposite results 
as previously shown in Table~\ref{Tab:Th_estimate_f_2}. 
In addition, the observed size of the deviation from unity 
in Eq.~(\ref{Eq:f_2}) may also be compared with the Cabibbo-model prediction 
as $(M_{\Xi}+M_{\Sigma})/(2M_N)=1.338$~\cite{Cabibbo:2003cu}, 
which corresponds to a factor accounted for the mass difference between
the $\XtoS$ and $\ntop$ decays. We again observe a 15\%
deviation from the Cabibbo model.

%
%	Q2-extrapolation for f2f1	
%

%
%	Table for renormalized f2
%
\begin{table}[htbp]
\caption{Results for 
renormalized $\left[f_2(0)\right]_{\XtoS}$
is evaluated by the $q^2$ extrapolation with the dipole and
quadratic functional form.
}
\begin{ruledtabular}
\begin{tabular}{cc| l l}
&& $\left[f^{\rm ren}_2(0)\right]_{\XtoS}$ &\\
$m_s$  & $m_{ud}$ 
& dipole & quadratic
\\
\hline
 0.08 
 &0.06 
 & 3.64(11)
 & 3.25(12)
 \\
 &0.05 
 &  3.60(13)
 &  3.20(14)
 \\
 &0.04 
 &  3.54(15)
 &  3.14(16)
 \\
 0.10
 &0.05 
 &  3.63(11)
 &  3.23(12)
 \\
 &0.04 
 &  3.57(13)
 &  3.17(14)
 \\
  \hline
---
 &  phys. point 
 & 3.30(24)
 & 2.92(25) 
 \\
\hline
\end{tabular}
\end{ruledtabular}
\label{Tab:renorm_f2}
\end{table}
%

%%%%% 
\begin{table}[htbp]
\caption{Summary for the $n\rightarrow p$ transition process.
The values of $f_2(0)/f_1(0)$ determined by the dipole 
form for the $q^2$ extrapolation, while the values $g_3(0)/g_1(0)$ are 
evaluated by the pion-pole fit. }
\begin{ruledtabular}
\begin{tabular}{c| l l l}
\hline
$m_{ud}$ & $g_1(0)/f_1(0)$ & $f_2(0)/f_1(0)$ & $g_3(0)/g_1(0)$
\\
\hline
 0.10 
 &1.240(12)
 &3.80(8)
 &12.06(32)
 \\
 0.08
 &1.226(11)
 & 3.71(10)
 &13.52(38)
 \\
0.06
&1.214(14)
&3.57(13)
&15.35(49)
\\
0.05
&1.209(18)
&3.47(17)
&16.73(59)
\\
0.04
&1.202(24)
&3.35(24)
&18.68(79)
\\
  \hline
phys. point &1.191(49) 
&3.18(29) 
&151.0(6.9)\\
 \hline
\end{tabular}
\end{ruledtabular}
\label{Tab:nbdecay}
\end{table}
%

%\clearpage
\subsection{Induced pseudo-scalar form factor $g_{3}(q^2)$}
\begin{figure}[htbp]
\bc
\includegraphics[width=.47\textwidth,clip]{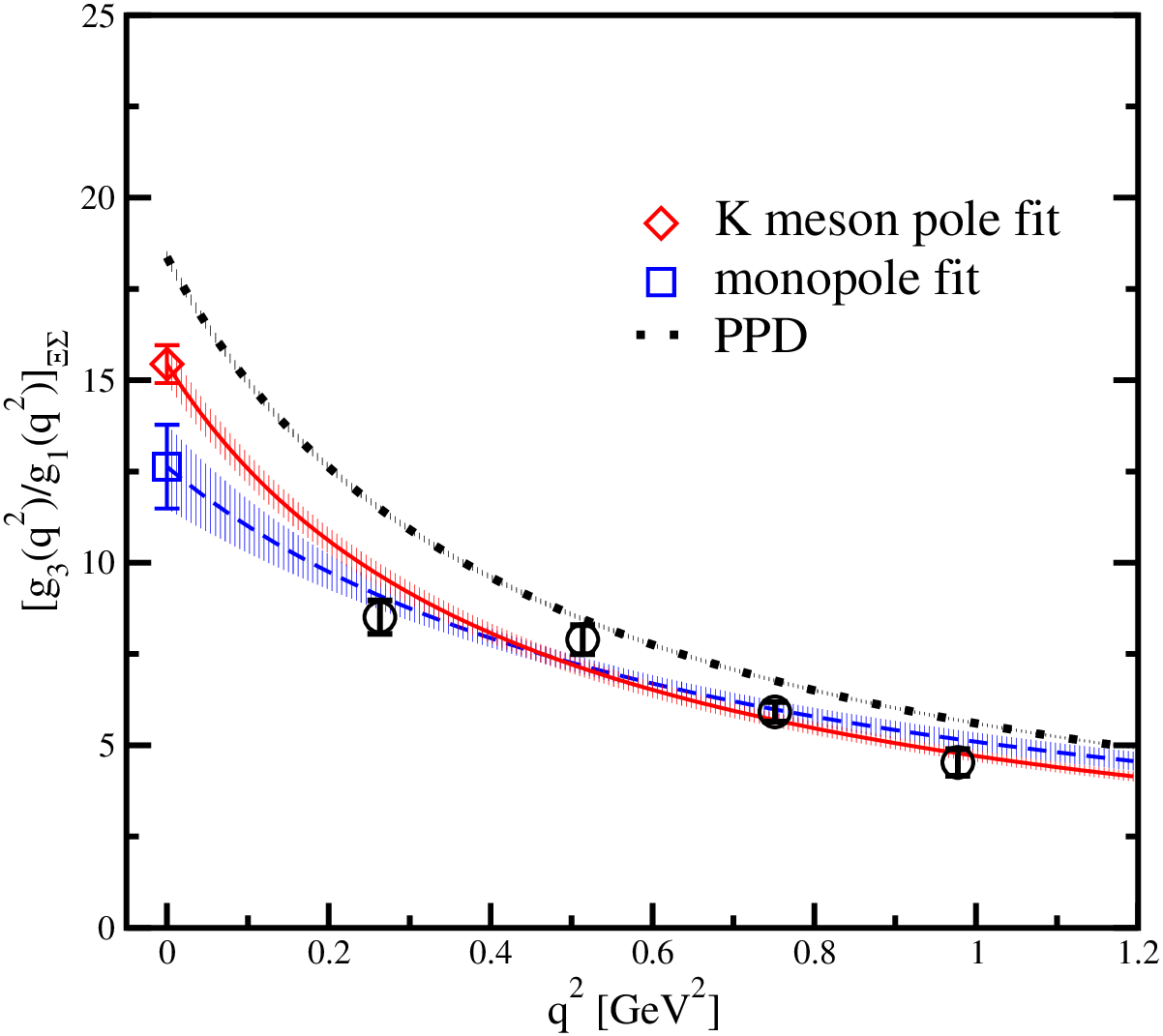}
\includegraphics[width=.47\textwidth,clip]{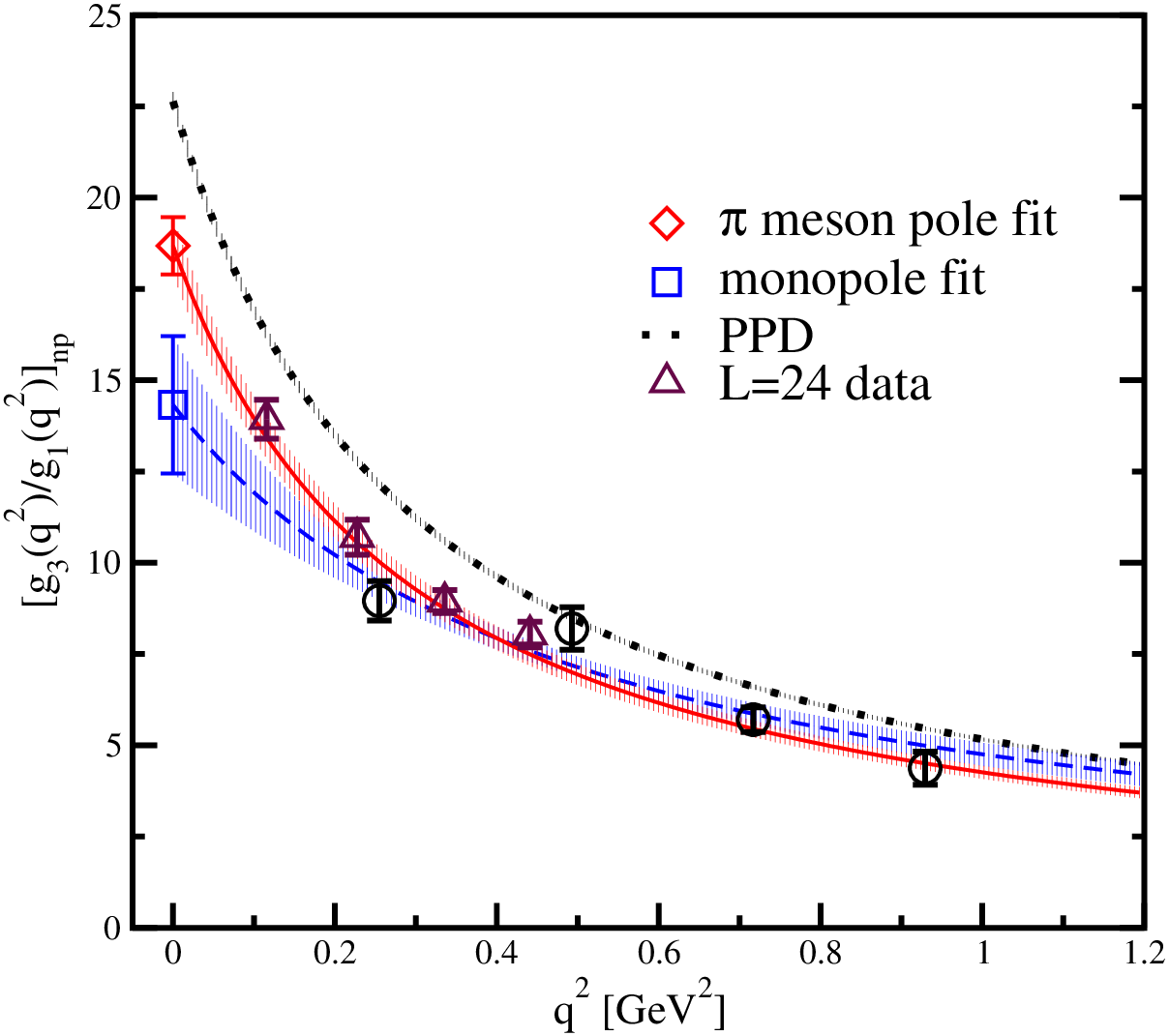}
\caption{The ratio $g_3(q^2)/g_1(q^2)$ as functions of four-momentum 
squared $q^2$ for $\XZtoSP$ (left panel) and $\ntop$ (right panel)
at $(m_{ud}, m_s)=(0.04, 0.08)$. Symbols are defined as in Fig.~\ref{FIG:f2_Qdep}.
The highest $q^2$ data point is not included in our fits.
}
\label{FIG:g3_Qdep}
\ec
\end{figure}

We next show the $q^2$ dependence of the ratio of the induced pseudo-scalar
form factor $g_3(q^2)$ and the
axial-vector form factor $g_1(q^2)$ at $(m_{ud}, m_s)=(0.04, 0.08)$
in Fig.~\ref{FIG:g3_Qdep}. It is phenomenologically known that
the $q^2$ dependence of $g_3(q^2)$ in the low $q^2$ region are well fitted 
by the pion-pole dominance (PPD) form, $g_3(q^2)=4M_N^2 g_1(q^2)/(q^2+M_{\pi}^2)$~\cite{{Nambu:1960xd},{Choi:1993vt},{Gorringe:2002xx}}. 
As shown in Ref.~\cite{Sasaki:2007gw}, 
the PPD form provides a good description of the $q^2$ dependence of
the nucleon's induced pseudo-scalar form factor measured in
quenched DWF simulations at least at low $q^2$. On the other hand, 
there is no direct experimental information in the case of $g_3(q^2)$ 
for any other hyperon decays. 

Under the partial conserved axial-vector current (PCAC) 
hypothesis~\footnote{
%%%%%%% footnote %%%%%%%%
Note that the presence of the second-class form factor $g_2$ does not modify 
the generalized Goldberger-Treiman relation~\cite{Weisberger:1966ip}, which 
is essential for the applicability of the PPD form~\cite{Sasaki:2007gw}.},
%%%%%%%%%%%%%%%%%%%%
an extension of PPD to $\Delta S=1$ decays predicts 
that the induced pseudo-scalar form factor for the hyperon beta decays, such 
as the $\XtoS$ decay, the $K$ meson pole, instead of the pion pole. 
In the generalized PPD form, the ratio of $g_3(q^2)/g_1(q^2)$ for 
the $\XtoS$ transition is given by a simple monopole form
\be
\left[\frac{g_3(q^2)}{g_1(q^2)}\right]^{\rm PPD}_{\XtoS}
=\frac{(M_{\Xi}+M_{\Sigma})^2}{q^2+M_{K}^2}
\ee
where a monopole mass corresponds to the Kaon mass.

In Fig.~\ref{FIG:g3_Qdep}, we include the predicted $q^2$ dependence 
(dotted curve) evaluated by the generalized PPD form 
with simulated baryon masses ($M_{N}$, $M_{\Sigma}$, $M_{\Xi}$) and meson masses ($M_{\pi}$, $M_{K}$). Three larger $q^2$ data points are quite close to predicted curves in either $\XtoS$ or $n\rightarrow p$, while the lowest $q^2$ data points
are underestimated in comparison with the PPD prediction. 
To extrapolate the value of $g_3(q^2)/g_1(q^2)$ to zero
four-momentum transfer, we first adopt the monopole form, which is 
inspired by the PPD form.  
In Fig.~\ref{FIG:g3_Qdep}, the dashed curves are fitting results 
by the monopole form with two parameters $\lambda_{3}^{(0)}$ 
and $\lambda_{3}^{(1)}$,
\be
\frac{g_3(q^2)}{g_1(q^2)}=\frac{\lambda_{3}^{(0)}}{1+\lambda_{3}^{(1)}q^2}
\ee
where $1/\sqrt{\lambda_{3}^{(1)}}$ corresponds to a monopole mass.
The obtained values of $g_3(0)/g_1(0)$ are significantly smaller than 
those of the PPD prediction as $(M_{\Xi}+M_{\Sigma})^2/M_K^2$ 
for the $\XtoS$ transition and $4M_N^2/M_K^2$ for the $\ntop$ transition. 
The resulting monopole masses are also bigger than the expected Kaon 
and pion masses. This substantial deviation from the PPD form is mainly 
caused by an unexpected reduction of the lowest $q^2$ data points.

In Ref.~\cite{Sasaki:2007gw}, we have reported that the lowest $q^2$
point in the nucleon's induced pseudo-scalar form factor at $L=16$,
which is now utilized in our current calculations, may suffer from the finite
volume effect. Indeed, the $q^2$ dependence of the results obtained 
from the larger lattice ($L=24$) was well fitted by the pion-pole structure. 
In this context, we may have an alternative way to evaluate the value 
of $g_3(0)/g_1(0)$ by a one-parameter fit in the monopole form 
with the monopole mass constrained by simulated $M_K$ or $M_{\pi}$.
We simply refer to such a fit as the ``Kaon-pole fit" or the ``pion-pole fit"
respectively. All extrapolated values of $g_3(q^2)/g_1(q^2)$ to zero 
four-momentum transfer with both determinations of monopole and 
Kaon-pole fits are listed in Table~\ref{Tab:g3g1_zeroQ}.

The solid curves in Fig.~\ref{FIG:g3_Qdep} are given by the Kaon- and 
pion-pole fits. For a justification of this analysis, see the right
panel of Fig.~\ref{FIG:g3_Qdep}. In this figure, four data points (open triangles) obtained from the larger lattice ($L=24$) are additionally included. 
One can easily see that these data points quite follow the solid curve,
which is determined by the pion-pole fit of the lowest three $q^2$ data points 
obtained from the lattice size of $L=16$. It is found that resulting values 
of $g_3(0)/g_1(0)$ are still slightly smaller than the PPD values 
in either case of $\XtoS$ or $\ntop$. 
The similar quenching was observed in our previous detailed study 
of the nucleon's induced pseudo-scalar form factor using the 
larger lattice ($L=24$)~\cite{Sasaki:2007gw}. 
It is worth mentioning that the size of this quenching for $\XtoS$ 
is similar to that of $\ntop$. Therefore, ratios of $[g_3(0)/g_1(0)]_{\XtoS}$ and $[g_3(0)/g_1(0)]_{\ntop}$ exhibit remarkable consistency with the PPD prediction as shown in Table~\ref{Tab:Ratio_g3g1}. Even if we adopt the monopole form for 
the $q^2$ extrapolation, the resulting ratios still barely agree with the PPD values.
Our results strongly suggest that the following relation is well fulfilled at least
in the simulated region
\be
\left[\frac{g_3(0)}{g_1(0)}\right]_{\XtoS}\approx
\left(\frac{M_\Xi+M_\Sigma}{2M_N}\right)^2\frac{M_{\pi}^2}{M_K^2}
\left[\frac{g_3(0)}{g_1(0)}\right]_{\ntop}.
\label{Eq:GPPDrelation}
\ee

As for the chiral extrapolation of $g_3(0)/g_1(0)$, the linear
fit in two mass combinations $M_{K}^2+M_{\pi}^2$ and $M_{K}^2-M_{\pi}^2$
(the same as method-B in Sec.~\ref{Sec:other_FF_2ndclass_chiral}) was utilized in  Ref.~\cite{Guadagnoli:2006gj}. However, this extrapolation doesn't 
take into account the expected large quark-mass dependence
of $g_3(0)/g_1(0)$ in the vicinity of the chiral limit 
like a divergent $1/M_{\pi}^2$ term for $\ntop$ or a $1/M_{K}^2$ 
term for $\XtoS$. This implies that the extrapolated values should be 
considerably underestimated especially for the case of $\ntop$.
For the ratio of $[g_3(0)/g_1(0)]_{\XtoS}$ and $[g_3(0)/g_1(0)]_{\ntop}$ at the physical point, the naive chiral extrapolation indeed yields a large value of 0.67(16) for the monopole fit or 0.52(4) for the Kaon(pion)-pole fit,
which should be compared with the PPD value of $M_K^2(M_{\Xi}+M_{\Sigma})^2/(4M_N^2M_{\pi}^2)=0.1430$ at the physical point.
%%%%%%%% footnote %%%%%%
%~\footnote{
%We use $M_N=938.9$ MeV, $M_{\pi}=139.6$ MeV, $M_{K}=493.7$ MeV
%$M_{\Xi}=1318.7$ MeV and $M_{\Sigma}=1193.2$ MeV, all of which 
%are the isospin-averaged masses, for an evaluation. 
%}.
%%%%%%%%%%%%%%%%%%%
This result is clearly contradicted with the finding expressed 
by Eq.~(\ref{Eq:GPPDrelation}) fulfilled in the simulated region. 
 
The simple linear fit in two mass combinations $M_{K}^2+M_{\pi}^2$ and $M_{K}^2-M_{\pi}^2$ is instead applied to the ratio of $g_3(0)/g_1(0)$ 
and its PPD value, which has a very mild quark-mass dependence in either 
case of $\XtoS$ or $\ntop$. The value of $[g_3(0)/g_1(0)]_{\XtoS}$ 
at the physical point is evaluated from the extrapolated value of this ratio
with a multiplicative factor of the physical PPD value. We obtain
$[g_3(0)/g_1(0)]_{\XtoS}=21.58(98)$ 
and $[g_3(0)/g_1(0)]_{\ntop}=151.0(6.9)$ at the physical point
for the Kaon(pion)-pole fit. The ratio of those values, which are determined to be 0.1429(2), correctly reproduces the PPD value. This determination is rather phenomenological. However, it is hard to perform the chiral extrapolation of 
$g_3(0)/g_1(0)$ without any assumption within our limited data sets.
Thus, instead of quoting any final value, we would like to stress that 
the expected relation between $[g_3(0)/g_1(0)]_{\XtoS}$
and $[g_3(0)/g_1(0)]_{\ntop}$ as in Eq.~(\ref{Eq:GPPDrelation}) is 
confirmed in our simulations.

%
%	Q2-extrapolation for g3g1	
%

%
%	Table for Ratio g3g1
%
\begin{table}[htbp]
\caption{
Ratio $\left[g_3(0)/g_1(0)\right]_{\XtoS}$
is evaluated by the $q^2$ extrapolation with the monopole and 
Kaon-pole form.
}
\begin{ruledtabular}
%\begin{tabular}{cc|ccc}
\begin{tabular}{cc| l l}
\hline
%&
&&$[g_3(0)/g_1(0)]_{\XtoS}$ & \\
$m_s$  & $m_{ud}$ 
&  monopole & Kaon pole
\\
\hline
 0.08 
 &0.06 
 &12.08(87)
 &14.40(42)
 \\
 &0.05 
 & 12.36(98)
 & 14.90(46)
\\
 &0.04 
 & 12.63(1.15)
 & 15.44(52)
 \\
 0.10
 &0.05 
 & 11.77(86)
 & 13.95(42)
 \\
 &0.04 
 & 11.97(1.00)
 & 14.34(47)
\\
\hline
 &phys. point
 & 16.73(2.51)
 & 21.58(98)
 \\
\hline
\end{tabular}
\end{ruledtabular}
\label{Tab:g3g1_zeroQ}
\end{table}
%

%
%	Table for (g3g1)/(g3g1)_SU3
%
\begin{table}[htbp]
\caption{
Comparison of measured ratios of $\left[g_3(0)/g_1(0)\right]_{\XtoS}$ and $\left[g_3(0)/g_1(0)\right]_{\ntop}$ to their PPD value 
given by $M_K^2(M_{\Xi}+M_{\Sigma})^2/(4M_N^2M_{\pi}^2)$.
}
\begin{ruledtabular}
\begin{tabular}{cc| l l l}
\hline
$m_s$  & $m_{ud}$ &  From monopole fit & From Kaon(pion)-pole fit
& PPD value 
\\
\hline
 0.08 
 &0.06 
 &0.957(17)
 &0.938(6)
 &0.930(1)
 \\
 &0.05 
& 0.925(33)
& 0.891(10)
& 0.878(2)
\\
 &0.04 
& 0.882(58)
& 0.827(15)
& 0.811(3)
 \\
 0.10
&0.05 
& 0.881(43)
& 0.834(14)
& 0.819(4)
\\
 &0.04 
& 0.836(68)
& 0.768(18)
& 0.751(5) 
 \\
\hline
\end{tabular}
\end{ruledtabular}
\label{Tab:Ratio_g3g1}
\end{table}
%

%
%	Table for (g3g1)/(g3g1)_SU3
%
\begin{table}[htbp]
\caption{
Summary of the $SU(3)$-breaking pattern observed 
in the $\XZtoSP$ decay. Here, $\Delta_{SU(3)}$ is defined
as $([X]_{\XtoS}-[X]_{\ntop})/[X]_{\ntop}$
for each quantity $X$.
We also evaluate the scaled values of $\Delta_{SU(3)}$ by
the expected size of the leading-order corrections.}
\begin{ruledtabular}
\begin{tabular}{c|llc}
$X$ & $\Delta_{SU(3)}$ & $\Delta_{SU(3)}/(2\delta)^n$ & $n$ \\
\hline
$f_1(0)$ & $-0.013(19)$ & $-1.3(1.9)$ & 2\\
$f_2(0)$ & $+0.16(11)$ & $+1.6(1.1)$ & 1 \\
$f_3(0)/f_1(0)$  & $+0.137(94)$~\footnote{
Because $f_3(0)=0$ and $g_2(0)=0$ for $\ntop$, 
$[f_3(0)/f_1(0)]_{\XtoS}$ and $[g_2(0)/g_1(0)]_{\XtoS}$
are instead quoted respectively.} & $+1.4(9)$ & 1\\
$g_1(0)/f_1(0)$ & $+0.051(35)$ & $+0.51(35)$ & 1\\
$g_2(0)/g_1(0)$ & $+0.677(177)$ $^{a}$ & $+6.8(1.8)$ & 1\\
\hline
\end{tabular}
\end{ruledtabular}
\label{Tab:Summary}
\end{table}
%

%%%%%%%%%%%%%%  SEC 4  %%%%%%%%%%%%%%%%%%%%%%%%
%\clearpage
\section{Summary}
\label{Sec:summary}

In this paper, we have studied flavor $SU(3)$-breaking effects
in the hyperon semileptonic decay, $\Xi^0 \rightarrow \Sigma^+ l\bar{\nu}_l$ 
using quenched DWF simulations with the lattice 
size $L^3\times T=16^3 \times 32$.
The spatial extent $La \approx 2.4\;{\rm fm}$ was large enough to
calculate all six form factors describing the beta-decay matrix element 
without a serious finite volume effect on the axial-vector coupling $g_1(0)$. 
From phenomenological point of view, the significance of this subject is twofold: 
(1) to extract the element $V_{us}$ of the Cabibbo-Kobayashi-Maskawa 
mixing matrix from the $\Delta S=1$ decay process, 
and (2) to provide vital information to analysis of 
the strange quark fraction of the proton spin with
the polarized deep inelastic scattering data.
Our particular choice of the $\XZtoSP$ decay process is 
highly sensitive to the flavor $SU(3)$ breaking, since this decay process is 
nothing but the direct analogue of neutron beta decay under the exchange 
of the down quark with the strange quark. The $SU(3)$-breaking pattern 
observed in this study is summarized in Table~\ref{Tab:Summary}.

The vector form factor at zero four-momentum transfer, $f_1(0)$ is protected by
the Ademollo-Gatto theorem against corrections at first order in symmetry breaking.
However, a sign of the second-order correction is somewhat controversial 
among various theoretical studies at present. Our estimate of renormalized
$[f_1(0)]_{\XtoS}$ at the physical point from quenched lattice QCD simulation 
is $0.989(19)$, which indicates that the second-order correction on 
$f_1(0)$ is likely negative. This leads to the closer value of $|V_{us}|$ to the 
value obtained from $K_{l3}$ decays. 
Although both the $1/N_c$ expansion analysis and 
the full one-loop ${\cal O}(p^4)$ calculation in HBChPT favor positive corrections, our observed tendency 
for the $SU(3)$ breaking correction agrees with predictions of quark models 
and CBChPT up to complete ${\cal O}(p^4)$.

The leading correction of the flavor $SU(3)$ breaking to $g_1(0)$ 
starts at first order in symmetry breaking. Although sizable breaking 
corrections, which is the order of 10\% estimated from the mass splitting 
in the octet baryons, is expected, we found relatively small and 
positive correction to $g_1(0)/f_1(0)$ as $[g_1(0)/f_1(0)]_{\XtoS}=1.051(33)
\times [g_1(0)/f_1(0)]_{\ntop}$ in contrast to the model predictions 
where large and negative correction is preferable.   
Unfortunately, the first and single experiment of the $\XZtoSP$ decay 
done by the KTeV Collaboration is not yet sufficiently precise to determine 
either the size, or the sign, of the $SU(3)$ breaking correction to 
$g_1(0)/f_1(0)$.

The advantages of lattice QCD studies of the hyperon beta decay 
are further demonstrated in determinations of the other beta-decay
form factors, while it is difficult to determine each form factor 
separately in experiments. 
Especially, information of the second-class form factors $g_2$ is required 
since linear combinations of $g_1(0)$ and $g_2(0)$ are actually 
measured in the experiments from the Dalitz plot that reflects 
the electron-neutrino angular correlation~\cite{Gaillard:1984ny}.
Furthermore, the nonzero value of the weak electricity form factor
$g_2$ as well as that of the induced scalar form factor $f_3$ is 
the direct evidence of the $SU(3)$ breaking effect in the hyperon beta decays. 
We obtain the ratios of $[g_2(0)/g_1(0)]_{\XtoS}=0.68(18)$ 
and $[f_3(0)/f_1(0)]_{\XtoS}=0.14(9)$. Although both values are roughly comparable to the expected size of the leading first-order corrections 
of the flavor $SU(3)$ breaking, the former is much larger than the latter.

A remarkable observation is that a relatively small first-order correction to 
$g_1(0)$ is accidentally compensated for flavor $SU(3)$-breaking effects 
on $\tilde{g}_1(0)$ with an opposite contribution due to the relatively 
large and positive value of $g_2(0)$. This may suggest why the conventional 
analysis of the hyperon beta decays based on the Cabibbo hypothesis works 
well, though the effects due to the $SU(3)$ breaking observed 
in the octet baryon masses and magnetic moments 
are expected to considerably affect the axial-vector part of 
the weak matrix elements. 

We have also found that the weak magnetism $f_2(0)$ receives positive 
corrections of order 16 \% for the flavor $SU(3)$ breaking, measuring $[f_2(0)/f_1(0)]_{\XtoS}=1.16(10)\times [f_2(0)/f_1(0)]_{\ntop}$.
Our result is {\it not} in agreement with either the generalized CVC hypothesis or 
the Cabibbo-model prediction. On the other hand, as for the induced pseudo-scalar form factor $g_3$, the generalized PPD form, which is extended even 
in $\Delta S=1$ decays under the strong assumption of PCAC,
provides a good prediction of the ratio of $[g_3(0)/g_1(0)]_{\XtoS}$ 
and $[g_3(0)/g_1(0)]_{\ntop}$ at the physical point as $M_K^2(M_\Xi+M_\Sigma)^2/(4M_N^2M_\pi^2)=0.1430$.
This indicates that the large $SU(3)$-breaking effects on $g_3(0)/g_1(0)$
is attributed to the Kaon(pion)-pole structure of the $g_3$ form factor.

In this study, we have focused only on the specific beta-decay process, $\XZtoSP$, 
However, if the flavor $SU(3)$-breaking pattern observed here would commonly 
appear in other beta-decay processes, our results call for non-negligible $SU(3)$-breaking effects in all hyperon beta decays. We also believe that 
the quenched approximation is not problematic for the determination of 
flavor $SU(3)$-breaking effects in the hyperon beta decays 
in similar to what was observed in calculations 
of $K_{l3}$ decays~\cite{{Boyle:2007qe},{Becirevic:2004ya}, {Dawson:2006qc}}.
Nevertheless, the simulation with dynamical 2+1 flavor quarks is an 
important future direction to be explored for full knowledge of the 
$SU(3)$-breaking pattern in the hyperon beta decays. 
Especially, in order to settle the signs of the leading order correction on $f_1(0)$,
$f_2(0)$ and $g_1(0)/f_1(0)$, more extensive lattice study is required. 
We plan to extend the present calculation 
to include other relevant hyperon beta-decay processes such as
$\Sigma^-\rightarrow n$ and $\Lambda \rightarrow p$
using dynamical $N_f=2+1$ flavor DWF lattice
configurations generated by the RBC and UKQCD 
Collaborations~\cite{{Allton:2007hx},{Allton:2008pn}}.
Such planning is now underway~\cite{Progress}

%%%%%%%%%%%%%%  APPENDIX  %%%%%%%%%%%%%%%%%%%%%%%%
\clearpage
\section*{Appendix A: Scalar function $f_S$}

Let us consider the matrix element of the divergence of the vector current:
\bea
\langle b(p')| \partial_{\alpha}V_{\alpha}(0)|B(p)\rangle
&=&
\bar{u}_{b}(p^{\prime})[i(\Pslash-{\Pslash}{}^{\prime})f^{\Btob}_{1}(q^2)
-\frac{q^2}{M_B+M_b}f^{\Btob}_{3}(q^2)]u_{B}(p) \nonumber\\
&=&
\left[(M_b- M_B)f^{\Btob}_{1}(q^2)
-\frac{q^2}{M_B+M_b}f^{\Btob}_{3}(q^2)\right]\bar{u}_{b}(p^{\prime})u_{B}(p).
\eea
Here, we have used the Dirac equation for both initial ($B$) and final ($b$) 
baryon states, 
$(i\Pslash+M_{B})u_{B}(p)=\bar{u}_{b}(p^{\prime})(i\Pslash{}^{\prime}+M_{b})=0$
to get from the first line to the second line. Combined with Eq.~(\ref{Eq:ScalarF}), one finds the following relation
\be
\langle b(p')| \partial_{\alpha}V_{\alpha}(0)|B(p)\rangle
=(M_b - M_B) f_S^{\Btob}(q^2)\bar{u}_{b}(p^{\prime})u_{B}(p),
\label{Eq:VWT}
\ee
where an overall factor $M_b-M_B$ on the right hand side is responsible for
the current conservation when the flavor $SU(3)$ symmetry is exact ($M_b=M_B$).

\section*{Appendix B: Other parametrization of the baryon weak matrix element}
Instead of the standard parametrization of Eqs.~(\ref{Eq:VcMat}) and (\ref{Eq:AxMat}), the following equivalent form~\cite{Linke:1969aa}
is more useful to derive all of Eqs.~(\ref{Eq:3pt_vec_time})-(\ref{Eq:3pt_axial_spatial}), which are considered
at the rest flame of the final ($b$) state (${\bf p}'={\bf 0}$):
%
% eq.
%
\bea
{\cal O}^{V}_{\alpha}(q) 
&=& \gamma_{\alpha} \tilde{f}_1^{\Btob}(q^2) + ip_{\alpha} \frac{\tilde{f}_2^{\Btob}(q^2)}{M_{B}+M_{b}}+iq_{\alpha}\frac{\tilde{f}_3^{\Btob}(q^2)}{M_{B}+M_{b}},
\label{Eq:VcMat2} \\
{\cal O}^{A}_{\alpha}(q)
&=& \gamma_{\alpha}\gamma_5 \tilde{g}_1^{\Btob}(q^2) 
+ ip_{\alpha}\gamma_5\frac{\tilde{g}_2^{\Btob}(q^2)}{M_{B}+M_{b}} 
+iq_{\alpha} \gamma_5 \frac{\tilde{g}_3^{\Btob}(q^2)}{M_{B}+M_{b}}.
\label{Eq:AxMat2}
\eea
The two sets of form factors are connected by
\be
\begin{array}{lcl}
\tilde{f}^{\Btob}_1(q^2)=f^{\Btob}_1(q^2)+f^{\Btob}_2(q^2), & & \tilde{g}^{\Btob}_1(q^2)=g^{\Btob}_1(q^2)-\frac{M_B-M_b}{M_B+M_b} g^{\Btob}_2(q^2), \\
\tilde{f}^{\Btob}_2(q^2)=2f^{\Btob}_2(q^2),  & &\tilde{g}^{\Btob}_2(q^2)=2g^{\Btob}_2(q^2), \\
\tilde{f}^{\Btob}_3(q^2)=f^{\Btob}_3(q^2)-f^{\Btob}_2(q^2), & & \tilde{g}^{\Btob}_3(q^2)=g^{\Btob}_3(q^2)-g^{\Btob}_2(q^2),
\end{array}
\ee
One can easily check above relations using the Gordon identity.

%%%%%%%%%%%%%%%%%%%%%%%%%%%%%%%%%%%%%%%%%%%

%--- acknowledgments ------------------------------------------------  
\begin{acknowledgments}
We would like to thank our colleagues in the RBC collaboration and especially  
T. Blum for helpful suggestions. % and his careful reading of the manuscript.
We also thank RIKEN, Brookhaven National Laboratory and the U.S. DOE
for providing the facilities essential for the completion of this work. The results
of calculations were performed by using of QCDOC at RIKEN BNL Research Center.
S.S. is supported by the JSPS for a Grant-in-Aid for Scientific Research (C)
(No. 19540265). 
T.Y. was supported by the U.S.\ DOE under contract DE-FG02-92ER40716,
and is the Yukawa Fellow supported by Yukawa Memorial Foundation.
\end{acknowledgments}

%--- bibliography ---------------------------------------------------  

%\input{figures}

\end{document}